\documentclass[a4paper,USenglish,cleveref,autoref]{lipics-v2021}

\newif\iflong
\longtrue %

\newcommand{\nsingleshot}{A}
\newcommand{\nproofsync}{B}
\newcommand{\npbftsafety}{C.1}
\newcommand{\npbftliveness}{C.2}
\newcommand{\npbftlatency}{C.3}
\newcommand{\npbftspace}{C.4}
\newcommand{\npbftrotation}{D}
\newcommand{\nhotstuff}{E}
\newcommand{\nbug}{F}

\usepackage{amsmath}
\usepackage{stmaryrd}
\usepackage{xspace}
\usepackage{amssymb}
\usepackage{graphicx}
\usepackage{color}
\usepackage{comment}
\usepackage{wrapfig}
\usepackage{tikz}
\renewcommand{\_}{\texttt{\textunderscore}}

\usepackage{cite}

\newcommand\marginA{.4em} %
\newcommand\marginB{.9em} %

\newtheorem*{proposition*}{Proposition}
\newtheorem*{theorem*}{Theorem}

\theoremstyle{claimstyle}
\newtheorem{myclaim}{Claim}

\usepackage[vlined,linesnumbered,noresetcount]{algorithm2e}
\SetKwBlock{SubAlgoBlock}{}{end}
\newcommand{\SubAlgo}[2]{#1 \SubAlgoBlock{#2}}
\let\oldnl\nl
\newcommand{\nonl}{\renewcommand{\nl}{\let\nl\oldnl}}

\newcommand{\ms}{\\[2pt]}

\def\be{\begin{equation}}
\def\ee{\end{equation}}

\newcommand{\ag}[1]{{{\bf AG:} {\em #1}}}
\newcommand{\mb}[1]{{{\bf MB:} {\em #1}}}
\newcommand{\gc}[1]{{{\bf GC:} {\em #1}}}
\renewcommand{\ag}[1]{}
\renewcommand{\mb}[1]{}
\renewcommand{\gc}[1]{}

\makeatletter
\newcommand{\removelatexerror}{\let\@latex@error\@gobble}
\makeatother

\SetKw{Trigger}{trigger}
\SetKw{WhenReceived}{when received}
\SetKw{WhenAccepted}{when accepted}
\SetKw{WhenDelivered}{when delivered}
\SetKw{KwTo}{to}
\SetKw{Send}{send}
\SetKw{Broadcast}{broadcast}
\SetKw{KwFrom}{from}
\SetKw{Upon}{upon}
\SetKw{Fun}{function}
\SetKw{WhenTimerFires}{when the timer expires}

\newcommand{\pbft}{PBFT-light\xspace}
\newcommand{\pbftr}{PBFT-rotation\xspace}
\newcommand{\hotstuff}{HotStuff-light\xspace}

\newcommand{\DECISION}{{\tt DECISION}}

\newcommand{\NEWLEADER}{{\tt NEW\_LEADER}}

\newcommand{\NEWVIEW}{{\tt NEW\_STATE}}

\newcommand{\BROADCAST}{{\tt BROADCAST}}
\newcommand{\FORWARD}{{\tt FORWARD}}

\newcommand{\PREPARE}{{\tt PREPARE}}

\newcommand{\PREPREPARE}{{\tt PREPREPARE}}

\newcommand{\COMMIT}{{\tt COMMIT}}
\newcommand{\WISH}{{\tt WISH}}

\newcommand{\PRECOMMIT}{{\tt PRECOMMIT}}

\newcommand{\wf}{{\sf wf}}

\newcommand{\batch}{B}
\newcommand{\queue}{{\sf queue}}
\newcommand{\krecovery}{{\sf init\_log\_length}}
\newcommand{\append}{{\tt append}}

\newcommand{\remove}{{\tt remove}}
\newcommand{\advanced}{{\sf advanced}}

\newcommand{\nextv}{{\sf next}}
\newcommand{\currview}{{\sf curr\_view}}

\newcommand{\correct}{\mathcal{C}}
\newcommand{\status}{{\sf status}}

\newcommand{\prepcmd}{{\sf prep\_log}}
\newcommand{\comcmd}{{\sf commit\_log}}
\newcommand{\vprepcmd}{\mathit{prep\_log}}
\newcommand{\vcmd}{\mathit{log}}

\newcommand{\phase}{{\sf phase}}

\newcommand{\prepview}{{\sf prep\_view}}
\newcommand{\lockview}{{\sf lock\_view}}
\newcommand{\vprepview}{\mathit{prep\_view}}

\newcommand{\noop}{{\tt nop}}

\newcommand{\leader}{{\sf leader}}

\newcommand{\cert}{{\sf cert}}

\newcommand{\vcert}{\mathit{cert}}

\newcommand{\valid}{{\sf valid}}
\newcommand{\validity}{{\sf valid}}

\newcommand{\ValidNewLeader}{{\sf ValidNewLeader}}

\newcommand{\ValidNewState}{{\sf ValidNewState}}

\newcommand{\quorum}{{\sf quorum}}
\newcommand{\accepted}{{\sf prepared}}

\newcommand{\lastdelivered}{{\sf last\_delivered}}

\newcommand{\hash}{{\sf hash}}

\newcommand{\view}{{\sf view}}
\newcommand{\viewp}{{\sf view}^+}
\newcommand{\prevv}{\textit{prev\_v}}
\newcommand{\prevvp}{\textit{prev\_v}^+}

\newcommand{\te}[2]{E_{#1}(#2)}
\newcommand{\tm}[1]{E_{\rm first}(#1)}

\newcommand{\tl}[1]{E_{\rm last}(#1)}

\newcommand{\ta}[2]{A_{#1}(#2)}
\newcommand{\sse}[2]{\mathbb{E}_{#1}(#2)}
\newcommand{\ssm}[1]{\mathbb{E}_{\rm first}(#1)}
\newcommand{\ssl}[1]{\mathbb{E}_{\rm last}(#1)}

\newcommand{\tam}[1]{A_{\rm first}(#1)}
\newcommand{\talast}[1]{A_{\rm last}(#1)}

\newcommand{\tae}[2]{T_{#1}(#2)}

\newcommand{\taelast}[1]{T_{\rm last}(#1)}

\newcommand{\B}{\mathcal{V}}

\newcommand{\GST}{{\sf GST}}
\newcommand{\GSTP}{\ensuremath{\overline{{\sf GST}}}}

\newcommand{\all}{\ {\bf all}}

\newcommand{\timerproposal}{{\sf timer\_proposal}}
\newcommand{\timerexecute}{{\sf timer\_delivery}}
\newcommand{\timerrecovery}{{\sf timer\_recovery}}
\newcommand{\timerbroadcast}{{\sf timer\_broadcast}}
\newcommand{\durationrecovery}{{\sf dur\_recovery}}

\newcommand{\durationexecution}{{\sf dur\_delivery}}
\newcommand{\timer}{{\sf timer}}
\newcommand{\timerview}{{\sf timer\_view}}

\newcommand{\padvance}{{\tt advance}}

\newcommand{\newview}{{\tt new\_view}}
\newcommand{\newconsview}{{\tt new\_consensus\_view}}
\newcommand{\View}{{\sf View}}
\newcommand{\Time}{{\sf Time}}
\newcommand{\lastViews}{{\sf max\_views}}

\newcommand{\GV}[1]{{\sf GV}(#1)}
\newcommand{\LV}[2]{{\sf LV}_{#1}(#2)}

\newcommand{\cmd}{{\sf log}}

\newcommand{\BLOCKED}{\textsc{advanced}}
\newcommand{\NORMAL}{\textsc{normal}}
\newcommand{\RECOVERING}{\textsc{initializing}}

\newcommand{\START}{\textsc{start}}
\newcommand{\PREPREPARED}{\textsc{preprepared}}

\newcommand{\COMMITTED}{\textsc{committed}}

\newcommand{\PREPARED}{\textsc{prepared}}
\newcommand{\PRECOMMITTED}{\textsc{precommitted}}

\newcommand{\TRUE}{\text{\sc true}}
\newcommand{\FALSE}{\text{\sc false}}

\newcommand{\committed}{\ensuremath{{\sf committed}}}

\newcommand{\val}{x}

\newcommand{\starttimer}{\ensuremath{{\tt start\_timer}}}
\newcommand{\stoptimer}{\ensuremath{{\tt stop\_timer}}}
\newcommand{\timeout}{\ensuremath{F}}

\newcommand{\fdef}{\mathpunct{\downarrow}}
\newcommand{\fundef}{\mathpunct{\uparrow}}

\newcommand{\tr}[2]{\iflong{}\S#1\else{}\cite[\S#2]{ext}\fi}

\newcommand{\monotonicity}{Monotonicity\xspace}
\newcommand{\validitysync}{Validity\xspace}
\newcommand{\entry}{Bounded Entry\xspace}
\newcommand{\startup}{Startup\xspace}
\newcommand{\progress}{Progress\xspace}

\makeatletter
\newcommand\myparagraph{\@startsection{subparagraph}{5}{\z@}%
                                       {1.5ex \@plus1ex \@minus .2ex}%
                                       {-1em}%
                                      {\sffamily\normalsize\bfseries}}
\makeatother

\title{Liveness and Latency of\\ Byzantine State-Machine Replication}

\author{Manuel Bravo}{Informal Systems, Madrid, Spain}{}{}{}
\author{Gregory Chockler}{University of Surrey, UK}{}{}{}
\author{Alexey Gotsman}{IMDEA Software Institute, Madrid, Spain}{}{}{}

\EventEditors{Christian Scheideler}
\EventNoEds{1}
\EventLongTitle{36th International Symposium on Distributed Computing (DISC 2022)}
\EventShortTitle{DISC 2022}
\EventAcronym{DISC}
\EventYear{2022}
\EventDate{October 25--27, 2022}
\EventLocation{Augusta, Georgia, USA}
\EventLogo{}
\SeriesVolume{246}
\ArticleNo{8}

\titlerunning{Liveness and Latency of Byzantine State-Machine Replication}
\authorrunning{M. Bravo, G. Chockler, and A. Gotsman}

\Copyright{Manuel Bravo, Gregory Chockler, and Alexey Gotsman}



\funding{This work was partially supported by an ERC Starting Grant RACCOON
and by a research grant from Nomadic Labs and the Tezos Foundation.}

\acknowledgements{We thank the following people for comments that helped improve
  the paper:
L\u{a}cr\u{a}mioara A\c{s}tef\u{a}noaei,
Hagit Attiya,
Alysson Besani, 
Armando Castañeda,
Peter Davies,
Dan O'Keeffe,
Idit Keidar,
Giuliano Losa,
Alejandro Naser, and 
Eugen Z\u{a}linescu.
}

\nolinenumbers

\begin{document}

\hideLIPIcs

\maketitle

\bigskip

\begin{abstract} 
  Byzantine state-machine replication (SMR) ensures the consistency of
  replicated state in the presence of malicious replicas and lies at the heart
  of the modern blockchain technology. Byzantine SMR protocols often guarantee
  safety under all circumstances and liveness only under synchrony. However,
  guaranteeing liveness even under this assumption is nontrivial. So far we have
  lacked systematic ways of incorporating liveness mechanisms into Byzantine SMR
  protocols, which often led to subtle bugs. To close this gap, we introduce a
  modular framework to facilitate the design of provably live and efficient
  Byzantine SMR protocols. Our framework relies on a {\em view} abstraction
  generated by a special {\em SMR synchronizer} primitive to drive the agreement
  on command ordering. We present a simple formal specification of an SMR
  synchronizer and its bounded-space implementation under partial synchrony. We
  also apply our specification to prove liveness and analyze the latency of
  three Byzantine SMR protocols via a uniform methodology. In particular, one of
  these results yields what we believe is the first rigorous liveness proof for
  the algorithmic core of the seminal PBFT protocol.

\end{abstract}

\bigskip

\section{Introduction}

{\em Byzantine state-machine replication (SMR)}~\cite{smr} ensures the
consistency of replicated state even when some %
of the replicas are malicious.
It lies at the heart of the modern blockchain technology and is closely related
to the classical Byzantine consensus problem.
Unfortunately, no deterministic protocol can guarantee both safety and liveness
of Byzantine SMR when the network is asynchronous~\cite{flp}. A common way to
circumvent this while maintaining determinism
is to guarantee safety under all circumstances and liveness only under
synchrony.
This is formalized by the {\em partial
  synchrony} model~\cite{dls,CT96},
which stipulates that after some unknown {\em Global Stabilization Time} (\GST)
the system becomes synchronous, with message delays bounded by an unknown
constant $\delta$ and process clocks tracking real time. Before $\GST$ %
messages can be lost or %
delayed, and clocks at different processes can drift apart.

Historically, researchers have paid more attention to safety of Byzantine SMR
protocols than their liveness. For example, while the seminal PBFT protocol came
with a detailed safety proof~\cite[\S{}A]{castro-thesis}, the nontrivial
mechanisms ensuring its liveness were only given a brief informal
justification~\cite[\S{}4.5.1]{castro-tocs}, which did not cover their most
critical properties. %
However, ensuring liveness under partial synchrony is far from trivial, as
illustrated by the many liveness bugs found in existing
protocols~\cite{wild,casper-bug,alysson-reads,tendermint-opodis,zyzzyva-bug}. In
particular, classical failure detectors and leader
oracles~\cite{CT96,petr-survey} are of little help: while they have been widely
used under benign failures~\cite{R00,MR99,GR04}, their implementations under
Byzantine failures are either impractical~\cite{kihlstrom} or detect only
restricted failure types~\cite{MR97,modular-smr,HK09}. As an
alternative, a textbook by Cachin et al.~\cite{cachin-book} proposed a leader
oracle-like abstraction that accepts hints from the application to identify
potentially faulty processes. However, as we explain in \S\ref{sec:related}
and~\tr{\ref{sec:bug}}{\nbug},
their specification of the abstraction is impossible to implement, and in fact,
the consensus algorithm constructed using it in~\cite{cachin-book} also suffers
from a liveness bug.

Recent work on ensuring liveness %
has departed from failure detectors and instead revisited the approach of the
original DLS paper~\cite{dls}. This exploits the common structure of Byzantine
consensus and SMR protocols under partial synchrony: such protocols usually
divide their execution into {\em views}, each with a designated leader process
that coordinates the protocol execution. If the leader is faulty, the processes
switch to another view with a different leader.
To ensure liveness, an SMR protocol needs to spend sufficient time in views that
are entered by all correct processes and where the leader correctly follows the
protocol. The challenge of achieving such {\em view synchronization} is that,
before $\GST$, clocks can diverge and messages that could be used to synchronize
processes can get lost or delayed; even after $\GST$, Byzantine processes may
try to disrupt attempts to bring everybody into the same view.  {\em View
  synchronizers}~\cite{NK20,bftlive,hotstuff,lumiere} encapsulate mechanisms for
dealing with this challenge, allowing them to be reused across protocols.

View synchronizers have been mostly explored in the context of (single-shot)
Byzantine consensus. In this case a synchronizer can just switch processes
through an infinite series of views, %
so that eventually there is a view with a correct leader that is long enough to
reach a decision~\cite{NK20,bftlive}. However, using such a synchronizer for SMR
results in suboptimal solutions. For example, one approach is to use the
classical SMR construction where each command is decided using a separate
black-box consensus instance~\cite{smr}, implemented using a view
synchronizer. However, this would force the processes in every instance to
iterate over the same sequence of potentially bad views until the one with a
correct leader and sufficiently long duration could be reached.  As we discuss
in~\S\ref{sec:related}, other approaches for implementing SMR based on this type
of synchronizers also suffer from drawbacks.

To minimize the overheads of view synchronization, 
instead of automatically switching processes through views based on a fixed
schedule, %
implementations such as PBFT allow processes to stay in the same view for as
long as they are happy with its %
performance. The processes can then reuse a single view to decide multiple
commands, usually with the same leader.
To be useful for such SMR protocols, a synchronizer needs to allow the processes
to control when they want to switch views via a special $\padvance$ call. We
call such a primitive an {\em SMR synchronizer}, to distinguish it from the less
flexible {\em consensus synchronizer} introduced above. This kind of
synchronizers was first introduced in~\cite{NK20,lumiere}, but only used as an
intermediate module to implement a consensus synchronizer.

In this paper we show that SMR synchronizers can be {\em directly} exploited to
construct efficient and provably live SMR protocols and 
develop a general blueprint that enables such constructions. In more detail:
\begin{itemize}
\item We propose a formal specification of an SMR synchronizer
  (\S\ref{sec:sync}), which is simpler and 
  more general than prior proposals~\cite{lumiere,NK20}.
  It is also strictly stronger than the consensus synchronizer
  of~\cite{bftlive}, which can be obtained from the SMR synchronizer at no extra
  cost.  Informally, our specification guarantees that {\em (a)} the system will
  move to a new view if enough correct processes call $\padvance$, and {\em (b)}
  all correct processes will enter the new view, provided that for long enough,
  no correct process that enters this view asks to leave it. These properties
  enable correct processes to iterate through views in search of a well-behaved
  leader, and to synchronize in a view they are happy with.
  
\item We give an SMR synchronizer implementation and prove that it satisfies our
  specification (\S\ref{sec:sync-impl}). Unlike prior
  implementations~\cite{NK20}, ours tolerates message loss before $\GST$ while
  using only bounded space; in practice, this feature is essential to defend
  against denial-of-service attacks. We also provide a precise latency analysis
  of our synchronizer, quantifying how quickly all correct processes enter the
  next view after enough of them call $\padvance$.
  
\item We demonstrate the usability of our synchronizer specification by applying
  it to construct and prove the correctness of several SMR protocols. First, we
  prove the liveness of a variant of PBFT using an SMR synchronizer
  (\S\ref{sec:pbft}-\ref{sec:liveness}): to the best of our knowledge, this is
  the first rigorous proof of liveness for PBFT's algorithmic core. The proof
  establishes a strong liveness guarantee that implies censorship-resistance:
  every command submitted by a correct process will be executed. The use of the
  synchronizer specification in the proof allows us to abstract from
  view synchronization mechanics and %
  focus on protocol-specific reasoning. This reasoning is done using a reusable
  methodology based on showing that the use of timers in the SMR protocol and
  the synchronizer together establish properties similar to those of failure
  detectors. The methodology also handles the realistic ways in which protocols
  such as PBFT adapt their timeouts to the unknown message delay $\delta$. We
  demonstrate the generality of our methodology by also applying it to a version
  of PBFT with periodic
  leader changes~\cite{aardvark,spinning,mirbft} and a HotStuff-like
  protocol~\cite{hotstuff} (\S\ref{sec:xbft}).

\item We exploit the latency bounds for our synchronizer to establish both
  bad-case and good-case bounds for variants of PBFT implemented on top of it
  (\S\ref{sec:latency}). Our bad-case bound assumes that the protocol starts
  before $\GST$; it shows that after $\GST$ all correct processes synchronize in
  the same view within a bounded time. This time is proportional to a
  conservatively chosen constant $\Delta$ that bounds post-$\GST$ message delays
  in all executions~\cite{pass-shi,HK89}.
  Our good-case bound quantifies decision latency when the protocol starts after
  $\GST$ and matches the lower bound of~\cite{ittai-good-case}.

\end{itemize}

\section{System Model}
\label{sec:model}

We consider a system of $n = 3f+1$ processes. At most $f$ of these can be
Byzantine (aka {\em faulty}), i.e., can behave arbitrarily. The rest of the
processes are {\em correct} and we denote their set by $\correct$. We call a set
$Q$ of $2f+1$ processes a {\em quorum} and write $\quorum(Q)$. %
We assume standard cryptographic primitives~\cite[\S{}2.3]{cachin-book}:
processes can communicate via authenticated point-to-point links, sign messages
using digital signatures, and use a collision-resistant hash function
$\hash()$. We denote by $\langle m \rangle_i$ a message $m$ signed by process
$p_i$.

We consider a {\em partial synchrony} model~\cite{dls,CT96}: for each execution
of the protocol, there exist a time $\GST$ and a duration $\delta$ such that
after $\GST$ message delays between correct processes are bounded by $\delta$;
before $\GST$ messages can get arbitrarily delayed or lost.
As in~\cite{CT96}, we assume that the values of $\GST$ and $\delta$ are unknown
to the protocol. This reflects the requirements of practical systems, whose
designers cannot accurately predict when network problems leading to asynchrony
will stop and what the latency will be during the following synchronous period.
We also assume that processes have %
hardware clocks that can drift unboundedly from real time before $\GST$, but do
not drift thereafter.

\section{SMR Synchronizer Specification and Implementation}
\label{sec:sync}

\begin{figure}[t]
\small
\setlength{\leftmargini}{12pt}
\renewcommand{\labelenumi}{\theenumi.}
\begin{enumerate}
\item\label{sync:local-order}
{\bf Monotonicity.} A process enters increasing views:\\[1pt]
  $\forall i, v, v'.\, {\te{i}{v}\fdef} \,\wedge\,
  {\te{i}{v'}\fdef} \,{\implies}\, (v < v' \,{\iff}\, \te{i}{v} < \te{i}{v'})$

\item\label{sync:enter-attempt}
  {\bf Validity.} A process only enters 
  $v+1$ if some correct process
  has attempted to advance from $v$:\\[1pt]
  $\forall i, v.\, {\te{i}{v+1}\fdef} \,{\implies}\, {\tam{v}\fdef} \,\wedge\,
  \tam{v} < \te{i}{v+1}$

\item\label{sync:2delta} {\bf Bounded Entry.}  For some $\B$ and
  $d$, %
  if a process enters a view $v\ge\B$ and no process attempts to advance to a
  higher view within time $d$,
  then all correct processes will enter $v$ within $d$:\\[1pt]
  \mbox{$\exists \B, d.\hspace{1pt} \forall v \hspace{1pt}{\ge}\hspace{1pt} \B.\hspace{1pt} 
  {\tm{v}\fdef} \wedge \neg(\tam{v} \hspace{1pt}{<}\hspace{1pt} 
  \tm{v} \hspace{1pt}{+}\hspace{1pt}  d) {\implies} (\forall p_i \in \correct. \hspace{1pt} {\te{i}{v}\fdef})
  \wedge \tl{v} \le \tm{v} \hspace{1pt}{+}\hspace{1pt}  d$}

\item\label{sync:enter1}
  {\bf Startup.} Some correct process will enter view $1$ if $f+1$ processes call $\padvance$:\\[1pt]
  $(\exists P \subseteq \correct.\, |P| = f+1 \,\wedge\, (\forall p_i \in P.\,
  {\ta{i}{0}\fdef})) \,{\implies}\, {\tm{1}\fdef}$

\item\label{sync:liveness} {\bf Progress.}  If a correct process enters a view
  $v$ and, for some set $P$ of $f+1$ correct processes, any process in $P$ that
  enters $v$ eventually calls $\padvance$, then some correct process will
  enter
 $v+1$:\\[1pt]
  $\forall v.\, {\tm{v}\fdef} \,\wedge\, (\exists P \subseteq \correct.\, |P| =
  f+1 \,\wedge\, (\forall p_i \in P.\, {\te{i}{v}\fdef} \,{\implies}\,
  {\ta{i}{v}\fdef})) \,{\implies}\, {\tm{v+1}\fdef}$
\end{enumerate}

\caption{SMR synchronizer specification.}
\label{fig:multi-sync-properties}
\end{figure}

We consider a synchronizer interface defined in~\cite{lumiere,NK20}, which here
we call an {\em SMR synchronizer}. Let $\View = \{1, 2, \ldots\}$ be the set of
{\em views}, ranged over by $v$; we use $0$ to denote an invalid initial view.
The synchronizer produces notifications $\newview(v)$ at a process, telling it
to {\em enter} view $v$. To trigger these, %
the synchronizer allows a process to call a function $\padvance()$, which
signals that the process wishes to {\em advance} to a higher view. 
We assume that a correct process does not call $\padvance$ twice
without an intervening $\newview$ notification.

Our first contribution is the SMR synchronizer specification in
Figure~\ref{fig:multi-sync-properties}, which is simpler and more general than
prior proposals~\cite{lumiere,NK20} (see \S\ref{sec:related} for a
discussion). The specification relies on the following notation. Given a view
$v$ entered by a correct process $p_i$, we denote by $\te{i}{v}$ the time when
this happens; we let $\tm{v}$ and $\tl{v}$ denote respectively the earliest and
the latest time when some correct process enters $v$. %
We denote by $\ta{i}{v}$ the time when a correct process $p_i$ calls $\padvance$
while in %
$v$, and let $\tam{v}$ and $\talast{v}$ denote respectively the earliest and the
latest time when this happens.
Given a partial function $f$, we write $f(x)\fdef$ if $f(x)$ is defined, and
$f(x)\fundef$ if $f(x)$ is undefined. 

The \monotonicity property in Figure~\ref{fig:multi-sync-properties} ensures
that views can only increase at a given process. \validitysync ensures that a
process may only enter a view $v+1$ if some correct process has called
$\padvance$ in $v$. This prevents faulty processes from disrupting the system by
forcing view changes. As a corollary of \validitysync we can prove that, if a
view $v'$ is entered by some correct process, then so are all the views $v$
preceding $v'$.
\begin{proposition}\label{sync:noskip}
  $\forall v, v'.\, 0 < v < v' \wedge {\tm{v'}\fdef} {\implies} {\tm{v}\fdef}
  \wedge \tm{v} < \tm{v'}$.
\end{proposition}
\begin{proof}
Fix $v' \ge 2$ and assume that a correct process enters $v'$, so that
$\tm{v'}\fdef$. We prove by induction on $k$ that
$\forall k=0..(v'-1).\, {\tm{v'-k}\fdef} \wedge \tm{v'-k} \le \tm{v'}$. The base
case of $k=0$ is trivial. For the inductive step, assume that the required holds
for some $k$. Then by \validitysync there exists a time
$t < \tm{v'-k}$ at which some correct process $p_j$ attempts to advance from
$v'-k-1$. But then $p_j$'s view at $t$ is $v'-k-1$. Hence, $p_j$ enters $v'-k-1$
before $t$, so that $\tm{v'-k-1} < t < \tm{v'-k} \le \tm{v'}$, as required.
\end{proof}

\entry ensures that, if some process enters view $v$, then all correct
processes will do so within at most $d$ time units of each other ($d = 2\delta$
for our %
implementation). This only holds if within $d$ no process attempts to advance to
a higher view, as this may make some processes skip $v$ and enter a higher view
directly.  \entry also holds only starting from some view $\B$, since we
may not be able to guarantee it for views entered before $\GST$.

\begin{figure}[t]
\small
\centering 
\begin{minipage}{7.5cm}
\begin{algorithm}[H]
  \setcounter{AlgoLine}{0}

  \SubAlgo{\textbf{when the process starts or \timer\ expires}}{\label{line3:timer-exp1}  
    \padvance();
  }

  \smallskip

  \SubAlgo{{\bf upon} $\newview(v)$}{
      $\stoptimer(\timer)$\; \label{line3:timer-stop}
      $\starttimer(\timer, \tau)$\;\label{line3:start-timer}
    }
\end{algorithm}
\end{minipage}
\caption{A simple client of the SMR synchronizer.}
\label{fig:toy}
\end{figure}
\startup ensures that some correct process enters view $1$ if $f+1$ processes
call $\padvance$. Given a view $v$ entered by a correct process, \progress
determines conditions under which some correct process will enter the next view
$v+1$. This will happen if for some set $P$ of $f+1$ correct processes, any
process in $P$ entering $v$ eventually calls $\padvance$. Note that even a
single $\padvance$ call at a correct process {\em may} lead to a view switch
(reflecting the fact that in implementations faulty processes may help this
correct process). \startup and \progress ensure that the synchronizer {\em must}
switch if at least $f+1$ correct processes ask for this. We now illustrate a
typical pattern of their use, which we later apply to PBFT
(\S\ref{sec:liveness}). To this end, we consider a simple client in
Figure~\ref{fig:toy}, where in each view a process sets a timer for a fixed
duration $\tau$ and calls $\padvance$ when the timer expires. Using \startup and
\progress we prove that this client keeps switching views forever as follows.
\begin{proposition}
  In any execution of the client in Figure~\ref{fig:toy}:
  $\forall v.\, \exists v'.\, v'>v \wedge {\tm{v'}\fdef}$.
\label{lem:live-toy}
\end{proposition}
\begin{proof}
Since all correct processes initially call $\padvance$, by
\startup some correct process eventually enters view $1$. Assume now that the
proposition is false, so that there is a maximal view $v$ entered by any correct
process. Let $P$ be any set of $f+1$ correct processes and consider an arbitrary
process $p_i \in P$ that enters $v$. When this happens, $p_i$ sets the $\timer$
for the duration $\tau$. The process then either calls $\padvance$ when $\timer$
expires, or enters a new view $v'$ before this. In the latter case $v' > v$ by
\monotonicity, which is impossible. Hence, $p_i$ eventually calls
$\padvance$ while in $v$. Since $p_i$ was chosen arbitrarily, 
$\forall p_i \in P.\, {\te{i}{v}\fdef} {\implies} {\ta{i}{v}\fdef}$. Then by
\progress we get $\tm{v+1}\fdef$: 
a contradiction.
\end{proof}

Similarly to Figure~\ref{fig:toy}, we can use an SMR synchronizer satisfying the
properties in Figure~\ref{fig:multi-sync-properties} to implement a {\em
  consensus synchronizer}~\cite{bftlive,NK20} without extra overhead. This lacks
an $\padvance$ call and provides only the $\newview$ notification, which it
keeps invoking at increasing intervals so that eventually the there is a view
long enough for the consensus protocol running on top to decide. We obtain a
consensus synchronizer if in Figure~\ref{fig:toy} we propagate the $\newview$
notification to the consensus protocol and set the $\timer$ to an unboundedly
increasing function of views instead of a constant $\tau$. In
\tr{\ref{sec:single-shot}}{\nsingleshot} we show that the resulting consensus
synchronizer satisfies the specification proposed in~\cite{bftlive}.

\subsection{A Bounded-Space SMR Synchronizer}
\label{sec:sync-impl}

We now present a bounded-space algorithm that implements the %
specification in Figure~\ref{fig:multi-sync-properties} under partial synchrony
for $d=2\delta$.
Our implementation reuses algorithmic techniques from a consensus synchronizer
of Bravo et al.~\cite{bftlive}. However, it supports a more general abstraction,
and thus requires a more intricate correctness proof and latency analysis
(\S\ref{sec:sync-bounds}).

\begin{figure}[t]
\vspace{3pt}
\begin{tabular}{@{\!\!\!}l@{\!\!\!}|@{\ \ }l@{}}
\scalebox{0.96}{%
\begin{minipage}[t]{6.8cm}
\removelatexerror
\vspace*{-12pt}
 \begin{algorithm*}[H]
  \setcounter{AlgoLine}{0}

  \SubAlgo{\Fun{} \padvance()}{\label{line:timer-exp1}  
    \Send $\WISH(\max(\view+1, \viewp))$ \quad \KwTo \all\;\label{line:send2}
    $\advanced \leftarrow \TRUE$\; \label{line:advance-true}
  }

  \smallskip
  \smallskip

 \SubAlgo{\textbf{periodically}\hfill $\triangleright$ every $\rho$ time units}{\label{line:retransmit-start} 
    \uIf{$\advanced$}{\label{line:start-guard}
      \Send $\WISH(\max(\view+1, \viewp))$ \KwTo \all\;\label{line:send4}
    }
    \ElseIf{$\viewp > 0$}{
      \Send $\WISH(\viewp)$ \KwTo \all\;\label{line:send3}
    }
  }
\end{algorithm*}
\vspace*{-3pt}
\end{minipage}}
&
\scalebox{0.95}{%
\begin{minipage}[t]{9cm}
\removelatexerror
\vspace*{-12pt}
\begin{algorithm*}[H]

  \SubAlgo{\WhenReceived $\WISH(v)$ {\bf from $p_j$}}{
    $\prevv, \prevv^+ \leftarrow \view, \viewp$\;
    \lIf{$v > \lastViews[j]$}{$\lastViews[j] \leftarrow v$\label{line:update-maxviews}}
    $\view \ \ \leftarrow \max\{v \mid \exists k.\, \lastViews[k] = v \wedge{}$\label{line:update-view}\\
    \nonl\hspace{2cm}$|\{j \mid \lastViews[j] \ge v\}| \ge 2f+1\}$\;
    $\viewp  \leftarrow \max\{v \mid \exists k.\, \lastViews[k] = v \wedge {}$\label{line:update-viewp}\\
    \nonl\hspace{2cm}$|\{j \mid \lastViews[j] \ge v\}| \ge f+1\}$\; 
    \If{$\viewp = \view \wedge \view > \prevv$}{\label{line:enter-condition}
      \Trigger $\newview(\view)$\; \label{line:trigger-newview}
      $\advanced \leftarrow \FALSE$\; \label{line:advance-false}
    }
    \If{$\viewp > \prevv^+$\label{line:cond-send5}}
    {\Send $\WISH(\viewp)$ \KwTo \all\label{line:send5}}
  }
\end{algorithm*}
\vspace*{-3pt}
\end{minipage}}
\end{tabular}
\vspace*{-3pt}
\caption{A bounded-space SMR synchronizer. All counters are initially $0$.}
\label{fig:sync}
\end{figure}

When a process calls $\padvance$ (line~\ref{line:timer-exp1}), the synchronizer
does not immediately move to the next view $v'$, but disseminates a $\WISH(v')$
message announcing its intention. A process enters a new view once it
accumulates a sufficient number of $\WISH$ messages supporting this. A naive
synchronizer design could follow %
Bracha broadcast~\cite{random-bracha}: enter a view $v'$ upon receiving $2f+1$
$\WISH(v')$ messages, and echo $\WISH(v')$ upon receiving $f+1$ copies thereof;
the latter is needed to combat equivocation by Byzantine processes. However, in
this case the process would have to track all newly proposed views for which
$< 2f+1$ $\WISH$es have been received.
Since messages sent before $\GST$ can be lost or delayed, this would require
unbounded space. To reduce the space complexity, in our algorithm a process only
remembers the highest view received from each process, kept in an array
$\lastViews$
(line~\ref{line:update-maxviews}). Variables $\view$ and $\viewp$ respectively
hold the $(2f\,{+}\,1)$st highest and the $(f\,{+}\,1)$st highest views in
$\lastViews$ (lines~\ref{line:update-view}-\ref{line:update-viewp}).
These variables never decrease and always satisfy $\view \le \viewp$.

The process enters the view stored in
$\view$
when this variable increases
(line~\ref{line:trigger-newview}).
A process\linebreak thus enters %
$v$ only if it receives $2f+1$ $\WISH$es for views $\ge v$,
and a process may be forced to switch views even if it did not call $\padvance$;
the latter helps lagging processes to catch up.
The variable $\viewp$ increases when the process receives $f+1$ $\WISH$es for
views $\ge\viewp$, and thus some correct process wishes to enter a new view
$\ge\viewp$. In this case we %
echo $\viewp$ (line~\ref{line:send5}), to help other processes switch views and
satisfy Bounded Entry and Progress.

The guard $\viewp = \view$ in line~\ref{line:enter-condition} ensures that a
process does not enter a ``stale'' view such that another correct process
already wishes to enter a higher one. Similarly, when the process calls
$\padvance$, it sends a $\WISH$ for the maximum of $\view+1$ and $\viewp$
(line~\ref{line:send2}). Thus, if $\view = \viewp$, so that the values of the
two variables have not changed since the process entered the current view, then
the process sends a $\WISH$ for the next view ($\view+1$). Otherwise,
$\view < \viewp$, and the process sends a $\WISH$ for the higher view $\viewp$.
Finally, to deal with message loss before $\GST$, a process retransmits the
highest $\WISH$ it sent every $\rho$ time units, according to its local clock
(line~\ref{line:retransmit-start}). Depending on whether the process has called
$\padvance$ in the current view (tracked by
\advanced),
the $\WISH$
is computed as in lines~\ref{line:send5} or~\ref{line:send2}.

Our SMR synchronizer requires only $O(n)$ variables for storing
views. Proposition~\ref{sync:noskip} also ensures that views entered by correct
processes %
do not skip values, which limits the power of the adversary to exhaust their
allocated space (similarly to~\cite{BazziD04}).

\subsection{SMR Synchronizer Correctness and Latency Bounds}
\label{sec:sync-bounds}

The following theorem (proved in \tr{\ref{sec:proof-sync}}{\nproofsync})
states the correctness of our synchronizer as well as and its performance
properties. In \S\ref{sec:latency} we apply the latter to bound the latency of
Byzantine SMR protocols.
Given a view $v$ that was entered by a correct process $p_i$, we let
$\tae{i}{v}$ denote the time at which $p_i$ either attempts to advance from $v$
or enters a view $>v$; we let $\taelast{v}$ denote the latest time when a
correct process does so.
We assume that every correct process eventually attempts to advance
from view $0$ unless it enters a view $>0$, i.e., 
$\forall p_i \in \correct.\, \tae{i}{0}\fdef$.
\begin{theorem}
\label{thm:bounds}
  Consider an execution with an eventual message delay $\delta$. 
  The algorithm in Figure~\ref{fig:sync} satisfies the properties in
  Figure~\ref{fig:multi-sync-properties} for $d=2\delta$ and
$\B = \max\{v \mid ({\tm{v}\fdef} \,\wedge\, \tm{v} < \GST + \rho) \vee v=0\} + 1$
if $\tam{0} < \GST$, and $\B = 1$, otherwise.
Furthermore:

{\renewcommand{\theenumi}{\Alph{enumi}}
\renewcommand{\labelenumi}{\theenumi.}
\begin{enumerate}
\item \label{eq:gen-bounded-entry:main}
  $\forall v.\, {\tm{v}\fdef} \,\wedge\, \tam{0} < \GST \,{\implies}\,
  \tl{v} \le \max(\tm{v}, \GST+\rho) + 2\delta$.

\vspace{5pt}

\item \label{eq:lat-bound1:main}
 $\forall v.\, {\tm{v+1}\fdef} \,{\implies}\,
\tl{v+1} \le
\begin{cases}
    \max(\taelast{v}, \GST + \rho) + \delta,& \text{if } \tam{0} < \GST;\\[-1pt]
    \taelast{v} + \delta,& \text{otherwise.}
\end{cases}
$
\end{enumerate}
} 
\label{thm:smr-sync-correct}
\end{theorem}

\vspace{1pt}

The theorem gives a witness for $\B$ in \entry: it is the
next view after the highest one entered by a correct process at or before
$\GST+\rho$ (or $1$ if no view was entered).
Property~\ref{eq:gen-bounded-entry:main} bounds the latest time any correct
process can enter a view that has been previously entered by a correct
process. It is similar to \entry, but also handles views $< \B$.
Property~\ref{eq:lat-bound1:main} refines \progress: while the latter guarantees
that the synchronizer will enter $v+1$ if enough processes ask for this, the
former bounds the time by which this will happen.

\section{PBFT Using an SMR Synchronizer}
\label{sec:pbft}

We now demonstrate how an SMR synchronizer can be used to implement Byzantine
SMR. More formally, we implement {\em Byzantine atomic
  broadcast}~\cite{cachin-book}, from which SMR can be implemented in the
standard way~\cite{smr}. This %
allows processes to broadcast {\em values}, and we assume an
application-specific predicate to indicate whether a value is
valid~\cite{cachin-crypto01} (e.g., a block in a blockchain is invalid if
it lacks correct signatures).
We assume that all values broadcast by
correct processes in a single execution are valid and unique. Then Byzantine
atomic broadcast is defined by the following properties:
\begin{itemize}
\item {\bf Integrity.} Every process delivers a value at most once.

\item {\bf External Validity.} A correct process delivers only values satisfying $\valid()$.

\item \textbf{Ordering.} If a correct process $p$ delivers $\val_1$ before $\val_2$,
  then another correct process $q$ cannot deliver $\val_2$ before $\val_1$.

\item {\bf Liveness.} If a correct process broadcasts or delivers $\val$, then
  eventually all correct processes will deliver $\val$. (Note that this implies
  {\em censorship-resistance}: the service cannot selectively omit values
  submitted by correct processes.)
\end{itemize}

\myparagraph{The \pbft protocol.}
We implement Byzantine atomic broadcast in a {\em \pbft} protocol
(Figures~\ref{fig:pbft-castro-normal}-\ref{fig:pbft-castro-preds}), which
faithfully captures the algorithmic core of the seminal Practical Byzantine
Fault Tolerance protocol (PBFT)~\cite{pbft}. Whereas PBFT integrated view
synchronization functionality with the core SMR protocol, \pbft delegates
this %
to an SMR synchronizer, and in \S\ref{sec:liveness} we rigorously prove its
liveness when using any synchronizer satisfying our specification. When using
the synchronizer in Figure~\ref{fig:sync}, the protocol also incurs only bounded
space overhead (see \tr{\ref{sec:pbft-space}}{\npbftspace} for details).

We base \pbft on the PBFT protocol with signatures and, for simplicity, omit the
mechanisms for managing checkpoints and watermarks; these can be easily added
without affecting liveness. The protocol works in a succession of views
produced by the synchronizer. A process stores its current view in %
$\currview$.  Each view $v$ has a fixed {\em leader}
$\leader(v) = p_{((v-1)\ \mathrel{\rm mod}\ n) + 1}$ that is responsible for
totally ordering values submitted for broadcast; the other processes are {\em
  followers}, which vote on proposals made by the leader. Processes store
the sequence of (unique) values proposed by the leader in a $\cmd$ array; at
the leader, a $\nextv$ counter points to the first free slot in the array.
Processes monitor the leader's behavior and ask the synchronizer to advance
to another view if they suspect that the leader is faulty. A $\status$ variable
records whether the process is operating as normal in the current view
($\NORMAL$) or is changing the view.

\begin{figure}[t]
\vspace{3pt}
\begin{tabular}{@{\!}l@{\!\!\!\!\!\!\!}|@{\ \ }l@{}}
\scalebox{0.96}{%
\begin{minipage}[t]{6.8cm}
\removelatexerror
\vspace*{-11pt}
\begin{algorithm*}[H]
\SetInd{\marginA}{\marginB}
  \setcounter{AlgoLine}{0}

\SetInd{\marginA}{\marginB}
  \SubAlgo{\Fun ${\tt start}()$}{\label{line:castro:start0}
    \lIf{$\currview = 0$}{\padvance()\label{line:castro:start}}
  }

  \smallskip

  \SubAlgo{\textbf{when a timer expires}\label{alg:castro:expire-timerexecute}}{ 
     ${\tt stop\_all\_timers}()$\; \label{line:castro:stoptimers}
      $\padvance()$\;
      $\status \leftarrow \BLOCKED$\;\label{line:castro:block}
    $\durationexecution \leftarrow \durationexecution + \tau$\; \label{line:castro:timer-inc1}
    $\durationrecovery \leftarrow \durationrecovery + \tau$\; \label{line:castro:timer-inc2}
  }

  \smallskip

  \SubAlgo{\Fun ${\tt broadcast}(\val)$\label{alg:castro:broadcast}}{
    \textbf{pre:} $\valid(\val)$\;
    \Send $\langle \BROADCAST(\val) \rangle_i$ \KwTo \all{}
    {\bf periodically until $x$ is delivered}\label{alg:castro:send-to-all}
  }

  \smallskip

  \SubAlgo{\WhenReceived $\BROADCAST(\val)$\label{alg:castro:broadcast-msg}}{
    \textbf{pre:} $\valid(\val)\wedge \status=\NORMAL \wedge{}$\\
    \nonl $\phantom{\text{{\bf pre:} }} (\timerexecute[\val]\mbox{ not active})
    \wedge{} $\\
    \nonl $\phantom{\text{{\bf pre:} }} (\forall k.\, k \le \lastdelivered {\implies} {}$\\
     \nonl $\phantom{\text{{\bf pre:} }(} \comcmd[k]\not=\val)$\;
     $\starttimer(\timerexecute[\val],$
     $\durationexecution)$\label{alg:castro:start-timerexecute}\;
     \Send $\langle \FORWARD(\val) \rangle_i$ \KwTo
     $\leader(\currview)$\;   
  }

  \smallskip

  \SubAlgo{\WhenReceived $\FORWARD(\val)$\label{alg:castro:forward}}{
    \textbf{pre:} $\valid(\val) \wedge \status=\NORMAL \wedge{}$\\
    \nonl $\phantom{\text{{\bf pre:} }} p_i = \leader(\currview)
    \wedge{}$\\
    \nonl $\phantom{\text{{\bf pre:} }}\forall k.\, \cmd[k]\not=\val$\;
    \Send $\langle \PREPREPARE(\currview, $ $\nextv, \val) \rangle_i$
    \KwTo \all\;
    $\nextv \leftarrow \nextv + 1$\;\label{alg:castro:increase-next}
  }
\end{algorithm*}
\vspace*{-4pt}
\end{minipage}}
&
\scalebox{0.96}{%
\begin{minipage}[t]{8.5cm}
\removelatexerror
\vspace*{-11pt}
\begin{algorithm*}[H]
  \SubAlgo{\WhenReceived $\langle \PREPREPARE(v, k, \val) \rangle_{j}$\label{alg:castro:receive-propose}}{
    \textbf{pre:} $p_j = \leader(v) \wedge \currview = v \wedge{}$\\\label{alg:castro:safety-check}
    \nonl$\phantom{\text{{\bf pre:} }} \status =\NORMAL \wedge \phase[k] =
    \START \wedge{}$\\
     \nonl$\phantom{\text{{\bf pre:} }} \valid(\val) \wedge (\forall k'.\, \cmd[k']\not=\val)$\\
    $(\cmd, \phase)[k] \leftarrow (\val, \PREPREPARED)$\;
    \Send $\langle \PREPARE(v, k, \hash(\val)) \rangle_i$ \KwTo \all\;
  }

  \smallskip
  
  \SubAlgo{\WhenReceived $\{\langle \PREPARE(v, k, h) \rangle_j \mid p_j
    \in Q\} = C$ {\bf for a quorum $Q$}\label{alg:castro:receive-prepared}}{
    \textbf{pre:} $\currview = v \wedge \phase[k] = \PREPREPARED \wedge {}$\\
    \nonl$\phantom{\text{{\bf pre:} }}{} \status = \NORMAL \wedge \hash(\cmd[k]) = h$\;
    $(\prepcmd, \prepview, \cert, \phase)[k] \leftarrow {}$\\
    \nonl$\quad (\cmd[k], \currview, C, \PREPARED)$\label{alg:castro:assign-cballot}\;
    \Send $\langle \COMMIT(v, k, h) \rangle_i$ \KwTo \all\;
  }

  \smallskip

  \SubAlgo{\WhenReceived $\{\langle \COMMIT(v, k, h) \rangle_j \mid p_j
    \in Q\} = C$ {\bf for a quorum $Q$}\label{alg:castro:receive-committed}}{
    \textbf{pre:} $\currview = v \wedge \phase[k] = \PREPARED \wedge {}$\\
    \nonl$\phantom{\text{{\bf pre:} }}\status = \NORMAL \wedge \hash(\prepcmd[k]) = h$\;
    $(\comcmd,\phase)[k] \leftarrow (\cmd[k], \COMMITTED)$\; 
     \Broadcast $\langle \DECISION(\comcmd[k], k, C)$\; \label{alg:castro:send-decision}
  }

  \smallskip

  \SubAlgo{{\bf when} $\comcmd[\lastdelivered+1] \not= \bot$\label{alg:castro:deliver}}{
    $\lastdelivered \leftarrow \lastdelivered+1$\; \label{alg:castro:update-last}
    \If{$\comcmd[\lastdelivered] \not= \noop$}{${\tt
        deliver}(\comcmd[\lastdelivered])$}\label{alg:castro:deliver-value}
        $\stoptimer($\quad $\timerexecute[\comcmd[\lastdelivered]])$\; \label{alg:castro:stop-timerexecute2}
    \If{$\lastdelivered = \krecovery\wedge{}$\hspace{2cm}$\status={\normalfont \NORMAL}$}{
        $\stoptimer(\timerrecovery)$\;\label{alg:castro:stop-timerrecovery}
        }
  }

\SubAlgo{\WhenReceived $\DECISION(\val, k, C)$ \label{alg:castro:receive-decision}}{
  \textbf{pre:} $\comcmd[k] \not= \bot \wedge{}$\\
  \nonl $\phantom{\text{{\bf pre:} }}\exists v.\, \committed(C, v, k,
    \hash(\val))$\;  \label{alg:castro:decision-safetycheck} 
    $\comcmd[k] \leftarrow \val$\;
  }

\end{algorithm*}
\vspace*{-4pt}
\end{minipage}}
\end{tabular}
\vspace*{-4pt}
\caption{Normal operation of \pbft at a process $p_i$.}
\label{fig:pbft-castro-normal}
\end{figure}

\myparagraph{Normal protocol operation.}
A process broadcasts a valid value $\val$ using a ${\tt broadcast}$ function
(line~\ref{alg:castro:broadcast}). This keeps sending the value to all
processes in a $\BROADCAST$ message until the process delivers the value, to 
tolerate message loss before $\GST$.
When a process receives a $\BROADCAST$ message with a new value
(line~\ref{alg:castro:broadcast-msg}), it forwards the value to the leader in a
$\FORWARD$ message. This ensures that the value reaches the leader even when
broadcast by a faulty process, which may withhold the $\BROADCAST$ message from
the leader. (We explain the timer set in
line~\ref{alg:castro:start-timerexecute} later.)
When the leader receives a new value $\val$ in a $\FORWARD$ message
(line~\ref{alg:castro:forward}), it sends a $\PREPREPARE$ message to all
processes (including itself) that includes $\val$ and its position in the log,
generated from the $\nextv$ counter.
Processes vote on the leader's proposal in two phases. Each process keeps track
of the status of values going through the vote in an array $\phase$, whose
entries initially store $\START$.

When a process receives a proposal $\val$ for a position $k$ from the leader of
its view $v$ (line~\ref{alg:castro:receive-propose}), it first checks that
$\phase[k] = \START$, so that it has not yet accepted a proposal for the
position $k$ in the current view. It also checks that the value is valid and
distinct from all values it knows about. The process then stores $\val$ in
$\cmd[k]$ and advances $\phase[k]$ to $\PREPREPARED$. Since a faulty leader may
send different proposals for the same position to different processes, the
process next communicates with others to check that they received the same
proposal. To this end, it disseminates a $\PREPARE$ message with the position
and the hash of the value $x$ it received. The process handles $x$ further once
it gathers a set $C$ of $\PREPARE$ messages from a quorum matching the value
(line~\ref{alg:castro:receive-prepared}), which we call a {\em prepared
  certificate} and check using the $\accepted$ predicate in
Figure~\ref{fig:pbft-castro-preds}.
In this case the process stores the value in $\prepcmd[k]$, the certificate in
$\cert[k]$, and the view in which it was formed in $\prepview[k]$. At this point
we say that the process {\em prepared} the proposal, as recorded by
setting its $\phase$ to $\PREPARED$. It is easy to show that processes
cannot prepare
different values at the same position and view, since 
each correct process can send only one corresponding $\PREPARE$ message.

Having prepared a value, the process
disseminates a $\COMMIT$ message with its hash. Once the process
gathers a quorum of matching $\COMMIT$ messages
(line~\ref{alg:castro:receive-committed}), it stores the value in a $\comcmd$
array and advances its $\phase$ to $\COMMITTED$: %
the value is now {\em committed}. The protocol ensures that correct processes cannot commit
different values at the same position, even in different views. We call a quorum
of matching $\COMMIT$ messages a {\em commit certificate} and check it using the
$\committed$ predicate in Figure~\ref{fig:pbft-castro-preds}. A process delivers
committed values
in the $\comcmd$ order, with $\lastdelivered$ tracking the position last 
delivered position.

To satisfy the Liveness property of atomic broadcast, similarly
to~\cite{alysson-reads}, \pbft allows a process to find out about committed
values from other processes directly. When a process commits a value
(line~\ref{alg:castro:receive-committed}), it disseminates a $\DECISION$ message
with the value, its position $k$ in the log and the commit certificate
(line~\ref{alg:castro:send-decision}). A process receiving a $\DECISION$ with a
valid certificate saves the value in $\comcmd[k]$, which allows it to be
delivered (line~\ref{alg:castro:deliver}).
The $\DECISION$ messages are disseminated via %
reliable broadcast ensuring that, if one correct process delivers the value,
then so do all others. %
To implement this, each process could periodically resend the $\DECISION$
messages it has (omitted from the pseudocode). A more practical implementation
would only resend information that other processes are
missing. %
As proved in~\cite{dls}, such periodic resends are unavoidable in the presence
of message loss.

\myparagraph{View initialization.}
When the synchronizer tells a process to move to a new view $v$
(line~\ref{alg:castro:newview}), the process sets $\currview$ to $v$, which
ensures that it will no longer accept messages from prior views. It also sets
$\status$ to $\RECOVERING$, which means that the process is not yet ready to
order values in the new view. It then sends a $\NEWLEADER$ message to the leader
of $v$ with the information about the values it has prepared so far and their
certificates%
\footnote{In PBFT this information is sent in {\tt VIEW-CHANGE} messages, which
  also play a role similar to $\WISH$ messages in our synchronizer
  (Figure~\ref{fig:sync}). In \pbft we opted to eschew {\tt VIEW-CHANGE}
  messages to maintain a clear separation between view synchronization internals
  and the SMR protocol.}.

\begin{figure}[t]
\vspace{2pt}
\begin{tabular}{@{}l@{\!\!\!\!\!\!}|@{\ \ }l@{}}
\scalebox{0.96}{%
\begin{minipage}[t]{7.7cm}
\removelatexerror
\vspace*{-10pt}
\begin{algorithm*}[H]
\SetInd{\marginA}{\marginB}
\SubAlgo{\Upon $\newview(v)$\label{alg:castro:newview}}{
     ${\tt stop\_all\_timers}()$\; \label{line:castro:stoptimers-enterview}
    $\currview \leftarrow v$\;\label{line:castro:set-currentview}
    $\status \leftarrow \RECOVERING$\;
    \Send $\langle \NEWLEADER(\currview, \prepview, $ $\prepcmd, \cert) \rangle_i$
    \KwTo $\leader(\currview)$\; 
    $\starttimer(\timerrecovery,$ $\durationrecovery)$\;\label{line:castro:start-timerrecovery}
  }

\smallskip

\SubAlgo{\WhenReceived $\{\langle \NEWLEADER(v, \vprepview_j,$ $\vprepcmd_j,
    \vcert_j) \rangle_j \mid$ $p_j \in Q\} = M$ \qquad\qquad {\bf for a quorum
      $Q$}\label{alg:castro:receive-newleader}}{
    \textbf{pre:} $p_i = \leader(v) \wedge \currview = v \wedge {}$\\
    \nonl $\phantom{\text{{\bf pre:} }} \status = \RECOVERING \wedge{}$\\
    \nonl $\phantom{\text{{\bf pre:} }} \forall m \in M.\, \ValidNewLeader(m)$\; 
    \ForAll{$k$\label{alg:castro:select-proposal}}{
      \lIf{$\exists p_{j'} \in Q.\, \vprepview_{j'}[k] \not= 0 \wedge{}$ $\forall p_{j} \in Q.\,
        \vprepview_{j}[k] \le \vprepview_{j'}[k]$}{%
        $\vcmd'[k] \leftarrow \vprepcmd_{j'}[k]$\label{alg:castro:prepared-entries}}
    }
    $\nextv \leftarrow \max\{k \mid \vcmd'[k]\neq \bot\}$\;\label{alg:castro:set-next}
}
 
\end{algorithm*}
\vspace*{-5pt}
\begin{tikzpicture}[overlay]
\node[draw=none, fill=white, thick,minimum width=.5cm,minimum height=.2cm] (b) at (.743,0){};%
\end{tikzpicture}
\end{minipage}}
&
\scalebox{0.96}{%
\begin{minipage}[t]{7.6cm}
\removelatexerror
\vspace*{-21pt}
\begin{algorithm*}[H]
\nonl\SubAlgo{}{
    \ForAll{$k = 1..(\nextv-1)$\label{alg:castro:clean-entries}}{
      \If{$\vcmd'[k] = \bot \vee \exists k'.\, k' \not= k \wedge {}$
        $\vcmd'[k'] = \vcmd'[k] \wedge \exists p_{j'} \in Q.\, \forall p_{j} \in
        Q. $
        $\vprepview_{j'}[k'] \,{>}\, \vprepview_{j}[k]$}{%
        $\vcmd'[k] \leftarrow \noop$\label{alg:castro:newview-end}%
      }
    }
    \Send $\langle \NEWVIEW(v, \vcmd', M) \rangle_i$ \KwTo \all\; \label{alg:castro:send-newview}
  }

\smallskip

  \SubAlgo{\WhenReceived $\langle \NEWVIEW(v, \vcmd', M) \rangle_j \hspace{1pt}{=}\hspace{1pt}m$\label{alg:castro:receive-newview}}{ 
    \textbf{pre:} $\status = \RECOVERING \wedge {}$\\
    \nonl $\phantom{\text{{\bf pre:} }} \currview = v \wedge \ValidNewState(m)$\; 
    $\cmd \leftarrow \vcmd'$\;
    \ForAll{$\{k \mid \cmd[k] \not= \bot\}$}{\label{alg:castro:forallnewview}
      $\phase[k] \leftarrow \PREPREPARED$\;
      \Send $\langle \PREPARE(v, k, \hash(\cmd[k])) \rangle_i$\quad \KwTo
      \all\; \label{alg:castro:preparenewview}
    }
    $\status \leftarrow \NORMAL$\; \label{alg:castro:status-normal}
    $\krecovery \leftarrow \max\{k \mid \cmd[k]\neq
    \bot\}$\;\label{alg:castro:set-krecovery}
    \If{$\krecovery\leq\lastdelivered$}
    {$\stoptimer(\timerrecovery)$\;\label{alg:castro:stop-timerrecovery2}}
}

\end{algorithm*}
\vspace*{-5pt}
\end{minipage}}
\end{tabular}

\vspace{-3pt}
\caption{View-initialization protocol of \pbft at a process $p_i$.}
\label{fig:pbft-castro-recovery}
\end{figure}

\begin{figure}[t]
\vspace{-10pt}
\small
\begin{minipage}{15cm}
\begin{gather*}
\accepted(C, v, k, h)
\iff
\exists Q.\, 
\quorum(Q) \wedge C = \{\langle \PREPARE(v, k, h) \rangle_j \mid p_j \in Q\}
\\[1pt]
\committed(C, v, k, h)
\iff
\exists Q.\, 
\quorum(Q) \wedge C = \{\langle \COMMIT(v, k, h) \rangle_j \mid p_j \in Q\}
\\[1pt]
\begin{array}{@{}l@{}}
\ValidNewLeader(\langle \NEWLEADER(v, \vprepview, \vprepcmd, \vcert) \rangle_{\_}) \iff {}
\\[1pt]
\quad \forall k.\, (\vprepview[k] > 0 {\implies} 
\vprepview[k] < v \wedge 
\accepted(\vcert[k], \vprepview[k], k, \vprepcmd[k]))
\end{array}
\\[1pt]
\begin{array}{@{}l@{}}
\ValidNewState(\langle \NEWVIEW(v, \vcmd', M) \rangle_i) \iff
 p_i = \leader(v) \wedge
  \exists Q, \vprepview, \vprepcmd, \vcert.\, 
\\[1pt]
\quad \quorum(Q) \wedge 
 M = \{\langle \NEWLEADER(v, \vprepview_j, \vcmd_j, \vcert_j)
  \rangle_j \mid p_j \in Q\} \wedge {} 
\\[1pt]
\quad  (\forall m \in C.\, \ValidNewLeader(m)) \wedge
(\vcmd' \mbox{ is computed from $M$ as
  per lines~\ref{alg:castro:select-proposal}-\ref{alg:castro:newview-end}})
\end{array}
\end{gather*}
\end{minipage}%
\vspace*{-2pt}
\caption{Auxiliary predicates for \pbft.}
\label{fig:pbft-castro-preds}
\end{figure}

The new leader waits until it receives a quorum of well-formed $\NEWLEADER$
messages, as checked by the predicate $\ValidNewLeader$
(line~\ref{alg:castro:receive-newleader}). Based on these, the leader computes
the initial log of the new view, stored in $\vcmd'$. Similarly to
Paxos~\cite{paxos}, for each index $k$ the leader puts at the $k$th position in
$\vcmd'$ the value prepared in the highest view
(line~\ref{alg:castro:select-proposal}).
The resulting array may contain empty or duplicate entries. To resolve this, the
leader writes $\noop$ into empty entries and those entries for which there is a
duplicate prepared in a higher view (line~\ref{alg:castro:clean-entries}). The
latter is safe because one can show that no value could have been committed in
such entries in prior views. Finally, the leader sends a $\NEWVIEW$ message to
all processes, containing the initial log and the $\NEWLEADER$ messages from
which it was computed (line~\ref{alg:castro:send-newview}).

A process receiving a $\NEWVIEW$ first checks its correctness by redoing the
leader's computation ($\ValidNewState$,
line~\ref{alg:castro:receive-newview}). If the check passes, the process
overwrites its log with the new one and sets $\status$ to $\NORMAL$.
It also sends $\PREPARE$ messages for all $\cmd$ entries,
to commit them in the new view. A more practical implementation would
include a checkpointing mechanism, so that a process restarts committing
previous $\cmd$ entries only from the last stable checkpoint~\cite{pbft}; this
mechanism can be easily added to \pbft.

\myparagraph{Triggering view changes.}
We now describe when a process calls $\padvance$, which is key to ensure
liveness (\S\ref{sec:liveness}). This happens either
on start-up (line~\ref{line:castro:start}) or when 
the process 
suspects that the current leader is faulty. To this end, the process monitors
the leader's behavior using timers; if one of these expires, the process calls
$\padvance$ and sets $\status$ to $\BLOCKED$
(line~\ref{alg:castro:expire-timerexecute}). First, the process checks that each
value it receives is delivered promptly: e.g., to guard against a faulty leader
censoring certain values.
For a value $x$ this is done using $\timerexecute[x]$, set for a
duration $\durationexecution$ when the process receives $\BROADCAST(x)$
(lines~\ref{alg:castro:start-timerexecute}). The timer is stopped when the
process delivers $x$ (line~\ref{alg:castro:stop-timerexecute2}).
A process also checks that the leader initializes a view quickly enough:
e.g., to guard against the leader crashing during the initialization.
Thus, when a process enters a view it starts $\timerrecovery$ for a
duration $\durationrecovery$ (line~\ref{line:castro:start-timerrecovery}).
The process stops the timer when it
delivers all values in the initial log
(lines~\ref{alg:castro:stop-timerrecovery}
and~\ref{alg:castro:stop-timerrecovery2}).
The above checks may make a process suspect a correct leader if the timeouts are
initially set too small with respect to the message delay $\delta$, unknown to
the process. To deal with this, a process increases $\durationexecution$ and
$\durationrecovery$ each time a timer expires, which signals that the current
view is not operating normally
(lines~\ref{line:castro:timer-inc1}-\ref{line:castro:timer-inc2}).

\section{Proving the Liveness of PBFT}
\label{sec:liveness}

Assume that \pbft is used with a synchronizer satisfying the specification in
Figure~\ref{fig:multi-sync-properties}; to simplify the following latency
analysis we let $d = 2\delta$, as for the synchronizer in
Figure~\ref{fig:sync}. We now prove that the protocol satisfies the Liveness
property of Byzantine atomic broadcast; we defer the proof of the other
properties to \tr{\ref{sec:pbft-safety}}{\npbftsafety}. To the best of our
knowledge, this is the first rigorous proof of liveness for the algorithmic core
of PBFT: as we elaborate in \S\ref{sec:related}, the liveness mechanisms of PBFT
came only with a brief informal justification, which did not cover their most
critical properties~\cite[\S{}4.5.1]{castro-tocs}. Our proof is simplified by
the use of the synchronizer specification, which allows us to abstract from
view synchronization mechanics.

We prove the liveness of \pbft by showing that the protocol establishes
properties reminiscent of those of failure detectors~\cite{CT96}. First,
similarly to their completeness property, %
we prove that 
every correct process eventually attempts to advance from a {\em bad}\/ view in
which no progress is possible (e.g., because the leader is faulty).
\begin{lemma}
  Assume that a correct process $p_i$ receives $\BROADCAST(\val)$ for a valid
  value $\val$ while in a view $v$. If $p_i$ never delivers $\val$ and never enters
  a view higher than $v$, then it eventually calls $\padvance$ in $v$.
\label{thm:castro:completeness}
\end{lemma}

The lemma holds because in \pbft each process monitors the leader's behavior
using timers, and we defer its easy proof to
\tr{\ref{sec:pbft-liveness}}{\npbftliveness}.
Our next lemma is similar to the eventual accuracy property of failure
detectors. It stipulates that if the timeout values are high enough, then
eventually any correct process that enters a {\em good}\/ view (with a correct
leader) will never attempt to advance from it.
Let $\durationrecovery_i(v)$ and $\durationexecution_i(v)$ denote respectively
the value of $\durationrecovery$ and $\durationexecution$ at a
correct process $p_i$ while in view $v$.
\begin{lemma}
  Consider a view $v\geq\B$ such that $\tm{v}\geq\GST$ and $\leader(v)$ is
  correct. If $\durationrecovery_i(v) > 6\delta$ and
  $\durationexecution_i(v) > 4\delta$ at each correct process $p_i$ that enters
  $v$, then no correct process calls $\padvance$ in $v$.
\label{thm:castro:all-good}
\end{lemma}

\begin{figure}[t]
\centerline{\includegraphics[width=0.5\textwidth]{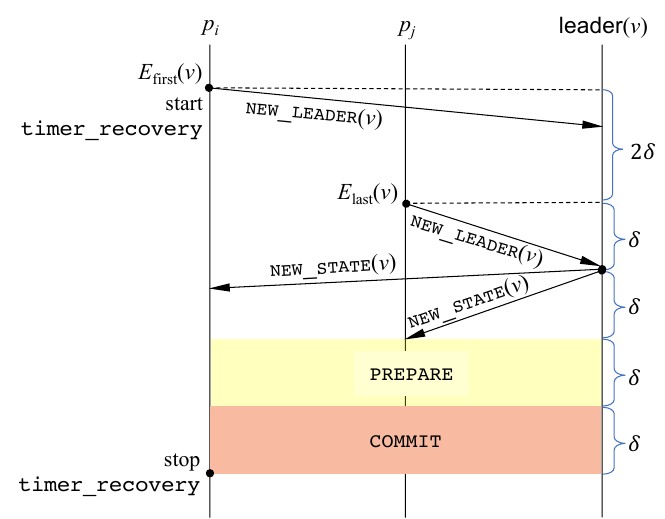}}
\caption{An illustration of the proof of  Lemma~\ref{thm:castro:all-good}.}
\label{fig:bounds}
\end{figure}

Before proving the lemma, we informally explain the rationale
for the bounds on timeouts in it, using the example of
$\durationrecovery$.  The timer $\timerrecovery$ is started at a process $p_i$
when this process enters a view $v$
(line~\ref{line:castro:start-timerrecovery}), and is stopped when the process
delivers all values inherited from previous views
(lines~\ref{alg:castro:stop-timerrecovery}
or~\ref{alg:castro:stop-timerrecovery2}). The two events are separated by $4$
communication steps of \pbft, exchanging messages of the types $\NEWLEADER$,
$\NEWVIEW$, $\PREPARE$ and $\COMMIT$ (Figure~\ref{fig:bounds}).
However, $4\delta$ would be too small a value for $\durationrecovery$. This is
because the leader of $v$ sends its $\NEWVIEW$ message only after receiving a
quorum of $\NEWLEADER$ messages, and different processes may enter $v$ and send
their $\NEWLEADER$ messages at different times (e.g., $p_i$ and $p_j$ in
Figure~\ref{fig:bounds}). Hence, $\durationrecovery$ must additionally
accommodate the maximum discrepancy in the entry times, which is $d = 2\delta$
by the Bounded Entry property. Then to ensure that $p_i$ stops the timer before
it expires, we require $\durationrecovery_i(v) > 6\delta$. 
As the above reasoning illustrates, Lemma~\ref{thm:castro:all-good} is more
subtle than Lemma~\ref{thm:castro:completeness}: while the latter is ensured
just by the checks in the SMR protocol, the former relies on the Bounded Entry
property of the synchronizer.

Another subtlety about Lemma~\ref{thm:castro:all-good} is that the $\delta$ used
in its premise is a priori unknown. Hence, to apply the lemma in the liveness
proof of \pbft, we have to argue that, if correct processes keep changing views
due to lack of progress, then all of them will eventually increase their
timeouts high enough to satisfy the bounds in Lemma~\ref{thm:castro:all-good}.
This is nontrivial due to the fact that, as in the original
PBFT~\cite[\S{}2.3.5]{castro-thesis}, in our protocol the processes update their
timeouts independently, and may thus disagree on their durations. For example,
the first correct process $p_i$ to detect a problem with the current view $v$
will increase its timeouts and call $\padvance$
(line~\ref{alg:castro:expire-timerexecute}). The synchronizer may then trigger
$\newview$ notifications at other correct processes before they detect the
problem as well, so that their timeouts will stay unchanged
(line~\ref{alg:castro:newview}).
One may think that this allows executions in which only some correct processes
keep increasing their timeouts until they are high enough, whereas others are
forever stuck with timeouts that are too low, invalidating the premise of
Lemma~\ref{thm:castro:all-good}. The following lemma rules out such scenarios
and also trivially implies Lemma~\ref{thm:castro:all-good}. It establishes that,
in a sufficiently high view $v$ with a correct leader, if the timeouts at a
correct process $p_i$ that enters $v$ are high enough, then this process cannot
be the first one to initiate a view change. Hence, for the protocol to enter
another view, some other process with lower timeouts must call $\padvance$ and
thus increase their durations (line~\ref{alg:castro:expire-timerexecute}).

\begin{lemma}
  Let $v\geq \B$ be such that $\tm{v}\geq\GST$ and $\leader(v)$ is
  correct, and consider a correct process $p_i$ that enters $v$. If
  $\durationrecovery_i(v) > 6\delta$ and $\durationexecution_i(v) > 4\delta$
  then $p_i$ is not the first correct process to call $\padvance$ in $v$.
\label{thm:castro:timers}
\end{lemma}
\begin{proof}
Since $\tm{v}\geq\GST$, messages sent by correct processes after $\tm{v}$ get
delivered to all correct processes within $\delta$ and process clocks track real
time. By contradiction, assume that $p_i$ is the first correct process to call
$\padvance$ in $v$. This happens because a timer expires at $p_i$.
Here we only consider the case when it is $\timerrecovery$, and handle $\timerexecute$ in
\tr{\ref{sec:pbft-liveness}}{\npbftliveness}.  A process starts $\timerrecovery$
when it enters the view $v$ (line~\ref{line:castro:start-timerrecovery}), and
hence, at $\tm{v}$ at the earliest (Figure~\ref{fig:bounds}). Because $p_i$ is
the first correct process to call $\padvance$ in $v$ and
$\durationrecovery_i(v)>6\delta$, no correct process calls $\padvance$ in $v$
until after $\tm{v}+6\delta$. Then by \entry all correct processes enter $v$ by
$\tm{v}+2\delta$. Also, by \validitysync no correct process can enter $v+1$
until after $\tm{v}+6\delta$, and by Proposition~\ref{sync:noskip} the same
holds for any view $>v$.  Thus, all correct processes stay in $v$ at least until
$\tm{v}+6\delta$.

When a correct process enters $v$, it sends a $\NEWLEADER$ message to the leader
of $v$, which happens by $\tm{v}+2\delta$. When the leader receives
such %
messages from a quorum of processes, it broadcasts a $\NEWVIEW$ message. Thus,
by $\tm{v}+4\delta$ all correct processes receive this message and set
$\status=\NORMAL$. If at that point $\krecovery\leq\lastdelivered$ at $p_i$,
then the process stops $\timerrecovery$
(line~\ref{alg:castro:stop-timerrecovery2}), which contradicts our
assumption. Hence, $\krecovery>\lastdelivered$. When a correct process receives
$\NEWVIEW$, it sends $\PREPARE$ messages for all positions $\leq \krecovery$
(line~\ref{alg:castro:preparenewview}). It then takes the correct processes at
most $2\delta$ to exchange the sequence of $\PREPARE$ and $\COMMIT$ messages
that commits the values at all positions $\leq \krecovery$. Thus, by
$\tm{v}+6\delta$ the process $p_i$ commits and delivers all these positions,
stopping $\timerrecovery$ (line~\ref{alg:castro:stop-timerrecovery}): 
a contradiction.
\end{proof}

\begin{theorem}
  \pbft satisfies the Liveness property of Byzantine atomic broadcast.
\label{thm:castro:liveness}
\end{theorem}
\begin{proof}
Consider a valid value $x$ broadcast by a correct process. We first prove that
$x$ is eventually delivered by some correct process. Assume the contrary. We show:

\begin{myclaim}
  Every view is entered by some correct process.
\end{myclaim}
\begin{claimproof}
Since all correct processes call {\tt start} (line~\ref{line:castro:start0}), by
\startup a correct process eventually enters some view. We now show that correct
processes keep entering new views forever (analogously to the proof of
Proposition~\ref{lem:live-toy} in \S\ref{sec:sync}). Assume that this is false,
so that there exists a maximal view $v$ entered by any correct process. Let $P$
be any set of $f+1$ correct processes and consider an arbitrary process
$p_i\in P$ that enters $v$. The process that broadcast $x$ is correct, and thus
keeps broadcasting $x$ until the value is delivered
(line~\ref{alg:castro:send-to-all}). Since $x$ is never delivered, $p_i$ is
guaranteed to receive $x$ while in $v$. Then by
Lemma~\ref{thm:castro:completeness}, $p_i$ eventually calls $\padvance$ while in
$v$. Since $p_i$ was picked arbitrarily, we have
$\forall p_i \in P.\, {\te{i}{v}\fdef} {\implies} {\ta{i}{v}\fdef}$. Then by
\progress we get $\tm{v+1}\fdef$, which yields a contradiction. Thus, correct
processes keep entering views forever. The claim then follows from
Proposition~\ref{sync:noskip}.
\end{claimproof}

Let view $v_1$ be the first view such that $v_1\ge \B$ and $\tm{v_1}\geq \GST$;
such a view exists by Claim 1. The next claim is needed to show that all correct
processes will increase their timeouts high enough to satisfy the bounds in
Lemma~\ref{thm:castro:all-good}. %

\begin{myclaim}
  Every correct process calls the timer expiration handler
  (line~\ref{alg:castro:expire-timerexecute}) infinitely often.
\end{myclaim}
\begin{claimproof}
Assume the contrary and let $C_{\rm fin}$ and
$C_{\rm inf}$ be the sets of correct processes that call the timer expiration
handler finitely and infinitely often, respectively. Then
$C_{\rm fin} \not= \emptyset$, and by Claim~1 and \validitysync, $C_{\rm inf}
\not= \emptyset$.  The values of 
$\durationexecution$ and $\durationrecovery$ increase unboundedly at processes
from $C_{\rm inf}$, and do not change after some view $v_2$ at processes from
$C_{\rm fin}$.  By Claim~1 and since leaders rotate round-robin, there is a
view $v_3\geq \max\{v_2, v_1\}$ with a correct leader such that any process
$p_i \in C_{\rm inf}$ that enters $v_3$ has
$\durationexecution_i(v_3)> 4\delta$ and $\durationrecovery_i(v_3)> 6\delta$. By
Claim~1 and \validitysync, at least one correct process
calls $\padvance$ in $v_3$; let $p_l$ be the first process to do so. Since
$v_3 \ge v_2$, $p_l$
cannot be in $C_{\rm fin}$ because none of these
processes increase their timers in $v_3$. Then $p_l\in C_{\rm inf}$, 
contradicting Lemma~\ref{thm:castro:timers}.
\end{claimproof}

By Claims~1 and~2, there exists a view $v_4\geq v_1$ with a correct leader such
that some correct process enters $v_4$, and for any correct process $p_i$ that
enters $v_4$ we have $\durationexecution_i(v_4) > 4\delta$ and
$\durationrecovery_i(v_4) > 6\delta$. By Lemma~\ref{thm:castro:all-good}, no
correct process calls $\padvance$ in $v_4$. Then, by \validitysync, no correct
process enters $v_4+1$, which contradicts Claim~1. This contradiction shows that
$x$ must be delivered by a correct process. Then, since the protocol reliably
broadcasts committed values (line~\ref{alg:castro:send-decision}), all correct
processes will also eventually deliver $x$.
\end{proof}

\section{Latency Bounds for PBFT}
\label{sec:latency}

Assume that \pbft is used with our SMR synchronizer in Figure~\ref{fig:sync}. We
now quantify its latency using the bounds for the synchronizer in
Theorem~\ref{thm:smr-sync-correct}, yielding the first detailed latency analysis
for a PBFT-like protocol. Due to space constraints we defer proofs to
\tr{\ref{app:pbftlatency}}{\npbftlatency}. To state our bounds, we assume the
existence of a known upper bound $\Delta$ on the maximum value of $\delta$ in
any execution~\cite{pass-shi,HK89}, so that we always have $\delta < \Delta$. In
practice, $\Delta$ provides a conservative estimate of the message delay during
synchronous periods, which may be much higher than the maximal delay $\delta$ in
a particular execution. We modify the protocol in
Figure~\ref{fig:pbft-castro-normal} so that in lines
\ref{line:castro:timer-inc1}-\ref{line:castro:timer-inc2} it does not increase
$\durationrecovery$ and $\durationexecution$ above $6\Delta$ and $4\Delta$,
respectively. This corresponds to the bounds in Lemma~\ref{thm:castro:all-good}
and preserves the protocol liveness. Finally, we assume that periodic
handlers %
(line~\ref{line:retransmit-start} in Figure~\ref{fig:sync} and
line~\ref{alg:castro:send-to-all} in Figure~\ref{fig:pbft-castro-normal}) are
executed every $\rho$ time units, and that the latency of reliable broadcast in
line~\ref{alg:castro:send-decision} under synchrony is $\le \delta+\rho$ (this
corresponds to an implementation that just periodically retransmits $\DECISION$
messages).

We quantify the latency of \pbft in both bad and good cases. For the bad case we
assume that the protocol starts during the asynchronous period.
Given a value $\val$ broadcast before $\GST$, we quantify how quickly after
$\GST$ all correct processes deliver $\val$.
For simplicity, we assume that timeouts are high enough at $\GST$ and that
$\leader(\B)$ is correct.
\begin{theorem}
  Assume that before $\GST$ all correct processes start executing the protocol
  and one of them broadcasts $\val$. Let $\B$ be defined as in
  Theorem~\ref{thm:smr-sync-correct} and assume that $\leader(\B)$ is correct
  and at $\GST$ each correct process has $\durationrecovery > 6\delta$ and
  $\durationexecution > 4\delta$. Then all correct processes deliver $\val$ by
  $\GST+\rho + \max\{\rho+\delta, 6\Delta\} + 4\Delta+\max\{\rho, \delta\} +
  7\delta$.
\label{thm:pbft-latency1}
\end{theorem}

Although the latency bound looks complex, its main message is simple: \pbft
recovers after a period of asynchrony in bounded time. This time is dominated by
multiples of $\Delta$; without the assumption that $\leader(\B)$ is correct it
would also be multiplied by $f$ due to going over up to $f$ views with faulty
leaders. In \tr{\ref{app:pbftlatency}}{\npbftlatency} we show the bound using
the latency guarantees of our synchronizer
(Properties~\ref{eq:gen-bounded-entry:main} and~\ref{eq:lat-bound1:main} in
Theorem~\ref{thm:smr-sync-correct}).

We now consider the case when the protocol starts during the synchronous period,
i.e., after $\GST$. The following theorem quantifies how quickly all correct
processes enter the first functional view, which in this case is view $1$.
If $\leader(1)$ is correct, it also quantifies how quickly a broadcast value
$\val$ is delivered by all correct processes. The bound takes into account the
following optimization: in view $1$ the processes do not need to exchange
$\NEWLEADER$ messages. Then, after the systems starts up, the protocol delivers
values within $4\delta$, which matches an existing lower bound of $3\delta$ for
the delivery time starting from the leader~\cite{ittai-good-case}. 
\begin{theorem}
  Assume that all correct processes start the protocol after\/ $\GST$ with
  $\durationrecovery > 5\delta$ and $\durationexecution > 4\delta$. Then the
  $\B$ defined in Theorem~\ref{thm:smr-sync-correct} is equal to $1$ and
  $\tl{1}\leq \taelast{0} + \delta$. Furthermore, if a correct process
  broadcasts $\val$ at $t \ge \GST$ and $\leader(1)$ is correct, then all correct
  processes deliver $\val$ by $\max\{t, \taelast{0}+\delta\} + 4\delta$.
\label{thm:pbft-latency:good}
\end{theorem}

\section{Additional Case Studies}
\label{sec:xbft}

To demonstrate the generality of SMR synchronizers,
we have also used it to ensure the liveness of two other protocols. First, we
handle a variant of PBFT that periodically forces a leader change, as is common
in modern Byzantine SMR~\cite{aardvark,spinning,mirbft}. In this protocol a
process calls $\padvance$ not only when it suspects the current leader to be
faulty, but also when it delivers $B$ values proposed by this leader (for a
fixed $B$).
Second, we have applied the SMR synchronizer %
to a variant of the above protocol that follows the approach of
HotStuff~\cite{hotstuff}. The resulting protocol adds an extra communication
step to the normal path of PBFT in exchange for reducing the communication
complexity of leader change. Due to space constraints, we defer the details
about these two protocols to \tr{\ref{sec:pbft-rotation}}{\npbftrotation} and
\tr{\ref{sec:hotstuff}}{\nhotstuff}. Their liveness proofs follow the
methodology we proposed for \pbft, establishing analogs of
Lemmas~\ref{thm:castro:completeness}-\ref{thm:castro:timers}.

For PBFT with periodic leader rotation we have also established latency bounds
when using the synchronizer in Figure~\ref{fig:sync} (see
\tr{\ref{sec:pbft-rotation}}{\npbftrotation}). The most interesting one
\iflong(Theorem~\ref{thm:pbftr-latency2})\else(Theorem~56)\fi{}
demonstrates the benefit of PBFT's mechanism
for adapting timeouts to an unknown $\delta$: recall that in PBFT a process only
increases its timeouts when a timer expires, which means that the current view
does not operate normally (\S\ref{sec:pbft}). We show that, since the protocol
does not increase its timeouts in good views (with correct leaders and under
synchrony), it pays a minimal latency penalty to recover the first time it
encounters a bad leader -- the initial value of $\durationrecovery$. This
contrasts with the simplistic way of adapting the timeouts to an unknown
$\delta$ by increasing them in every view: in this case, as the protocol keeps
changing views, the timeouts would eventually increase up to the maximum
(determined by $\Delta$), and the protocol would have to wait that much to
recover from a faulty leader.

\section{Related Work and Discussion}
\label{sec:related}

\myparagraph{Failure detectors.}
Failure detectors and leader oracles~\cite{CT96,petr-survey} have been 
widely used for implementing consensus and SMR 
under benign failures~\cite{MR99,GR04,R00},
but  their
implementations under Byzantine failures are either impractical~\cite{kihlstrom}
or detect only restricted failure types~\cite{MR97,modular-smr,HK09}. %
Another %
approach was proposed in a textbook by Cachin et
al.~\cite{cachin-book}. This relies on a leader-based Byzantine Epoch-Change
(BEC) abstraction,
which accepts ``complain'' hints from the \linebreak application %
suggesting that the trust in the current leader should be revoked.
However, like the classical leader oracles, BEC requires all correct processes
to eventually trust the same correct
leader, which is impossible to achieve in Byzantine settings.
In fact, the BEC-based Byzantine consensus algorithm in~\S{}5.6.4
of~\cite{cachin-book} suffers from a liveness bug, which we describe
in~\tr{\ref{sec:bug}}{\nbug}. The bug has been confirmed with one of the
textbook's authors~\cite{cristian-personal}.

Although our $\padvance$ is %
similar to ``complain'',
we use it to implement a weaker abstraction of an SMR synchronizer.
We then obtain properties similar to accuracy and completeness of failure
detectors by carefully combining SMR-level timers with uses of $\padvance$
(Lemmas~\ref{thm:castro:completeness}-\ref{thm:castro:all-good}). Also,
while~\cite{cachin-book} does not specify constraints on the use of ``complain''
(see~\tr{\ref{sec:bug}} {\nbug}),
we give a complete characterization of $\padvance$ and show its sufficiency for
solving SMR.

BFT-SMaRt~\cite{bftsmart,bftsmart-thesis} built on the ideas
of~\cite{cachin-book} to propose an abstraction of {\em validated and provable
  (VP) consensus}, which allows its clients to control leader changes.
Although the overall BFT-SMaRt protocol appears to be correct, its liveness
proof sketch suffers from issues with rigor similar to those of~\cite{cachin-book}. 
In particular, the conditions on how to change the leader in
VP-Consensus to ensure its liveness were underspecified (again, see~\tr{\ref
{sec:bug}}{\nbug}).

\myparagraph{Emulating synchrony.}
Alternative abstractions avoid dependency on the specifics of a failure model
by simulating synchrony~\cite{Gaf98,heardof,KS06,heardof-bft}.
The first such abstraction is due to Awerbuch~\cite{Awe85} who proposed a family
of synchronizer algorithms emulating a  round-based synchronous system of top of an
asynchronous network with reliable communication and processes. 
The first such emulation in a failure-prone partially synchronous
system was introduced in the DLS paper~\cite{dls}.
It relied on an expensive
clock synchronization protocol, which interleaved its messages with 
every step of a high-level consensus algorithm implemented on top of it.
Later work proposed more practical solutions, which reduce the synchronization
frequency by relying on either timers~\cite{cezara} or synchronized hardware
clocks~\cite{ADDNR19,indulgent1,indulgent2} (the latter can be obtained using
one of the existing fault-tolerant clock synchronization
algorithms~\cite{SWL86,DHSS95}).
However, the DLS model emulates communication-closed rounds, i.e., eventually, a
process in a round $r$ receives {\em all} messages sent by correct processes in
$r$. This property rules out {\em optimistically
  responsive}~\cite{optimistic-responsiveness,hotstuff} protocols such as PBFT,
which can make progress as soon as they receive messages from {\em any quorum}.

\myparagraph{Consensus synchronizers.}
To address the shortcoming of DLS rounds, recent work proposed a more flexible
abstraction (``consensus synchronizer'' in \S\ref{sec:sync}) that switches
processes through an infinite series of {\em
  views}~\cite{NK20,bftlive,hotstuff}. In contrast to rounds, each view may
subsume multiple communication steps.
Although consensus synchronizers can be used for efficient
single-shot Byzantine consensus~\cite{bftlive}, using them for SMR results in
suboptimal implementations.
A classical approach is to decide on each SMR command using a separate black-box
consensus instance~\cite{smr}.  However, implementing the latter using a
consensus synchronizer would force the processes in every instance to iterate
over the same sequence of potentially bad views until the one with a correct
leader and sufficiently long duration could be reached.

An alternative approach was proposed in HotStuff~\cite{hotstuff}. This SMR
protocol is driven by a {\em pacemaker}, which keeps generating views similarly
to a consensus synchronizer. Within each view HotStuff runs a voting protocol
that commits a block of client commands in a growing hash chain. Although the
voting protocol is optimistically responsive,
committing the next block is delayed until the pacemaker generates a new view,
which increases latency. The cost the pacemaker may incur to generate a view is
also paid for every single block.

\myparagraph{SMR synchronizers.}
In contrast to the above approaches, SMR synchronizers allow the application to
initiate view changes on demand via an $\padvance$ call.
As we show,
this affords SMR protocols the flexibility to judiciously manage their view
synchronization schedule: in particular, it prevents the timeouts from growing
unnecessarily (\S\ref{sec:xbft}) and
avoids the overheads of further view synchronizations once a stable view is
reached (Lemma~\ref{thm:castro:all-good}, \S\ref{sec:liveness}).

The first synchronizer with a $\newview$/$\padvance$ interface, which here we
call an SMR synchronizer, was proposed by Naor et al.~\cite{NK20,lumiere}. They
used it as an intermediate module in a communication-efficient implementation of
a consensus synchronizer. The latter is sufficient to ensure the liveness of
HotStuff~\cite{hotstuff} via either of the two straightforward SMR constructions
we described above. The specification of the $\newview$/$\padvance$ module of
Naor et al. was only used as a stepping stone in the proof of their consensus
synchronizer, and as a result, is more low-level and complex than our SMR
synchronizer specification. Naor et al. did not investigate the usability of the
SMR synchronizer abstraction as a generic building block applicable to a wide
range of Byzantine SMR protocols -- a gap we fill in this paper. Finally, they
only handled a simplified version of partial synchrony where messages are never
lost and $\delta$ is known a priori, whereas our SMR synchronizer implementation
handles partial synchrony in its full generality. This implementation builds on
the consensus synchronizer of Bravo et al.~\cite{bftlive}. However, its
correctness proof and performance analysis are more intricate,
since %
the timing of the view switches is not fixed a priori, but driven by external
$\padvance$ inputs.

A\c{s}tef\u{a}noaei et al.~\cite{tenderbake} proposed another framework for
implementing Byzantine SMR protocols, based on DLS rounds. This uses a simple
synchronizer that does not exchange any messages: it recovers from a period of
asynchrony by progressively increasing round durations until they are long
enough for all correct processes to overlap in the same round. This way of view
synchronization rules out optimistically responsive SMR protocols and does not
bound the time to reach a decision after $\GST$, as we do.

\myparagraph{SMR liveness proofs.}
PBFT~\cite{pbft,castro-thesis,castro-tocs} is a seminal protocol whose design
choices have been widely
adopted~\cite{zyzzyva,spinning,mirbft,sbft}. %
To the best of our knowledge, our proof in~\S\ref{sec:liveness} is the first one
to formally establish its liveness. An informal argument given
in~\cite[\S{}4.5.1]{castro-tocs} mainly justifies
liveness assuming all correct processes enter a view with a correct leader and
stay in that view for sufficiently long. It does not rigorously justify why such
a view will be eventually reached, and in particular, how this is ensured by the
interplay between SMR-level timeout management and view synchronization
(\S\ref{sec:liveness}). Liveness mechanisms were also omitted from the formal
specification of PBFT by an I/O-automaton~\cite{castro-tocs,castro-thesis}.

Bravo et al.~\cite{bftlive} have applied consensus synchronizers to several
consensus protocols, including a single-shot version of PBFT. These protocols
and their proofs and are much more straightforward than the full SMR protocols
we consider here. In particular, since a consensus synchronizer keeps switching
processes between views regardless of whether their leaders are correct, the
proof of the single-shot PBFT in~\cite{bftlive} does not need to establish
analogs of completeness and accuracy (Lemmas~\ref{thm:castro:completeness}
and~\ref{thm:castro:all-good}) or deal with the fact that processes may disagree
on timeout durations (Lemma~\ref{thm:castro:timers}).

Byzantine SMR protocols often integrate view synchronization into the core
protocol, enabling white-box
optimizations~\cite{tendermint-arxiv,tendermint-netys,libra,pbft}. Our work does
not rule out this approach, but allows %
making it more systematic: we can first develop efficient mechanisms for view
synchronization independently from SMR protocols, and do white-box optimizations
afterwards.

\bibliographystyle{plainurl}
\bibliography{biblio}

\iflong

\appendix
\clearpage

\section{Constructing a Consensus Synchronizer from an SMR Synchronizer}
\label{sec:single-shot}

\begin{figure}[h]
\small
  \begin{algorithm*}[H]
  \setcounter{AlgoLine}{0}

  \SubAlgo{\Fun $\texttt{start}$()}{\label{line2:start}
    \padvance();
  }

  \smallskip

  \SubAlgo{{\bf upon} $\newview(v)$}{\label{line2:newview}
      $\stoptimer(\timerview)$\; \label{line2:timer-stop}
      $\starttimer(\timerview, \timeout(v))$\; \label{line2:start-timer}
      \Trigger $\newconsview(v)$\;
    }

    \smallskip

  \SubAlgo{\textbf{when \timerview\ expires}}{\label{line2:timer-exp1}  
    \padvance();
  }
\end{algorithm*}
\caption{A consensus synchronizer from an SMR synchronizer.}
\label{fig:single-shot}
\end{figure}

\begin{figure}[h]
\center
\setlength{\leftmargini}{20pt}
\renewcommand{\theenumi}{\Roman{enumi}}
\renewcommand{\labelenumi}{\theenumi.}
\begin{enumerate}
\item
  \label{prop:local-order} $\forall i, v, v'.\, {\sse{i}{v}\fdef} \wedge
  {\sse{i}{v'}\fdef} {\implies} (v < v' {\iff} \sse{i}{v} < \sse{i}{v'})$
\item
  \label{prop:after-t} $\ssm{\B} \ge \GST$
\item
  \label{prop:no-skip-sync:1}
  $\forall i.\, \forall v\ge \B.\, p_i \in \correct {\implies} {\sse{i}{v}\fdef}$
\item
  \label{prop:no-skip-sync:2}
  $\forall v \ge \B.\, \ssl{v} \le \ssm{v}+d$
\item
\label{prop:no-skip-sync:3}
  $\forall v \ge \B.\, \ssm{v+1} > \ssm{v}+\timeout(v)$
\end{enumerate}
\caption{Consensus synchronizer specification~\cite{bftlive}, holding for some $\B \in \View$.}
\label{fig:sync-properties}
\end{figure}

A {\em consensus synchronizer} produces a stream of notifications
$\newconsview(v)$ at each correct process, telling it to {\em enter} a view
$v$. A process can ensure that the synchronizer has started operating by calling
a special ${\tt start}()$ function. We assume that each correct process
eventually calls ${\tt start}()$, unless it gets a $\newconsview$ notification
first. For a consensus protocol to terminate, its processes need to stay in the
same view for long enough to complete the message exchange leading to a
decision. Since the message delay $\delta$ after $\GST$ is unknown to the
protocol, we need to increase the view duration until it is long enough for the
protocol to terminate. To this end, the synchronizer is parameterized by a
function defining this duration -- $\timeout : \View \cup \{0\} \to \Time$,
which is monotone, satisfies $\timeout(0) = 0$, and increases unboundedly:
\begin{equation}
\label{prop:increasing}
\forall \theta.\,\exists v.\,\forall v'.\, v'\ge v \implies F(v')>\theta.
\end{equation}

Figure~\ref{fig:single-shot} shows how we can construct a consensus synchronizer
from an SMR synchronizer. Upon a ${\tt start}()$ call, the consensus
synchronizer just tells the underlying SMR synchronizer to advance
(line~\ref{line2:start}).  When the SMR synchronizer tells the process to enter
a view $v$ (line~\ref{line2:newview}), the consensus synchronizer produces the
corresponding $\newconsview(v)$ notification. It also sets a timer $\timerview$
for the duration $F(v)$. When the timer expires (line~\ref{line2:timer-exp1}),
the consensus synchronizer tells the SMR synchronizer to advance.

Figure~\ref{fig:sync-properties} presents the specification of a consensus
synchronizer proposed in~\cite{bftlive}. This relies on the following notation,
analogous to the one used for SMR synchronizers. Given a view $v$ for which a
correct process $p_i$ received a $\newconsview(v)$ notification, we denote by
$\sse{i}{v}$ the time when this happens; we let $\ssm{v}$ and $\ssl{v}$ denote
respectively the earliest and the latest time when some correct process receives
a $\newconsview(v)$ notification. Like an SMR synchronizer, a consensus
synchronizer must guarantee that views only increase at a given process
(Property~\ref{prop:local-order}). A consensus synchronizer ensures view
synchronization only starting from some view $\B$, entered after $\GST$
(Property~\ref{prop:after-t}). Starting from $\B$, correct processes do not skip
any views (Property~\ref{prop:no-skip-sync:1}), enter each view $v \ge \B$
within at most $d$ of each other (Property~\ref{prop:no-skip-sync:2}) and stay
there for a determined amount of time: until $\timeout(v)$ after the first
process enters $v$ (Property~\ref{prop:no-skip-sync:3}). We next prove the
following theorem, showing that the consensus synchronizer in
Figure~\ref{fig:single-shot} satisfies these properties.
\begin{theorem}
  The consensus synchronizer in Figure~\ref{fig:single-shot} satisfies
  Properties~\ref{prop:local-order}-\ref{prop:no-skip-sync:3} in
  Figure~\ref{fig:sync-properties}, provided the SMR synchronizer it uses
  satisfies the properties in Figure~\ref{fig:multi-sync-properties}.
\label{th:main-app}
\end{theorem}

The implementation in Figure~\ref{fig:single-shot} ensures the following
proposition, which we tacitly use in our proof.
\begin{proposition}
$\forall i, v.\, \sse{i}{v} = \te{i}{v}$.
\end{proposition}

\begin{proposition}
$\forall v.\, \exists v'.\, v'>v \wedge {\ssm{v'}\fdef}$.
\label{lem:live}
\end{proposition}
\begin{proof}
Analogous to Proposition~\ref{lem:live-toy}.
\end{proof}

\begin{lemma}
If a correct process enters a view $v>0$ and $\tm{v} \ge \GST$,
then for all $v' > v$, no correct process attempts to advance from $v'-1$ before
$\tm{v} + \timeout(v)$.
\label{lem:i-wont-try}
\end{lemma}
\begin{proof}
Suppose by contradiction that there exists a time 
$t' < \tm{v} + \timeout(v)$ and a correct process $p_i$
such that $p_i$ attempts to advance from $v'-1 > v-1$ at $t'$.
Since $v' \ge v + 1 > 1$, at $t'$ the process $p_i$ executes
the handler at line~\ref{line2:timer-exp1} and 
the last view it entered is $v' - 1$.
Since $p_i.\timerview$ is not enabled at $t'$, $p_i$ must have
entered $v' - 1$ at least $\timeout(v)$ before $t'$ according
to its local clock. Since 
$v' - 1 \ge v$, by Proposition~\ref{sync:noskip},
we have $\tm{v' - 1} \ge \tm{v} \ge \GST$. Therefore,
given that the clocks of all correct processes progress
at the same rate as real time after $\GST$, we get
$$
\tm{v} \le \tm{v' - 1} \le t' - \timeout(v' - 1).
$$
Hence,
$$
t' \ge \tm{v} + \timeout(v' - 1).
$$
Since $\timeout$ is non-decreasing and $v' - 1 \ge v$, 
we have $\timeout(v' - 1) \ge \timeout(v)$, so that
$$
t' \ge \tm{v} + \timeout(v),
$$
which contradicts our assumption that $t' < \tm{v} + \timeout(v)$.
This contradiction shows the required.
\end{proof}

\begin{proof}[Proof of Theorem~\ref{th:main-app}]
Property~\ref{prop:local-order} follows from \monotonicity of
the SMR synchronizer.  Let $\B$ be the view from \entry and
let $\B'$ be the minimal view such that $\B' \ge \B$, $\tm{\B'} \ge \GST$ and
$\timeout(\B') \ge 2\delta$. Such a view exists by~(\ref{prop:increasing}) and
Proposition~\ref{lem:live}. Then Property~\ref{prop:after-t} holds for
$\B = \B'$. By Propositions~\ref{sync:noskip} and~\ref{lem:live}, a correct
process enters every view $v \ge \B'$. By Proposition~\ref{sync:noskip},
$v \ge \B'$ implies
\begin{equation}
\tm{v} \ge \tm{\B'} \ge \GST.
\label{eq:v-after-gst}
\end{equation}
Since $\timeout$ is a non-decreasing function, $\timeout(v) \ge 2\delta$. Thus, by
Lemma~\ref{lem:i-wont-try} and \entry, all correct processes
enter $v$, and $\tl{v} \le \tm{v} + 2\delta$, which validates
Properties~\ref{prop:no-skip-sync:1} and~\ref{prop:no-skip-sync:2} for
$\B = \B'$.  To prove Property~\ref{prop:no-skip-sync:3}, fix a view $v\ge \B'$.
Since a correct process enters view $v+1$, by \validitysync,
there exist a time $t < \tm{v+1}$ at which some correct process attempts to
advance from $v$.  By~(\ref{eq:v-after-gst}), $\tm{v} \ge \GST$. Then by
Lemma~\ref{lem:i-wont-try} we get $t \ge \tm{v} + \timeout(v)$, so that
$\tm{v+1} > t \ge \tm{v} + \timeout(v)$, as required.
\end{proof}

\section{Correctness and Performance of the Synchronizer Algorithm}
\label{sec:proof-sync}

\subsection{Proof of the Synchronizer Correctness}

The {\em local view}\/ of a process $p_i$ at time $t$, denoted $\LV{i}{t}$, is
the latest view entered by $p_i$ at or before $t$, or $0$ if $p_i$ has not
entered any views by then.

\begin{lemma}
For all $t$ and $v\ge 0$, if a correct process sends
$\WISH(v+1)$ at $t$, then there exists a time $t' \le t$
such that some correct process attempts to advance from $v$ at $t'$.
\label{lem:wish-attempt}
\end{lemma}
\begin{proof}
We first prove the following auxiliary proposition:
\begin{multline} 
\forall p_i.\, \forall v.\, 
p_i \text{~is~correct} \wedge p_i \text{~sends~} \WISH(v+1)
\text{~at~} t \implies\\
\exists t' \le t.\, \exists v' \ge v.\, \exists p_j.\,
p_j \text{~is~correct} \wedge p_j 
\text{~attempts~to~advance~from~} v' \text{~at~} t'.
\label{eq:attempt-gtv}
\end{multline}
By contradiction, assume that a correct process $p_i$ sends $\WISH(v+1)$ at $t$,
but for all $t' \le t$ and all $v' \ge v$, no correct process attempts to
advance from $v'$ at $t'$. Consider the earliest time $t_k$ when some correct
process $p_k$ sends a $\WISH(v_k)$ with $v_k \ge v+1$, so that $t_k \le t$.

Since $p_k$ sends $\WISH(v_k)$ at $t_k$, either 
$v_k = p_k.\viewp(t_k)$ or $p_k.\view(t_k) = p_k.\viewp(t_k) = v_k - 1$, and in
the latter case $p_k$ executes either line~\ref{line:send2} or line~\ref{line:send4}.
If $p_k.\viewp(t_k) = v_k \ge v+1$, then $p_k.\lastViews(t_k)$ includes $f+1$ entries 
$\ge v_k \ge v+1$, and therefore, there exists a correct process $p_l$ that 
sent $\WISH(v')$ with $v' \ge v+1$ at $t_l < t_k$, contradicting the
assumption that $t_k$ is the earliest time when this can happen.
Suppose that $p_k.\view(t_k) = p_k.\viewp(t_k) = v_k - 1$ and 
at $t_k$, $p_k$ executes either line~\ref{line:send2}
or line~\ref{line:send4}. Then $\LV{k}{t_k} = v_k - 1$.
If $p_k$ executes line~\ref{line:send2} at $t_k$, then since $\LV{k}{t_k} = v_k - 1$,
$p_k$ attempts to advance from $v_k-1 \ge v$ at $t_k\le t$, contradicting our
assumption that no such attempt can occur. Suppose now that $p_k$ executes the code in
line~\ref{line:send4} at $t_k$. If $v_k > 1$, then 
since $p_k.\view(t_k) = p_k.\viewp(t_k) = v_k - 1$, we know that
$\te{k}{v_k-1}$ is defined and satisfies $\te{k}{v_k-1} < t_k$.
Let $t'_k = \te{k}{v_k-1}$ if $v_k > 1$, and $t'_k = 0$ otherwise.
Then $p_k.\view(t'_k) = p_k.\viewp(t'_k) = v_k - 1$ and 
$p_k.\advanced(t'_k) = \FALSE$.
Since $p_k.\advanced(t_k) = \TRUE$, there exists a time
$t''_k$ such that $t'_k < t''_k \le t_k$ and $p_k$ calls $\padvance()$ at $t''_k$.
Since both $p_k.\view$ and $p_k.\viewp$ are non-decreasing,
and both are equal to $v_k - 1$ at
$t''_k$ as well as $t_k$, $p_k.\view(t''_k) = p_k.\viewp(t''_k) = v_k -
1$. Thus, $\LV{k}{t''_k} = v_k-1$, which implies that at $t''_k < t_k \le t$,
$p_k$ attempts to advance from $v_k-1 \ge v$,
contradicting our assumption that no such attempt can happen. 
Thus,~(\ref{eq:attempt-gtv}) holds.

We now prove the lemma. Let $t$ and $v$ be such that
some correct process sends $\WISH(v+1)$ at $t$.
By~(\ref{eq:attempt-gtv}), there exists a correct process
that attempts to advance from a view $\ge v$ at or before $t$.
Let $t'$ be the earliest time when some correct process attempts to advance from 
a view $\ge v$, and let $p_j$ be this process and $v' \ge v$ be the view
from which $p_j$ attempts to advance at $t'$.
Thus, at $t'$, $p_j$ executes the code in line~\ref{line:send2}
and $\LV{j}{t'} = v' \ge v$.
Hence, there exists an earlier time at which 
$p_j.\viewp = p_j.\view = v'$. Since $p_j.\viewp$ is non-decreasing,
$p_j.\viewp(t') \ge v'$. If $p_j.\viewp(t') > v'$, then given that
$v' \ge v$, $p_j.\viewp(t') \ge v+1$. Thus, there exists a correct
process $p_k$ and time $t'' < t'$ such that $p_k$ sent 
$\WISH(v'')$ with $v'' \ge v+1$ to $p_j$ at $t''$. By~(\ref{eq:attempt-gtv}),
there exists a time $\le t'' < t'$ at which some correct process
attempts to advance from a view $\ge v''-1 \ge v$, which is impossible.
Thus, $p_j.\viewp(t') = v'$. Since $\LV{j}{t'} = v'$, we have
$p_j.\view(t') = p_j.\viewp(t') = v' \ge v$.
By the definitions of $\view$ and $\viewp$,
$v'$ is both the lowest view among the highest $2f+1$ views
in $p_j.\lastViews(t')$, and the lowest view among the highest
$f+1$ views in $p_j.\lastViews(t')$. 
Hence,  $p_j.\lastViews(t')$ includes $f+1$ entries equal to $v'$, and 
therefore, there exists a correct process $p_k$ such that
\begin{equation}\label{entry-eq-v-1}
p_j.\view(t') = p_j.\viewp(t') = p_j.\lastViews[k](t') = v'
\ge v-1.
\end{equation}
Also, for all correct processes $p_l$, $p_j.\lastViews[l](t') < v+1$:
otherwise, some correct process sent $\WISH(v'')$ with $v'' \ge v+1$
at $t'' < t'$, and therefore, by~(\ref{eq:attempt-gtv}),
some correct process attempted to advance from a view $\ge v$ earlier
than $t'$, which is impossible. Thus, 
$$
p_j.\view(t') = p_j.\viewp(t') = p_j.\lastViews[k](t') < v+1.
$$
Together with~(\ref{entry-eq-v-1}), this implies
$$
p_j.\view(t') = p_j.\viewp(t') = v.
$$
Hence, $\LV{j}{t'} = v$, and therefore, $p_j$ attempts to advance from $v$ at $t'$. Thus,
$v' = v$ and $t' \le t$, as required.
\end{proof}

\begin{lemma}
\label{lemma:enter-attempt}
\validitysync holds: 
$\forall i, v.\, {\te{i}{v+1}\fdef} {\implies} {\tam{v}\fdef} \wedge \tam{v} < \te{i}{v+1}$.
\end{lemma}
\begin{proof}
Since $p_i$ enters a view $v+1$, we have
$p_i.\view(\te{i}{v+1}) = p_i.\viewp(\te{i}{v+1})=v+1$. By the definitions of $\view$
and $\viewp$, $v+1$ is both the lowest view among the highest $2f+1$ views in
$p_i.\lastViews(\te{i}{v+1})$, and the lowest view among the highest $f+1$ views
in $p_i.\lastViews(\te{i}{v+1})$. Hence, $p_i.\lastViews(\te{i}{v+1})$ includes
$f+1$ entries equal to $v+1$. Then there exists a time $t' < \te{i}{v+1}$ at which
some correct process sends $\WISH(v+1)$. Hence, by Lemma~\ref{lem:wish-attempt},
there exists a time $t \le t' < \te{i}{v+1}$ at which some correct process
attempts to advance from $v$.
\end{proof}

\begin{lemma}
For all times $t$ and views $v>0$, if a correct process sends
$\WISH(v)$ at $t$, then there exists a time $t' \le t$
such that some correct process attempts to advance from view $0$ at $t'$.
\label{lem:wish-start}
\end{lemma}
\begin{proof}
Consider the earliest time $t_k \le t$ at which some
correct process $p_k$ sends $\WISH(v_k)$ for some view $v_k$. 
By Lemma~\ref{lem:wish-attempt}, there exists a time $t_j \le t_k$ 
at which some correct process attempts to advance from
$v_k-1 \ge 0$, and therefore, sends $\WISH(v_k)$ at $t_j$.
Since $t_k$ is the earliest time when this could happen, 
we have $t_j = t_k$. Also, if $v_k - 1 > 0$, 
then $\te{k}{v_k-1}$ is defined, and hence,
by Lemma~\ref{lemma:enter-attempt}, some correct 
process attempts to advance from $v_k - 2$ by 
sending $\WISH(v_k - 1)$ earlier than $t_j = t_k$, which
cannot happen. Thus, $v_k = 1$ and at $t_k$, $p_k$ attempts to advance from $0$, 
as required.
\end{proof}

\begin{proposition}
Let $p_i$ be a correct process. Then:
\begin{enumerate}

\item \label{prop:inv:4}
$\forall v.\forall t.\, p_i \text{~sends~} \WISH(v) \text{~at~} t {\implies} 
v \in \{p_i.\viewp(t), p_i.\viewp(t)+1\}$.

\item \label{prop:inv:41}
$\forall v.\forall t.\, p_i \text{~sends~} \WISH(v) \text{~at~} t \wedge 
v = p_i.\viewp(t)+1 {\implies}$\\
$p_i.\viewp(t)=p_i.\view(t) \wedge p_i.\advanced(t) = \TRUE$.

\end{enumerate}
\label{prop:inv}
\end{proposition}

\begin{lemma}
For all views $v, v' > 0$, if a correct process sends $\WISH(v)$ before
sending $\WISH(v')$, then $v \le v'$.
\label{lem:wish-grow} 
\end{lemma}
\begin{proof}
Let $s$ and $s'$ such that $s < s'$ be the times at which a correct process
$p_i$ sends $\WISH(v)$ and $\WISH(v')$ messages, respectively. We show that
$v' \ge v$. By Proposition~\ref{prop:inv}(\ref{prop:inv:4}),
$v \in \{p_i.\viewp(s), p_i.\viewp(s)+1\}$ and
$v' \in \{p_i.\viewp(s'), p_i.\viewp(s')+1\}$.  Hence, if $v = p_i.\viewp(s)$ or
$v' = p_i.\viewp(s')+1$, then we get $v \le v'$ from the fact that $p_i.\viewp$
is non-decreasing.  It thus remains to consider the case when
$v = p_i.\viewp(s)+1$ and $v' = p_i.\viewp(s')$. In this case by
Proposition~\ref{prop:inv}(\ref{prop:inv:41}), $p_i.\viewp(s)=p_i.\view(s)$ and
$p_i.\advanced(s) = \TRUE$. We now consider several cases depending on the
line at which $\WISH(v')$ is sent.
\begin{itemize}
\item $\WISH(v')$ is sent at lines~\ref{line:send2} or~\ref{line:send4}. Then
  $v'=p_i.\viewp(s')=\max(p_i.\view(s')+1, p_i.\viewp(s'))$.  Since
  $p_i.\view$ is non-decreasing, we get
  $p_i.\viewp(s') \ge p_i.\view(s')+1 > p_i.\view(s') \ge p_i.\view(s) =
  p_i.\viewp(s)$.  Hence, $p_i.\viewp(s') > p_i.\viewp(s)$, and therefore,
  $v'=p_i.\viewp(s') \ge p_i.\viewp(s)+1=v$, as required.
\item $\WISH(v')$ is sent at line~\ref{line:send3}. Then
  $p_i.\advanced(s') = \FALSE$. Since $p_i.\advanced(s) = \TRUE$, there exists a
  time $s''$ such that $s < s'' < s'$ and $p_i$ enters a view at $s''$. By the
  view entry condition $p_i.\view(s'') > p_i.\prevv(s'')$.
  Since $p_i.\view$ is non-decreasing, we get
  $p_i.\viewp(s') \ge p_i.\view(s') \ge p_i.\view(s'') > p_i.\view(s) =
  p_i.\viewp(s)$.  Thus, $p_i.\viewp(s') > p_i.\viewp(s)$ and therefore,
  $v' = p_i.\viewp(s') \ge p_i.\viewp(s)+1 = v$, as required.
\item $\WISH(v')$ is sent at line~\ref{line:send5}. Then
  $p_i.\viewp(s') > p_i.\prevvp(s') \ge p_i.\viewp(s)$, and therefore,
  $v' = p_i.\viewp(s') \ge p_i.\viewp(s) + 1 = v$, as required.
\end{itemize}
\end{proof}

In order to cope with message loss before $\GST$, every correct process
retransmits the highest $\WISH$ it sent every $\rho$ time units, according to
its local clock (lines~\ref{line:retransmit-start}-\ref{line:send3}).
Eventually, one of these retransmissions will occur after $\GST$, and therefore,
there exists a time by which all correct processes are guaranteed to send their
highest $\WISH$es at least once after $\GST$.  The earliest such time, $\GSTP$,
is defined as follows:
\begin{equation*}
\GSTP = 
\begin{cases}
    \GST + \rho,& \text{if } \tam{0} < \GST;\\
    \tam{0},& \text{otherwise}.
\end{cases}
\label{eq:gstp-def}
\end{equation*}
From this definition it follows that
\begin{equation}
\GSTP \ge \GST.
\label{eq:gstp-gst}
\end{equation}
Lemma~\ref{lem:postgst} below formalizes the key property of $\GSTP$.
\begin{lemma}
For all correct processes $p_i$, times $t \ge \GSTP$, and views $v$, 
if $p_i$ sends $\WISH(v)$ at a time $\le t$, then there exists a view $v' \ge v$
and a time $t'$ such that $\GST \le t' \le t$ and $p_i$ sends $\WISH(v')$ at $t'$.
\label{lem:postgst}
\end{lemma}
\begin{proof}
Let $s \le t$ be the time at which $p_i$ sends $\WISH(v)$. We consider two
cases. Suppose first that $\tam{0} \ge \GST$.  By Lemma~\ref{lem:wish-start}, $s
\ge \tam{0}$, and therefore, $\GST \le s \le t$. Thus, choosing $t'=s$ and $v' =
v$ validates the lemma.
Suppose next that $\tam{0} < \GST$. Then by the definition of $\GSTP$, $t \ge \GST + \rho$. 
If $s \ge \GST$, then $\GST \le s \le t$, and therefore, 
choosing $t'=s$ and $v' = v$ validates the lemma. Assume now that $s < \GST$.
Since after $\GST$ the $p_i$'s local clock advances at the same rate as real time, 
there exists a time $s'$ satisfying $\GST \le s' \le t$ such that
$p_i$ executes the periodic retransmission code 
in  lines~\ref{line:retransmit-start}-\ref{line:send3} 
at $s'$. We now show that 
\begin{equation}
p_i.\advanced(s') \vee p_i.\viewp(s') > 0.
\label{eq:retr-wish}
\end{equation}
Since $p_i$ already sent a $\WISH$ message at $s < \GST \le s'$,
by the structure of the code, 
\begin{equation*}
p_i.\advanced(s) \vee p_i.\viewp(s) > 0.
\end{equation*}
If $p_i.\viewp(s) > 0$, then since $p_i.\viewp$ is non-decreasing,
$p_i.\viewp(s') > 0$, and therefore,~(\ref{eq:retr-wish}) holds.
Assume now that $p_i.\advanced(s)$. If $p_i.\advanced(s')$,
then~(\ref{eq:retr-wish}) holds too. We therefore consider the case when 
$\neg p_i.\advanced(s')$. Then there exists a time $s \le s'' \le s'$ 
at which $p_i$ enters the view $p_i.\view(s'')>0$. 
Hence, $p_i.\viewp(s') \ge p_i.\viewp(s'') \ge p_i.\view(s'') > 0$, 
validating~(\ref{eq:retr-wish}). 
Thus,~(\ref{eq:retr-wish}) holds in all cases.
Therefore, at $s'$ the process $p_i$ sends $\WISH(v')$ for
some view $v'$. By Lemma~\ref{lem:wish-grow}, $v' \ge v$,
and above we established $\GST \le s' \le t$, as required.
\end{proof}

\begin{lemma}
Consider a view $v>0$ and assume that $v$ is entered by a correct process.
If $\tm{v} \ge \GSTP$, and 
no correct process attempts to advance from $v$ 
before
$\tm{v} + 2\delta$, then all correct processes enter $v$ and
$\tl{v} \le \tm{v} + 2\delta$.
\label{lem:GST-bound2}
\end{lemma}
\begin{proof}
If some correct process attempts to advance from a view $v' > v$
before $\tm{v} + 2\delta$, then 
by Proposition~\ref{sync:noskip}, some correct process must also enter the view
$v+1$. By Lemma~\ref{lemma:enter-attempt}, this implies that some 
correct process attempts to advance from $v$ before $\tm{v} + 2\delta$,
contradicting the lemma's premise. Thus, no correct process attempts 
to advance from any view $v' \ge v$
before $\tm{v} + 2\delta$, and therefore,
by Lemma~\ref{lem:wish-attempt}, 
no correct process can send $\WISH(v')$ with $v' > v$ earlier
than $\tm{v} + 2\delta$. Once any such $\WISH(v')$ is sent, it will take a non-zero time until
it is received by any correct process. Thus, we have:
\smallskip
\smallskip
{\setlength{\leftmargini}{22pt}
\begin{enumerate}
\item[(*)] no correct process receives $\WISH(v')$ with $v' > v$
from a correct process until after $\tm{v} + 2\delta$. \label{lem:GST-bound2-generic:3}
\end{enumerate}}
\smallskip
\smallskip

Let $p_i$ be a correct process that enters $v$ at $\tm{v}$.
By the view entry condition,
$p_i.\view(\tm{v}) = v$, and therefore
$p_i.\lastViews(\tm{v})$ includes $2f+1$ entries $\ge v$.
At least $f+1$ of these entries belong to correct processes,
and by~(*),
none of them can be $> v$.
Hence, there exists a set $C$ of $f+1$ correct processes, each of which sends 
$\WISH(v)$ to all processes before $\tm{v}$.

Since $\tm{v} \ge \GSTP$, by Lemma~\ref{lem:postgst},  every $p_j \in C$ also sends 
$\WISH(v')$ with $v'\ge v$ at some time $s_j$ such that $\GST \le s_j \le \tm{v}$. 
Then by~(*) we have
$v'=v$. It follows that each $p_j \in C$ is guaranteed to send 
$\WISH(v)$ to all correct processes between $\GST$ and $\tm{v}$. Since all
messages sent by correct processes after $\GST$ are guaranteed to be received by
all correct processes within $\delta$ of their transmission, by
$\tm{v} + \delta$ all correct processes will receive $\WISH(v)$ from at least
$f+1$ correct processes.

Consider an arbitrary correct process $p_j$ and let $t_j \le \tm{v} + \delta$ be
the earliest time by which $p_j$ receives $\WISH(v)$ from $f+1$
correct processes. By~(*), no correct process sends
$\WISH(v')$ with $v' > v$ before $t_j < \tm{v} + 2\delta$. Thus, $p_j.\lastViews(t_j)$
includes at least $f+1$ entries equal to $v$ and at most
$f$ entries $>v$, so that $p_j.\viewp(t_j) = v$.
Then $p_j$ sends $\WISH(v)$ to all processes no later than 
$t_j \le \tm{v} + \delta$.
Since $\tm{v} \ge \GSTP$, by Lemma~\ref{lem:postgst},
$p_j$ also sends $\WISH(v')$ with $v'\ge v$ in-between
$\GST$ and $\tm{v} + \delta$. By~(*), $v'=v$, and therefore,
$p_j$ must have sent $\WISH(v)$ to all processes sometime between 
$\GST$ and $\tm{v} + \delta$. Hence, all correct processes are guaranteed to 
send $\WISH(v)$ to all correct processes between $\GST$ and $\tm{v} + \delta$. 

Consider an arbitrary correct process $p_k$ and let $t_k \le \tm{v} + 2\delta$
be the earliest time by which $p_k$ receives $\WISH(v)$ from all correct
processes. Then by~(*), all entries of correct processes in
$p_k.\lastViews(t_k)$ are equal to $v$. Since there are at least $2f+1$ correct processes:
\emph{(i)} at least $2f+1$ entries in 
$p_k.\lastViews(t_k)$ are equal to $v$, and 
\emph{(ii)} one of the $f+1$ highest entries in $p_k.\lastViews(t_k)$ is equal to $v$.
From \emph{(i)}, $p_k.\viewp(t_k) \ge p_k.\view(t_k) \ge v$, 
and from \emph{(ii)}, $p_k.\view(t_k) \le p_k.\viewp(t_k) \le v$.
Therefore, $p_k.\view(t_k) = p_k.\viewp(t_k) = v$, so that $p_k$ enters $v$ no later than 
$t_k \le \tm{v}+2\delta$. We have thus shown that by $\tm{v}+2\delta$, all
correct processes enter $v$, as required.
\end{proof}

\begin{lemma}
\startup holds: 
suppose there exists a set $P$ of $f+1$ correct  processes such that 
$\forall p_i \in P.\, {\ta{i}{0}\fdef}$;
then eventually some correct process enters view $1$.
\label{lem:enter1}
\end{lemma}
\begin{proof}
Assume by contradiction that there exists a set $P$ of $f+1$ correct  processes such that 
$\forall p_i \in P.\, {\ta{i}{0}\fdef}$, and 
no correct process enters the view $1$. By Proposition~\ref{sync:noskip},
the latter implies 
\begin{equation}
\forall v' > 0.\, \tm{v'}\fundef.
\label{eq:enter-never1}
\end{equation}
Then by Lemma~\ref{lem:wish-attempt} we have
\begin{equation}
\forall t.\, \forall v' > 1.\, \forall p_i.\, 
\neg (p_i \text{~sends~} \WISH(v') \text{~at~} t \wedge 
p_i \text{~is~correct}).
\label{eq:upper-never1}
\end{equation}
Let $T_1 = \max(\GSTP, \talast{0})$.
Since there exists a set $P$ of $f+1$ correct processes
that attempt to advance from view $0$, each $p_i\in P$ sends 
$\WISH(v_i)$ with $v_i > 0$ before $T_1$. 
Since $T_1 \ge \GSTP$, by Lemma~\ref{lem:postgst}, there exists a view $v_i'\ge 1$
and a time $s_i$ such that $\GST \le s_i \le T_1$ and $p_i$ sends $\WISH(v_i')$ at $s_i$.
By~(\ref{eq:upper-never1}), $v_i'=1$. Since the links are reliable after
$\GST$, the $\WISH(1)$ sent by $p_i$ at $s_i$ will be received by all correct
processes.

Thus, there exists a time $T_2 \ge T_1 \ge \GSTP$ by which 
all correct processes
have received $\WISH(1)$ from all processes in $P$.
Fix an arbitrary correct process $p_j$. 
Since all process in $P$ are correct,
all entries in $p_j.\lastViews(T_2)$ associated with the processes in $P$
are equal to $1$. Since $|P|=f+1$, $p_j.\lastViews(T_2)$ 
includes at least $f+1$ entries $\ge 1$, and therefore, 
$p_j.\viewp(T_2)\ge 1$.
Hence, $p_j$ sends $\WISH(v_j)$ with $v_j \ge 1$ no later than $T_2$.
Since $T_2 \ge \GSTP$, by Lemma~\ref{lem:postgst} there exists
a view $v'_j\ge 1$ and a time $s_j$ such that $\GST \le s_j \le T_2$ and
$p_j$ sends $\WISH(v'_j)$ at $s_j$. By~(\ref{eq:upper-never1}), $v'_j = 1$.
Since the links are reliable after $\GST$, the $\WISH(1)$ 
sent by $p_j$ will be received by all correct processes.

Thus, there exists a time $T_3 \ge T_2 \ge \GSTP$ by which all correct processes 
have received $\WISH(1)$ from all correct processes. Fix an arbitrary correct process $p_k$. 
By~(\ref{eq:upper-never1}), all entries of correct processes in $p_k.\lastViews(T_3)$ 
are equal to $1$. Since there are at least $2f+1$ correct processes:
\emph{(i)} at least $2f+1$ entries in 
$p_k.\lastViews(T_3)$ are equal to $1$, and 
\emph{(ii)} one of the $f+1$ highest entries in $p_k.\lastViews(T_2)$ is equal to $1$.
From \emph{(i)}, $p_k.\viewp(T_2) \ge p_k.\view(T_2) \ge 1$, 
and from \emph{(ii)}, $p_k.\view(T_2) \le p_k.\viewp(T_2) \le 1$.
Hence, $p_k.\view(T_2) = p_k.\viewp(T_2) = 1$, and therefore, 
$p_k$ enters view $1$ by $T_2$, contradicting~(\ref{eq:enter-never1}).
\end{proof}

\begin{lemma}
\progress holds: 
consider a view $v>0$ that is entered by a correct process, and suppose
there exists a set $P$ of $f+1$ correct  processes such that 
\begin{equation}
\forall p_i\in P.\, {\te{i}{v}\fdef} \implies {\ta{i}{v}\fdef};
\label{eq:live-pre}
\end{equation}
then eventually some correct process enters $v+1$.
\label{lem:gv-live}
\end{lemma}
\begin{proof}
Assume by contradiction that the required does not hold. Then, there exists
a view $v > 0$ such that some correct process enters $v$,~(\ref{eq:live-pre})
holds, and no correct process enters the view $v+1$. By Proposition~\ref{sync:noskip},
the latter implies that 
\begin{equation}
\forall v' > v.\, \tm{v'}\fundef.
\label{eq:enter-never}
\end{equation}
Thus, by Lemma~\ref{lem:wish-attempt}, we have
\begin{equation}
\forall t.\, \forall v' > v + 1.\, \forall p_i.\,
\neg (p_i \text{~sends~} \WISH(v') \text{~at~} t \wedge 
p_i \text{~is~correct}).
\label{eq:upper-never}
\end{equation}
Let $T_1 = \max(\GSTP, \tm{v})$.
Since some correct process entered $v$ by $T_1$, there
exists a set $C$ consisting of $f+1$ correct processes all of which sent
$\WISH(v')$ with $v'\ge v$ before $T_1$. 
Consider $p_i\in C$ and let $t_i \le T_1$ be a time such that at $t_i$
the process $p_i$ sends $\WISH(v_i)$ with $v_i \ge v$.
Since $T_1 \ge \GSTP$, by Lemma~\ref{lem:postgst}, there exists a view $v_i'\ge v_i$
and a time $s_i$ such that $\GST \le s_i \le T_1$ and $p_i$ sends $\WISH(v_i')$ at $s_i$.
By~(\ref{eq:upper-never}), we have $v_i' \in \{v, v+1\}$.
Since the links are reliable after $\GST$, the $\WISH(v_i')$ sent by $p_i$ at
$s_i$ will be received by all correct processes.

Thus, there exists a time $T_2 \ge T_1 \ge \GSTP$  by which all correct
processes have received $\WISH(v')$ with $v' \in \{v, v+1\}$
from all processes in $C$. Consider an arbitrary correct process $p_j$.
By~(\ref{eq:upper-never}), the entry of every process in $C$ in
$p_j.\lastViews(T_2)$  is equal to either $v$ or $v+1$.
Since $|C|\ge f+1$ and all processes
in $C$ are correct, $p_j.\lastViews(T_2)$ includes at least $f+1$ entries $\ge v$.
Thus, $p_j.\viewp(T_2) \ge v$, and therefore, $p_j$ sends
$\WISH(v_j)$ with $v_j \ge v$ no later than at $T_2$. By~(\ref{eq:upper-never}),
$v_j \in \{v, v+1\}$.
Since $T_2 \ge \GSTP$, by Lemma~\ref{lem:postgst}, there exists a view $v_j'\ge v$ and a time
$s_j$ such that $\GST \le s_j \le t_j$ and $p_j$ sends $\WISH(v_j')$ at $s_j$.
By~(\ref{eq:upper-never}), $v_j' \in \{v, v+1\}$.
Since the links are reliable after
$\GST$, the $\WISH(v_j')$ sent by $p_j$ at $s_j$ will be received by all correct
processes.  

Thus, there
exists a time $T_3 \ge T_2 \ge \GSTP$  by which all correct processes have
received $\WISH(v')$ such that $v' \in \{v, v+1\}$ from
all correct processes. 
Consider an arbitrary correct process $p_k$, and suppose that $p_k$
is a member of the set $P$ stipulated by the lemma's premise.
Then at $T_3$, all entries of correct processes
in $p_k.\lastViews$ are $\ge v$.
By~(\ref{eq:upper-never}), each of these entries is equal to either $v$ or $v+1$.
Since at least $2f+1$ processes are correct:
\emph{(i)} at least $2f+1$ entries in $p_k.\lastViews(T_3)$ are $\ge v$, and 
\emph{(ii)} one of the $f+1$ highest entries in $p_k.\lastViews(T_3)$ is $\le v+1$. 
From \emph{(i)}, $p_k.\viewp(T_3) \ge p_k.\view(T_3) \ge v$, 
and from \emph{(ii)}, $p_k.\view(T_3) \le p_k.\viewp(T_3) \le v+1$.
Hence, $p_k.\view(T_3), p_k.\viewp(T_3) \in \{v, v+1\}$.
Since no correct process enters $v+1$, $p_k.\view(T_3)$ and $p_k.\viewp(T_3)$
cannot be both simultaneously equal to $v+1$. Thus,
$p_k.\view(T_3) = v$, and either $p_k.\viewp(T_3) = v$ 
or $p_k.\viewp(T_3)=v+1$. If $p_k.\viewp(T_3) = v+1$, then 
$p_k$ has sent $\WISH(v_k)$ with $v_k=v+1$ when $p_k.\viewp$
has first become equal to $v+1$ sometime before $T_3$. On the other hand,
if $p_k.\view(T_3) = p_k.\viewp(T_3) = v$, then $p_k$ has entered
$v$ at some time $t\le T_3$. 
Since $p_k\in P$, 
by~(\ref{eq:live-pre}), there exists a time $t' \ge t$ such that
$p_k$ attempts to advance from $v$ at $t'$, and therefore, sends 
$\WISH(v_k)$ with $v_k \ge v+1$ at $t'$. By~(\ref{eq:upper-never}), 
$v_k \le v + 1$, and therefore, $v_k = v+1$. 
Thus, there exists a time $t_k \ge T_3$ by which $p_k$ sends $\WISH(v+1)$ to all processes.
Since $t_k \ge T_3 \ge \GSTP$, by Lemma~\ref{lem:postgst}, there exists
a view $v'_k\ge v+1$ and a time $s_k$ such that $\GST \le s_k \le t_k$ and
$p_k$ sends $\WISH(v'_k)$ at $s_k$. By~(\ref{eq:upper-never}), $v'_k=v+1$.
Since the links are reliable after $\GST$, the $\WISH(v+1)$ 
sent by $p_k$ will be received by all correct processes.

Thus, there exists a time $T_4 \ge T_3 \ge \GSTP$ by which 
all correct processes
have received $\WISH(v+1)$ from all processes in $P$.
Fix an arbitrary correct process $p_l$. 
Since all process in $P$ are correct,
by~(\ref{eq:upper-never}),
all entries in $p_l.\lastViews(T_4)$ associated with the processes in $P$
are equal to $v+1$. Since $|P|=f+1$, $p_l.\lastViews(T_4)$ 
includes at least $f+1$ entries equal to $v+1$, and therefore, 
$p_l.\viewp(T_4)\ge v+1$.
Hence, $p_l$ sends $\WISH(v_l)$ with $v_l \ge v+1$ no later than $T_4$. 
Since $T_4 \ge \GSTP$, by Lemma~\ref{lem:postgst} there exists
a view $v'_l\ge v+1$ and a time $s_l$ such that $\GST \le s_l \le T_4$ and
$p_l$ sends $\WISH(v'_l)$ at $s_l$. By~(\ref{eq:upper-never}), $v_l' = v+1$.
Since the links are reliable after $\GST$, the $\WISH(v+1)$ 
sent by $p_l$ will be received by all correct processes.

Thus, there exists a time $T_5 \ge T_4 \ge \GSTP$ by which all correct processes 
have received $\WISH(v + 1)$ from all correct processes. Fix an arbitrary correct process $p_m$. 
By~(\ref{eq:upper-never}), all entries of correct processes in $p_m.\lastViews(T_5)$ 
are equal to $v+1$. Since there are at least $2f+1$ correct processes:
\emph{(i)} at least $2f+1$ entries in 
$p_m.\lastViews(T_5)$ are equal to $v+1$, and 
\emph{(ii)} one of the $f+1$ highest entries in $p_m.\lastViews(T_5)$ is equal to $v+1$.
From \emph{(i)}, $p_m.\viewp(T_5) \ge p_m.\view(T_5) \ge v+1$, 
and from \emph{(ii)}, $p_m.\view(T_5) \le p_l.\viewp(T_5) \le v+1$.
Hence, $p_m.\view(T_5) = p_m.\viewp(T_5) = v+1$, and therefore, 
$p_m$ enters $v+1$ by $T_5$, contradicting~(\ref{eq:enter-never}).
\end{proof}

\begin{theorem}
  Consider an execution with an eventual message delay $\delta$.
  Then %
  in this execution the algorithm in Figure~\ref{fig:sync} satisfies
  the properties in Figure~\ref{fig:multi-sync-properties} for $d=2\delta$.
\label{thm:smr-sync-correct2}
\end{theorem}
\begin{proof}
\monotonicity is satisfied trivially, and 
\validitysync, \startup, and \progress
are given by Lemmas~\ref{lemma:enter-attempt},~\ref{lem:enter1}, and~\ref{lem:gv-live},
respectively. To prove \entry, let
\begin{equation}
\B=\max\{v \mid {\tm{v}\fdef} \wedge \tm{v} < \GSTP\} + 1.
\label{eq:theview}
\end{equation}
Then $\forall v \ge \B.\, {\tm{v}\fdef} {\implies} \tm{v} \ge \GSTP$. Thus, by
Lemma~\ref{lem:GST-bound2}, \entry holds for $d=2\delta$, as
required.
\end{proof}

\subsection{Proof of the Synchronizer Performance Properties}

The following lemma bounds the latency of entering
$v$ as a function of the time by which all correct processes have sent such
$\WISH$es.

\begin{lemma}
For all views $v>0$ and times $s$, if all correct processes $p_i$
send $\WISH(v_i)$ with $v_i \ge v$ no later than at $s$, and
some correct process enters $v$, then $\tl{v} \le \max(s, \GSTP) + \delta$.\!\!\!\!
\label{lem:simple-bound-generic}
\end{lemma}
\begin{proof}
Fix an arbitrary correct process $p_i$ that sends
$\WISH(v_i)$ with $v_i \ge v$
to all processes at time $t_i \le s \le \max(s, \GSTP)$.
Since $\max(s, \GSTP) \ge \GSTP$, by Lemma~\ref{lem:postgst}
there exists a time $t_i'$ such that $\GST \le t_i' \le \max(s, \GSTP)$ and at $t_i'$, 
$p_i$ sends $\WISH(v_i')$ with $v_i' \ge v_i \ge v$
to all processes. Since $t_i' \ge \GST$, all correct processes 
receive $\WISH(v_i')$ from $p_i$ no later than at 
$t_i' + \delta \le \max(s, \GSTP) + \delta$.

Consider an arbitrary correct process $p_j$ and let 
$t_j \le \max(s, \GSTP) + \delta$ be the earliest time by which 
$p_j$ receives $\WISH(v_i')$ with with $v_i' \ge v$
from each correct processes $p_i$.
Thus, at $t_j$, the entries of all correct processes 
in $p_j.\lastViews$ are occupied by views $\ge v$.
Since at least $2f+1$ entries in $p_j.\lastViews$ belong to correct processes,
the $(2f+1)$th highest entry is $\ge v$. Thus,
$p_j.\view(t_j) \ge v$. Since $p_j.\view$ is non-decreasing,
there exists a time $t_j' \le t_j$ at which $p_j.\view$ first became
$\ge v$. If $p_j.\view(t_j') = p_j.\viewp(t_j') = v$, then 
$p_j$ enters $v$ at $t_j'$. Otherwise, either 
$p_j.\view(t_j') > v$ or $p_j.\viewp(t_j') > v$. 
Since both $p_j.\view$ and $p_j.\viewp$ are non-decreasing,
$p_j$ will never enter $v$ after $t_j'$. Thus, a correct process cannot
enter $v$ after $\max(s, \GSTP) + \delta$. Since
by the lemma's premise, some correct process does enter $v$, 
$\tl{v} \le \max(s, \GSTP) + \delta$, as needed.
\end{proof}

The next lemma gives an upper bound on the duration of time a correct process
may spend in a view before sending a $\WISH$ for a higher view.
\begin{lemma}
  Let $p_k$ be a correct process that enters a view $v$.
  Then $p_k$ sends $\WISH(v_k)$ with $v_k \ge v+1$ 
  no later than at $\taelast{v}$.
  \label{lem:one-proc-after-tl}
\end{lemma}
\begin{proof}
Suppose that $p_k$ enters a view $v > 0$ at time $\GST \le s_k \le \tl{v}$. Then
$$
p_k.\view(s_k) = p_k.\viewp(s_k) = v.
$$
By the definition of $\taelast{v}$, there exists a time $s_k'$ such that
$$
s_k \le s_k' \le \taelast{v},
$$
and at $s_k'$, $p_k$ either attempts to advance from $v$ or enters
a view $v' > v$. 
If $p_k$ attempts to advance from $v$ at $s_k'$, then 
$p_k$ sends $\WISH(v_k)$ with 
$v_k = \max(p_k.\view(s_k')+1, p_k.\viewp(s_k'))$.
Since both $p_k.\view$ and $p_k.\viewp$ are non-decreasing,
we have $p_k.\view(s_k') \ge v$ and $p_k.\viewp(s_k') \ge v$.
Thus, $v_k \ge v + 1$, as required.
On the other hand, if $p_k$ enters a view $v' > v$
at $s_k'$, then $v' = p_k.\view(s_k') > p_k.\view(s_k) = v$
and therefore, $p_k.\viewp(s_k') \ge p_k.\view(s_k') \ge v+1$.
Since $p_k.\viewp$ is non-decreasing and $p_k.\viewp(s_k)=v$,
$p_k.\viewp$ must have changed its value from $v$ to 
$v_k'' \ge v+1$ at some time $s_k''$ such that $s_k < s_k'' \le s_k'$. Thus, the condition
in line~\ref{line:cond-send5} holds at $s_k''$, which means 
that $p_k$ sends $\WISH(v_k)$ with $v_k \ge v+1$ at $s_k''$.
Thus, in all cases,
$p_k$ sends $\WISH(v_k)$ with $v_k \ge v+1$ no later than at 
$\max(\tl{v}, \GST) + \timeout(v)$, as required.
\end{proof}

The next lemma bounds the time by which every correct process
either enters a view $v > 0$, or sends a $\WISH$ messages with 
a view $> v$.

\begin{lemma}
Consider a view $v > 0$ such that some correct process enters $v$.
Then, for all times $t$, if $t \ge \max(\tm{v}, \GSTP)$, then
$\tl{v} \le t+2\delta$ and for all correct processes $p_k$,
if $p_k$ never enters $v$, then, by $\tl{v}$,
$p_k$ sends $\WISH(v_k)$ with $v_k \ge v+1$ to all processes.
\label{lem:all-send-wish1-generic:aux}
\end{lemma}
\begin{proof}
Since $v>0$, $\tm{v}\fdef$, and $t \ge \tm{v}$,
there exists a correct process $p_l$ such 
that $p_l$ entered $v$ and $\te{l}{v} \le t$. By the view entry
condition, $p_l.\view(\te{l}{v}) = v$, and therefore
$p_l.\lastViews(\te{l}{v})$ includes $2f+1$ entries 
$\ge v$. Since $f+1$ of these entries belong to correct processes, 
there exists a set $C$ of $f+1$ correct processes $p_i$, each of which sent 
$\WISH(v_i)$  with $v_i\ge v$ to all processes before 
$\te{l}{v} \le t$. 
Since  $t \ge \GSTP$, by Lemma~\ref{lem:postgst},
$p_i$ sends $\WISH(v_i')$ with $v_i' \ge v_i \ge v$ sometime
between $\GST$ and $t$. 
Since after $\GST$ every message sent by a correct process is received
by all correct processes within $\delta$ of its transmission, 
the above implies that by $t + \delta$ every correct process receives a
$\WISH(v_i')$ with $v_i' \ge v$ from each process $p_i \in C$. 

Consider an arbitrary correct process $p_j$ and let $t_j \le t + \delta$ be the
earliest time by which $p_j$ receives $\WISH(v_i)$ with 
$v_i \ge v$ from each process $p_i \in C$. 
Thus, for all processes $p_i\in C$, 
$p_j.\lastViews[i](t_j) \ge v$.
Since $|C|=f+1$, the $(f+1)$th highest entry in $p_j.\lastViews[i](t_j)$
is $\ge v$, and therefore, $p_j.\viewp(t_j) \ge v$.
Then each correct process $p_j$ sends $\WISH(v_j)$ with $v_j \ge v$ to all
correct processes no later than $t_j \le t + \delta$. Since
$t+\delta > t \ge \GSTP$ and, and
some correct process entered $v$, by Lemma~\ref{lem:simple-bound-generic},
\begin{equation}\label{simple-bound-generic-app}
\tl{v} \le t + 2\delta.
\end{equation}
In addition, by Lemma~\ref{lem:postgst}, there exists a time
$t_j'$ such that $\GST \le t_j' \le t+\delta$ and $p_j$ sends $\WISH(v_j')$
with $v_j' \ge v_j \ge v$ at $t_j'$. Since a message sent by a correct
process after $\GST$ is received by all correct processes within $\delta$
of its transmission, all correct processes must have
received $\WISH(v_j')$ with $v_j' \ge v$ from each correct process
$p_j$ in-between $\GST$ and $t + 2\delta$.

Suppose that $p_k$ never enters $v$, and let $t_k$ be
the earliest time $\ge \GST$ by which $p_k$ receives $\WISH(v_j')$ 
from each correct process $p_j$; we have $t_k \le t+2\delta$.
Since $v_j' \ge v$, and there are $2f+1$ correct processes,
$p_k.\lastViews(t_k)$ includes at least $2f+1$ entries $\ge v$.
Thus, $p_k.\view(t_k) \ge v$. Since $p_k$ never enters $v$,
we have either $p_k.\viewp(t_k) \ge p_k.\view(t_k) \ge v+1$
or $p_k.\view(t_k) = v \wedge p_k.\viewp(t_k) \ge v+1$.
Thus, $p_k.\viewp(t_k) \ge v + 1$ and 
therefore, $p_k$ sends $\WISH(v_k)$ with $v_k\ge v+1$
by $t_k \le t+2\delta$, which combined with~(\ref{simple-bound-generic-app})
validates the lemma.
\end{proof}

We are now ready to prove the SMR synchronizer performance
bounds.
\begin{theorem}
The SMR synchronizer in Figure~\ref{fig:sync} satisfies
Property~\ref{eq:gen-bounded-entry:main}.
\label{thm:gen-bounded-entry:main}
\end{theorem}
\begin{proof} 
Consider a view $v$ such that $\tm{v}\fdef$, and let  $t =
\max(\tm{v}, \GSTP)$. Since $\tam{0} < \GST$, by the definition of  $\GSTP$,
$\GSTP = \GST + \rho$. Thus, $t = \max(\tm{v}, \GST + \rho)$. By
Lemma~\ref{lem:all-send-wish1-generic:aux}, $\tl{v} \le t + 2\delta =
\max(\tm{v}, \GST + \rho) + 2\delta$, as needed.
\end{proof}

\begin{theorem}
The SMR synchronizer in Figure~\ref{fig:sync} satisfies
Property~\ref{eq:lat-bound1:main}.
\label{thm:lat-bound1:main}
\end{theorem}
\begin{proof}
Consider a view $v\ge 0$ such that $\tm{v+1}\fdef$. 
If $v=0$, then since we assume for all 
correct processes $p_i$, $\tae{i}{0}\fdef$, by Lemma~\ref{lem:one-proc-after-tl},
all correct processes send $\WISH(v')$ with $v' \ge 0$ to all processes
no later than at $\taelast{0}$. Thus, by Lemma~\ref{lem:simple-bound-generic},
$\tl{1} \le \max(\taelast{0}, \GSTP) + \delta$.
If $\tam{0} < \GST$, then $\GSTP = \GST + \rho$, and therefore,
$\tl{1} \le \max(\taelast{0}, \GST + \rho) + \delta$. Otherwise,
$\GSTP = \tam{0} \le \taelast{0}$, so that 
$\tl{1} \le \taelast{0} + \delta$. Thus, the theorem holds
for $v=0$.

Suppose that $v>0$. 
Since some correct process enters $v+1$, 
by Proposition~\ref{sync:noskip}, some correct process
enters view $v$ as well. 
Consider a correct process $p_k$. If $p_k$ enters $v$,
then by Lemma~\ref{lem:all-send-wish1-generic:aux},
$\te{k}{v} \le \max(\tm{v}, \GSTP)+2\delta$, and therefore,
by Lemma~\ref{lem:one-proc-after-tl}, $p_k$ sends
$\WISH(v_k)$ with $v_k \ge v+1$ no later than at
\begin{equation}
\taelast{v} > \max(\tm{v}, \GSTP)+2\delta.
\label{eq:taelast-max} 
\end{equation}
On the other hand, if $p_k$ never enters $v$, then 
by Lemma~\ref{lem:all-send-wish1-generic:aux}, $p_k$
sends $\WISH(v_k)$ with $v_k \ge v+1$ than 
at $\max(\tm{v}, \GSTP)+2\delta$. Thus, every correct process $p_k$
sends $\WISH(v_k)$ with $v_k \ge v+1$ no later than
$$
\max(\taelast{v}, \max(\tm{v}, \GSTP)+2\delta),
$$
which by~(\ref{eq:taelast-max}), implies that all correct
processes send a $\WISH$ message with a view $\ge v+1$ 
no later than $\taelast{v}$. Thus, 
by Lemma~\ref{lem:simple-bound-generic}, we have
\begin{equation}
\tl{v+1} \le \max(\taelast{v}, \GSTP) + \delta.
\label{eq:vplus1}
\end{equation}
If $\tam{0} < \GST$, then $\GSTP = \GST + \rho$, and 
therefore,~(\ref{eq:vplus1}) implies that
$\tl{v+1} \le \max(\taelast{v}, \GST + \rho) + \delta$, as required.
Otherwise, $\GSTP = \tam{v} \le \taelast{v}$, which
by~(\ref{eq:vplus1}) implies that  $\tl{v+1} \le \taelast{v} + \delta$,
validating the theorem.
\end{proof}

\begin{proof}[Proof of Theorem~\ref{thm:smr-sync-correct}]
Follows from Theorems~\ref{thm:smr-sync-correct2},~\ref{thm:gen-bounded-entry:main},
and~\ref{thm:lat-bound1:main}.
\end{proof}

\section{Additional Material about \pbft}

\subsection{Proof of Safety for \pbft}
\label{sec:pbft-safety}

Let us write $\wf(C)$ (for {\em well-formed}) if the set of correctly signed
messages $C$ were generated in the execution of the protocol.
In \pbft, committing a value requires preparing it, which implies
\begin{proposition}
  \label{lemma:pbft:committed-prepared}
$$
  \forall k, v, C, h.\, \committed(C, v, k, h) \wedge \wf(C) {\implies}
 \exists C'.\, \accepted(C', v, k, h) \wedge \wf(C').
$$
\end{proposition}
Furthermore, the validity checks in the protocol ensure that any prepared value is valid:
\begin{proposition}
\label{lemma:pbft:validityliveness}
$\forall k, v, C, \val .\, \accepted(C, v, k, \hash(\val)) \wedge \wf(C) {\implies} \validity(\val)$.
\end{proposition}
The above two propositions imply
\begin{corollary}
\pbft satisfies External Validity.
\end{corollary}

\begin{proposition}
\label{lemma:pbft:view-increase}
The variables $\currview$ and $\prepview[k]$ (for any $k$) at a correct process
never decrease and we always have $\prepview[k] \le \currview$.
\end{proposition}

\begin{proposition}
\label{lemma:pbft:singlecmd}
\begin{align*}
\forall k, v, C, C', \val, \val '.\, & \accepted(C, v, k, \hash(\val)) \wedge
\accepted(C', v, k, \hash(\val'))  \wedge {}
\\
 &  \wf(C) \wedge \wf(C') {\implies} \val = \val'.
\end{align*}
\end{proposition}
\begin{proof}
By contradiction, suppose that $\val\not=\val '$. Because a
$\accepted$ certificate consists of at least $2f+1$ $\PREPARE$ messages and
there are $3f+1$ processes in total, there must be a correct process that sent
two $\PREPARE$ messages with different hashes for the same position and
view. But this is impossible due to the check on the check on $\phase$ in
line~\ref{alg:castro:safety-check}.
\end{proof}

\begin{lemma}
\label{lemma:pbft:nodupl}
If $m = \langle\NEWVIEW(v', \vcmd', M)\rangle_{\leader(v')}$ is a sent message
such that $\ValidNewState(m)$, then
$$
\forall k, k'.\, \vcmd'[k] = \vcmd'[k'] \not\in \{\bot, \noop\} {\implies} k=k'.
$$
\end{lemma}
\begin{proof} 
We prove the statement of the lemma by induction on $v'$. Assume this holds for
all $v' < v^*$; we now prove it for $v' = v^*$. Let
$$
M = \{\langle \NEWLEADER(v', \vprepview_j, \vcmd_j, \vcert_j) \rangle_j \mid p_j
\in Q\}
$$
for some quorum $Q$. By contradiction, assume that for some $k$, $k'$ and $\val$
we have $k \not= k'$, $\vcmd'[k] = \vcmd'[k'] = \val \not\in \{\bot,
\noop\}$. Since $\ValidNewState(m)$, $\vcmd'$ is computed from $M$ as per
lines~\ref{alg:castro:select-proposal}-\ref{alg:castro:newview-end}. Then due to the loop at
line~\ref{alg:castro:clean-entries}, for some $i, i' \in Q$ we have
$\vcmd_i[k] = \vcmd_{i'}[k'] = \val$ and
$\vprepview_{i}[k] = \vprepview_{i'}[k'] = v$ for some $v$ such that
$0 < v < v'$. Hence, for some $C$ and $C'$ we have
$$
\accepted(C, v, k, \hash(\val))
\wedge
\accepted(C', v, k', \hash(\val))
\wedge
\wf(C) \wedge \wf(C').
$$
Because a $\accepted$ certificate consists of at least $2f+1$ $\PREPARE$
messages and there are $3f+1$ processes in total, there must be a correct process
that sent messages $\PREPARE(v, k, \hash(\val))$ and
$\PREPARE(v, k', \hash(\val))$. But this is impossible because by the induction
hypothesis, the process starts the view $v$ with a log without duplications
(except $\noop$s), and does not add duplicate entries due to the check at
line~\ref{alg:castro:safety-check}. This contradiction demonstrates the
required.
\end{proof}

\begin{corollary}
\label{cor:pbft:nodupl}
\begin{align*}
\forall \val, v, k, k', C, C'.\, &
\accepted(C, v, k, \hash(\val)) \wedge
\accepted(C', v, k', \hash(\val)) \wedge {}
\\
& \wf(C) \wedge \wf(C') \wedge x \not= \noop
{\implies} k = k'.
\end{align*}
\end{corollary}
\begin{proof} 
Assume the contrary. Because a $\accepted$ certificate consists of at least
$2f+1$ $\PREPARE$ messages and there are $3f+1$ processes in total, there must
be a correct process that sent messages $\PREPARE(v, k, \hash(\val))$ and
$\PREPARE(v, k', \hash(\val))$. But this is impossible because by
Lemma~\ref{lemma:pbft:nodupl}, the process starts the view $v$ with a log
without duplications (except $\noop$s), and does not add duplicate entries due
to the check at line~\ref{alg:castro:safety-check}. This contradiction demonstrates the
required.
\end{proof}

\begin{lemma}
\label{lemma:pbft:main}
Fix $k$, $v$, $v'$, $C$ and $\val$, and assume
$$
\committed(C, v, k, \hash(\val)) \wedge \wf(C) \wedge v' > v.
$$
\begin{itemize}
\item
$\forall C', \val'.\, \accepted(C', v', k, \hash(\val')) \wedge \wf(C') {\implies} \val = \val'$.
\item
$\forall C', k'.\, \val \not= \noop \wedge
\accepted(C', v', k', \hash(\val)) \wedge \wf(C') {\implies} k = k'$.
\end{itemize}
\end{lemma}
\begin{proof}
We prove the statement of the lemma by induction on $v'$. Assume this holds for
all $v' < v^*$; we now prove it for $v' = v^*$. Thus, we have 
\begin{equation}\label{pbft:hyp4}
\forall C'', k'', v''.\, v < v'' < v' \wedge \val \not= \noop \wedge \accepted(C'', v'',
k'', \hash(\val)) \wedge \wf(C'') {\implies} k = k''.
\end{equation}
The induction hypothesis also implies
$$
\forall C'', \val'', v''.\, v < v'' < v' \wedge \accepted(C'', v'',
k, \hash(\val'')) \wedge \wf(C'') {\implies} \val = \val''.
$$
Furthermore, by Propositions~\ref{lemma:pbft:singlecmd}
and~\ref{lemma:pbft:committed-prepared} we have
$$
\forall C'', \val''.\, 
\accepted(C'', v, k, \hash(\val'')) \wedge \wf(C'') {\implies} \val = \val'',
$$
so that overall we get
\begin{equation}\label{pbft:hyp2}
\forall C'', \val'', v''.\, v \le v'' < v' \wedge \accepted(C'', v'', k, 
\hash(\val'')) \wedge \wf(C'') \implies \val = \val''.
\end{equation}

Assume now that $\accepted(C', v', k, \hash(\val'))$ and $\wf(C')$. Then a
correct process that sent the corresponding $\PREPARE$ message must have
received a message $m =\NEWVIEW(v', \vcmd', M)$ from the leader of $v'$
satisfying $\ValidNewState(m)$. Let
$$
M = \{\langle \NEWLEADER(v', \vprepview_j, \vcmd_j, \vcert_j) \rangle_j \mid p_j
\in Q\}
$$
for some quorum $Q$. Since $\ValidNewState(m)$, we have
$\forall m' \in M.\, \ValidNewLeader(m')$, so that
\begin{multline*}
  \forall p_j \in Q.\, \vprepview_j < v' \wedge {}\\
  ({\vprepview_j \not= 0}
{\implies}
\accepted(\vcert_j, \vprepview_j, k, \hash(\vcmd_j[k])) \wedge
\wf(\vcert_j)).
\end{multline*}
From this and~(\ref{pbft:hyp2}) we get that
\begin{equation}\label{pbft:hyp3}
  \forall p_j \in Q.\, {\vprepview_j \ge v} {\implies}  \vcmd_j[k] = \val.
\end{equation}

Since $\committed(C, v, k, \hash(x))$, a quorum $Q'$ of processes sent
$\COMMIT(v, k, \hash(x))$.  The quorums $Q$ and $Q'$ have to intersect in some
correct process $p_{i}$, which has thus sent both $\COMMIT(v, k, \hash(\val))$ and
$\NEWLEADER(v', \vprepview_i, \vcmd_i, \vcert_i)$. Since $v< v'$, this process
$p_{i}$ must have sent the $\COMMIT$ message before the $\NEWLEADER$
message. Before sending $\COMMIT(v, k, \hash(\val))$ the process set
$\prepview[k]$ to $v$ (line~\ref{alg:castro:assign-cballot}). Then by
Proposition~\ref{lemma:pbft:view-increase} process $p_{i}$ must have had
$\prepview[k] \ge v$ when it sent the $\NEWLEADER$ message. Hence,
$\vprepview_{i}[k] \ge v > 0$ and $\max\{\vprepview_{j'}[k] \mid p_{j'} \in Q\} \ge v$.
Then from~(\ref{pbft:hyp3}) we get
\begin{equation}\label{pbft:concl1}
\forall p_j \in Q.\, \vprepview_j[k] = \max\{\vprepview_{j'}[k] \mid p_{j'} \in Q\} {\implies}
\vcmd_j[k] = \val.
\end{equation}

Assume now that $\val \not= \noop$, but $\vcmd'[k] = \noop$ due to
line~\ref{alg:castro:newview-end}. Then
$$
\exists k'.\, k'  \not= k \wedge \vcmd'[k'] = \val \wedge \exists p_j \in Q.\, 
\forall p_{j'} \in Q.\, \vprepview_j[k'] > \vprepview_{j'}[k]
$$
and $v' > \vprepview_j[k'] > \vprepview_{i}[k] \ge v$. Since
$\ValidNewState(m)$, for some $C''$ we have
$\accepted(C'', \vprepview_j[k'], k', \hash(\val))$ and $\wf(C'')$.  Then
by~(\ref{pbft:hyp4}) we have $k = k'$, which yields a contradiction. This
together with~(\ref{pbft:concl1}) and $\ValidNewState(m)$ implies
$\vcmd'[k] = \val$, as required.

Assume now $\val \not= \noop$, $\accepted(C', v', k', \hash(\val))$ and
$\wf(C')$. Then a correct process that sent the corresponding $\PREPARE$ message
must have received a message $m = \NEWVIEW(v', \vcmd', M)$ from the leader of
$v'$ satisfying $\ValidNewState(m)$. As before, we can show $\vcmd'[k] =
\val$. By Lemma~\ref{lemma:pbft:nodupl}, the process starts the view $v'$ with a
log without duplications (except $\noop$s), and does not add duplicate entries
due to the check at line~\ref{alg:castro:safety-check}. Hence, we must have
$k' = k$, as required.
\end{proof}

\begin{corollary}\label{thm:pbft:agreement}
\pbft satisfies Ordering.
\end{corollary}
\begin{proof}
By contradiction, assume that Ordering is violated. Then for some $k$, two
correct processes execute the handler in line~\ref{alg:castro:deliver} for
$\lastdelivered = k-1$ so that $\comcmd[k] = \val$ at one process and
$\comcmd[k] = \val'$ at the other, where $\val \not= \val'$. Then
$\committed(C, v, k, \hash(\val))$ and $\committed(C', v', k, \hash(\val'))$ for
some well-formed $C$ and $C'$. By
Proposition~\ref{lemma:pbft:committed-prepared} we have
$\accepted(C_0, v, k, \hash(\val))$ and $\accepted(C'_0, v', k, \hash(\val'))$
for some well-formed $C_0$ and $C'_0$. Without loss of generality assume
$v \le v'$.  If $v = v'$, then $\val = \val'$ by
Proposition~\ref{lemma:pbft:singlecmd}. If $v < v'$, then $\val = \val'$ by
Lemma~\ref{lemma:pbft:main}. In either case we get a contradiction.
\end{proof}

\begin{corollary}\label{thm:pbft-integrity}
\pbft satisfies Integrity.
\end{corollary}
\begin{proof} 
By contradiction, assume that Integrity is violated. Then for some $k$, $k'$
such that $k\not=k' $and $\val \not= \noop$, a correct process executes the
handler in line~\ref{alg:castro:deliver} first in a view $v$ for
$\lastdelivered = k-1$ and $\comcmd[k] = \val$ and then in a view $v'$ for
$\lastdelivered = k'-1$ and $\comcmd[k'] = \val$. We must have
$\committed(C, v, k, \hash(\val))$ and $\committed(C', v', k', \hash(\val'))$
for some well-formed $C$ and $C'$.  By
Proposition~\ref{lemma:pbft:committed-prepared} we have
$\accepted(C_0, v, k, \hash(\val))$ and $\accepted(C'_0, v', k', \hash(\val'))$
for some well-formed $C_0$ and $C'_0$. Without loss of generality assume
$v \le v'$. If $v = v'$, then we get a contradiction by
Corollary~\ref{cor:pbft:nodupl}. If $v < v'$, then we get a contradiction by
Lemma~\ref{lemma:pbft:main}.
\end{proof}

\subsection{Additional Details for the Proof of Liveness of \pbft}
\label{sec:pbft-liveness}

\begin{proof}[Proof of Lemma~\ref{thm:castro:completeness}]
We know that at some point $p_i$ enters view $v$, and at this moment it starts
$\timerrecovery$. If the timer expires, then $p_i$ calls $\padvance$ in $v$, as
required. Assume that $\timerrecovery$ does not expire at $p_i$. Then $p_i$
stops the timer at
lines~\ref{line:castro:stoptimers},~\ref{alg:castro:stop-timerrecovery},~\ref{alg:castro:stop-timerrecovery2}
or~\ref{line:castro:stoptimers-enterview}. The latter is impossible, as this
would imply that $p_i$ enters a higher view. If $p_i$ stops the timer at
line~\ref{line:castro:stoptimers}, then it calls $\padvance$ in $v$, as
required. Assume now that $p_i$ stops the timer at
lines~\ref{alg:castro:stop-timerrecovery}
or~\ref{alg:castro:stop-timerrecovery2}. This implies that $p_i$ sets
$\status=\NORMAL$ at some point while in $v$. If $p_i$ sets $\status=\BLOCKED$
while in $v$, then it calls $\padvance$ in $v$, as required. Thus, it remains to
consider the case when $p_i$ sets $\status=\NORMAL$ at some point while in $v$
and does not change it while in this view. Since $p_i$ receives
$\BROADCAST(\val)$ for a valid value $\val$ while in a view $v$, the handler at
line~\ref{alg:castro:broadcast-msg} is executed at some point. At this point
$p_i$ starts $\timerexecute[x]$. If the timer expires, then $p_i$ calls
$\padvance$ in $v$, as required. Otherwise $p_i$ stops the timer at
lines~\ref{line:castro:stoptimers},~\ref{alg:castro:stop-timerexecute2}
or~\ref{line:castro:stoptimers-enterview}. The last two are impossible, as this
would imply that $p_i$ enters a higher view or that $x$ is delivered. In the
remaining case $p_i$ calls $\padvance$ in $v$, as required.
\end{proof}

\begin{figure}[t]
  \centerline{\includegraphics[width=0.55\textwidth]{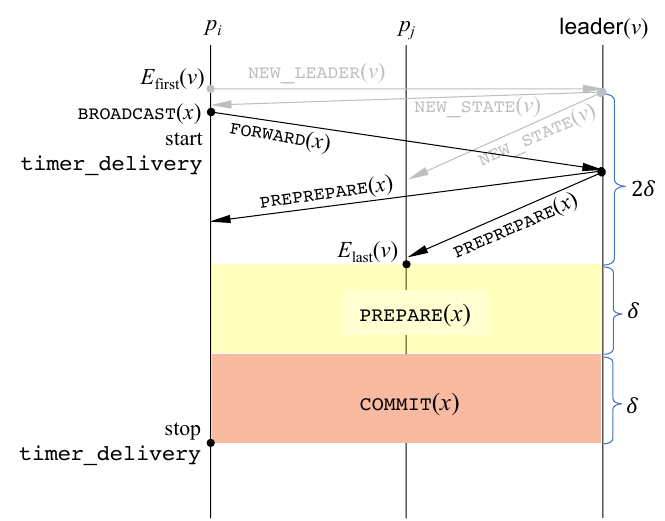}}
  \caption{An illustration of the bound on $\durationexecution$ in
    Lemma~\ref{thm:castro:all-good}.}
\label{fig:bounds2}
\end{figure}

\begin{proof}[The remaining case in the proof of Lemma~\ref{thm:castro:timers}]
Assume that for some value $\val$, $\timerexecute[\val]$ expires at $p_i$ in
$v$. The process starts $\timerexecute[\val]$ when it receives
$\BROADCAST(\val)$ and it has not yet delivered $\val$
(line~\ref{alg:castro:start-timerexecute}). Let $t$ be the time when this
happens; then $t\geq \tm{v}$ (Figure~\ref{fig:bounds2}). Because $p_i$ is the
first correct process to call $\padvance$ in $v$ and
$\durationexecution_i(v)>4\delta$, no correct process calls $\padvance$ in $v$
until after $t+4\delta$. Then by \entry all correct processes enter $v$ by
$\tm{v}+2\delta$.  Furthermore, by \validitysync no correct process can enter
$v+1$ until after $t+4\delta$, and by Proposition~\ref{sync:noskip} the same
holds for any view $>v$.  Thus, all correct processes stay in $v$ at least until
$t+4\delta$.

The process $p_i$ has $\status=\NORMAL$ at $t$, so that by this time $p_i$ has
handled the $\NEWVIEW$ message from the leader of $v$. Thus, all correct
processes receive $\NEWVIEW$ by $t+\delta$. Since $t\geq\tm{v}$ and all correct
processes enter $v$ by $\tm{v}+2\delta$, all correct processes handle $\NEWVIEW$
by $t+2\delta$. When a process handles $\NEWVIEW(v, \_, \_)$, it sends
$\PREPARE$ messages for all positions $\leq \krecovery$. Therefore, by
$t+2\delta$ all correct processes send $\PREPARE$ for all positions
$\leq \krecovery$.

When $p_i$ starts $\timerexecute[\val]$, it sends $\FORWARD(\val)$ to
$\leader(v)$, which receives the message no later than $t+\delta$. Consider
first the case when $\leader(v)$ has $\val$ in its $\cmd$ at position
$k\leq \krecovery$ when it receives $\FORWARD(\val)$. Then all
correct processes send $\PREPARE$ for all positions $\leq k$ by
$t+2\delta$. Assume now that, when the leader receives $\FORWARD(\val)$, either
$\val\not\in\cmd$ or for some $k >\krecovery$ we have $\cmd[k]=\val$. In the
former case the leader sends $\PREPREPARE(v, k, \val)$ to all processes.  In the
latter case, due to lines~\ref{alg:castro:set-next}
and~\ref{alg:castro:increase-next}, the leader has already sent
$\PREPREPARE(v, k, \val)$ to all processes. Thus, in either case the leader
sends $\PREPREPARE(v, k, \val)$ no later than $t+\delta$. Hence, due to
lines~\ref{alg:castro:set-next} and~\ref{alg:castro:increase-next}, the leader
sends a $\PREPREPARE$ for all positions from $\krecovery+1$ up to $k$ no later
than $t+\delta$, and all correct processes receive these messages no later than
$t+2\delta$. We have established that all correct processes have handled
$\NEWVIEW(v,\_,\_)$ by $t+2\delta$. Then all correct processes handle the
$\PREPREPARE$ messages for positions from $\krecovery+1$ up to $k$ by
$t+2\delta$, i.e., they send a $\PREPARE$ for each of these
positions. Furthermore, we have established that all correct processes send a
$\PREPARE$ message for each position $\leq \krecovery$ by
$t+2\delta$. Therefore, all correct processes send $\PREPARE$ for each position
$\leq k$ by $t+2\delta$. It then takes them at most $2\delta$ to exchange the
corresponding sequence of $\PREPARE$ and $\COMMIT$ messages. Hence, all correct
processes receive $\COMMIT$ for all positions $\leq k$ by $t+4\delta$.

Assume that when $p_i$ receives all these $\COMMIT$ messages, it has
$\lastdelivered<k$. Then $p_i$ delivers $\val$ by $t+4\delta$. Since $p_i$'s
$\timerexecute[\val]$ has not expired by then, the process stops the timer,
which contradicts our assumption. Therefore, when $p_i$ receives all the
$\COMMIT$ messages for positions $\leq k$, it has $\lastdelivered\geq k$, so
that $p_i$ has already delivered a value $x'$ at this position. Then $p_i$ must
have formed a certificate $C'$ such that $\committed(C', v', k,
\hash(\val'))$. By Proposition~\ref{lemma:pbft:committed-prepared}, for some
well-formed certificate $C''$ we have $\accepted(C'', v', k,
\hash(\val'))$. Since all correct processes stay in $v$ until $t+4\delta$, we
must have $v'\leq v$. If $v=v'$, then by Proposition~\ref{lemma:pbft:singlecmd}
we get $\val=\val'$. If $v'<v$, then by Lemma~\ref{lemma:pbft:main} we get
$\val=\val'$. Hence, $p_i$ delivers $\val$. By
line~\ref{alg:castro:broadcast-msg}, $p_i$ does not start $\timerexecute[\val]$
if $\val$ has already been delivered. Since $p_i$ started the timer at $t$, it
has to deliver $\val$ at some point after $t$ and no later than
$t+4\delta$. Since $p_i$'s $\timerexecute[\val]$ has not expired by then, the
process stops the timer, which contradicts our assumption. Therefore,
$\timerexecute$ cannot expire at $p_i$.
\end{proof}

\subsection{Proof of the Latency Bounds for \pbft}
\label{app:pbftlatency}

To show the bound, we take advantage of the latency guarantees of our
synchronizer (Theorem~\ref{thm:smr-sync-correct}).
In more detail, the bound has to account for an unfavorable scenario where the
last view $\B-1$ of the asynchronous period is not operational (e.g., not all
correct processes enter it). In this case, to deliver $\val$ the protocol first needs to
bring all correct processes into the same view $\B$.
To bound the time required for that, we first use
Property~\ref{eq:gen-bounded-entry:main} to determine the latest time when a
correct process can enter $\B-1$: $\GST+\rho+2\delta$. We then add the time
such a process may spend in $\B-1$ before it detects that $\val$ is taking too
long to get delivered and call $\padvance$:
$\max\{\rho+\delta,6\Delta\} +4\Delta$, where $6\Delta$ and $4\Delta$ come from
the maximal timeout values. %
The first clause of Property~\ref{eq:lat-bound1:main} then shows that all
correct processes will enter $\B$ within an additional $\delta$, i.e.,
$\tl{\B} \le \GST+\rho+\max\{\rho+\delta,6\Delta\}+4\Delta+3\delta$. 
Finally, we
add the time %
for $\val$ to be delivered in $\B$: $\max\{\rho, \delta\} + 4\delta$.

We now proceed with the formal proof.

\begin{lemma}
\label{lem:relbcast}
If a correct process delivers a value $\val$ at $t$, then all
correct processes deliver $\val$ by $\max\{t+\delta, \GST+\rho+\delta\}$.
\end{lemma}
We omit the easy proof of the lemma.

\begin{lemma}
\label{lem:earliestadv}
If a correct process $p_i$ enters a view $v$, then the earliest time it
may call $\padvance$ is $\te{i}{v}+\min\{\durationrecovery_i(v), \durationexecution_i(v)\}$.
\end{lemma}
\begin{proof}
A process calls $\padvance$ in a view if a timer expires. Assume first that
$\timerrecovery$ expires.  The process $p_i$ starts $\timerrecovery$ when it
enters the view $v$.  Thus, if $\timerrecovery$ expires, then $p_i$ calls
$\padvance$ at $\te{i}{v}+\durationrecovery_i(v)$. Assume now that
$\timerexecute[\val]$ expires for some value $\val$. The earliest time $p_i$ may
start this timer is $\te{i}{v}$. Then the earliest time $p_i$ may call
$\padvance$ in this case is $\te{i}{v}+\durationexecution_i(v)$. Thus, $p_i$
cannot call $\padvance$ in $v$ before
$\te{i}{v}+\min\{\durationrecovery_i(v), \durationexecution_i(v)\}$, as
required.
\end{proof}

\begin{lemma}
  Assume that a correct process $p_i$ that enters $v$ receives $\BROADCAST(\val)$ at
  $t\geq \te{i}{v}$ for a valid value $\val$. Assume that
  $p_i$ does not deliver $\val$ until after
  $\max\{t, \te{i}{v}+\durationrecovery_i(v)\}
  +\durationexecution_i(v)$. Then
  $\tae{i}{v}\fdef$ and
  $$
  \tae{i}{v}\leq\max\{t, \te{i}{v}+\durationrecovery_i(v)\} +
  \durationexecution_i(v).
  $$
\label{thm:castro:completeness:latency}
\end{lemma}
\begin{proof}
We know that at some point $p_i$ enters view $v$, and at this moment it starts
$\timerrecovery$. If the timer expires, then $p_i$ calls $\padvance$
in $v$ by $\te{i}{v}+\durationrecovery_i(v)$, as
required. Assume that $\timerrecovery$ does not expire at $p_i$. Then $p_i$
stops the timer at
lines~\ref{line:castro:stoptimers},~\ref{alg:castro:stop-timerrecovery},~\ref{alg:castro:stop-timerrecovery2}
or~\ref{line:castro:stoptimers-enterview}. If $p_i$ stops the timer at
line~\ref{line:castro:stoptimers}, then it calls $\padvance$ in $v$ by $\te{i}{v}+\durationrecovery_i(v)$, as
required. If $p_i$ stops the timer at
line~\ref{line:castro:stoptimers-enterview}, then it enters a higher view by $\te{i}{v}+\durationrecovery_i(v)$, as
required. Assume now that $p_i$ stops the timer at
lines~\ref{alg:castro:stop-timerrecovery}
or~\ref{alg:castro:stop-timerrecovery2}. This implies that $p_i$ sets
$\status=\NORMAL$ by $\te{i}{v}+\durationrecovery_i(v)$ while in
$v$. If $p_i$ calls $\padvance$ or enters a higher view by $\max\{t,
\te{i}{v}+\durationrecovery_i(v)\}$, we get the required. Assume that
this is not the case. Then $p_i$ sets $\status=\NORMAL$ and 
receives $\val$ by $\max\{t, \te{i}{v}+\durationrecovery_i(v)\}$.
Since $p_i$ has not delivered $\val$ by
$\max\{t, \te{i}{v}+\durationrecovery_i(v)\}$, by this point $p_i$ starts
$\timerexecute[\val]$.  If the timer expires, then $p_i$ calls $\padvance$ in
$v$ by $\max\{t, \te{i}{v}+\durationrecovery_i(v)\} + \durationexecution_i(v)$,
as required. Otherwise $p_i$ stops the timer at
lines~\ref{line:castro:stoptimers},~\ref{alg:castro:stop-timerexecute2}
or~\ref{line:castro:stoptimers-enterview}. If $p_i$ stops the timer at
line~\ref{line:castro:stoptimers-enterview}, then it enters a higher view by
$\max\{t, \te{i}{v}+\durationrecovery_i(v)\}+ \durationexecution_i(v)$, as
required. If $p_i$ stops the timer at line~\ref{alg:castro:stop-timerexecute2},
then it delivers $\val$ by
$\max\{t, \te{i}{v}+\durationrecovery_i(v)\}+ \durationexecution_i(v)$, which is
impossible.  In the remaining case $p_i$ calls $\padvance$ in $v$ by
$\max\{t, \te{i}{v}+\durationrecovery_i(v)\}+ \durationexecution_i(v)$, as
required.
\end{proof}

\begin{lemma}
  Assume that all correct processes start executing \pbft before $\GST$. If
  $\B = 1$ in Theorem~\ref{thm:bounds}, then $\tm{\B}\fdef$ and
  $\tl{\B}\leq \GST + \rho + \delta$.
\label{lem:pbft-view1-lat}
\end{lemma}
\begin{proof}
By Theorem~\ref{thm:bounds}, if $\B=1$, then $\GV{\GST}=0$. Since all correct
processes start executing the protocol before $\GST$, then all correct
processes attempt to advance from view $0$ and $\taelast{0}<\GST$. By \startup,
$\tm{\B}\fdef$. Applying the first clause of Property~\ref{eq:lat-bound1:main}, we get
$\tl{\B} \le \max\{\taelast{0},\GST+\rho + \delta\}$. Since $\taelast{0}<\GST$,
then $\tl{\B} \le \GST+\rho + \delta$,  as required.
\end{proof}

\begin{lemma}
  Consider a view $v\geq\B$ such that $\tm{v}\geq\GST$ and $\leader(v)$ is
  correct. Assume that a correct process $p_j$ broadcast a value
  $\val$ before $\tm{v}$. If $\durationrecovery_i(v) > 6\delta$ and
  $\durationexecution_i(v) > 4\delta$ at each correct process $p_i$ that enters
  $v$, then all correct processes deliver $\val$ by $\tl{v}+\max\{\rho, \delta\} + 4\delta$.
\label{lem:all-good:latency}
\end{lemma}
\begin{proof}
By Lemma~\ref{thm:castro:all-good}, no correct process calls $\padvance$
in $v$. Therefore, by \validitysync, no correct process enters
$v+1$, and by Proposition~\ref{sync:noskip} the same holds for any
view $>v$. By \entry, all correct processes enter $v$. Assume that $p_j$ delivers $\val$ by
$\tl{v}+\rho$. Then by Lemma~\ref{lem:relbcast} all correct processes
deliver $\val$ by $\max\{\tl{v}+\rho+\delta,
\GST+\rho+\delta\}$. Since $\tm{v}\geq\GST$, we get $\tl{v}+\rho+\delta
\geq \GST+\rho+\delta$. Thus, all correct processes
deliver $\val$ by $\tl{v}+\rho+\delta$, as required.

Consider now the case when $p_j$ does not deliver $\val$ by
$\tl{v}+\rho$. Then $p_j$ retransmits $\val$ between $\GST$ and $\GST+\rho \le
\tl{v}+\rho$, so that $\leader(v)$ receives $\val$ by $\tl{v}+\rho+\delta$. 
When a process enters $v$, it sends
$\NEWLEADER$ to $\leader(v)$. The leader receives $2f+1$ of these
messages by $\tl{v}+\delta$ and sends a $\NEWVIEW$ message to all
processes. A process handles $\NEWVIEW$ by $\tl{v}+2\delta$ and sets
its $\status$ to $\NORMAL$. Since no process calls $\padvance$ in $v$,
every correct process has $\status=\NORMAL$ after handling $\NEWVIEW$ onwards.
We have established that $\leader(v)$ receives $\val$ by $\tl{v}+\rho+\delta$.
Then $\leader(v)$ sets $\status=\NORMAL$ and receives $\val$ by $t\leq\tl{v} +
\max\{\rho, \delta\} + \delta$. Assume that $\leader(v)$ has 
delivered $\val$ by $\tl{v}+\max\{\rho, \delta\} + \delta$. 
Then by
Lemma~\ref{lem:relbcast} all correct process deliver $\val$ by
$\max\{\tl{v}+\max\{\rho, \delta\} + 2\delta,
\GST+\rho+\delta\}$. Since $\tm{v}\geq\GST$, we get $\tl{v}+\max\{\rho,
\delta\} + 2\delta \ge \GST+\rho+\delta$. Thus, in this case all
correct processes deliver $\val$ by $\tl{v}+\max\{\rho,
\delta\} + 2\delta$, as required. 

Assume now that $\leader(v)$ has not
delivered $\val$ by $t$. Consider
first then case when $\leader(v)$ already has $\val$ in its log at $t$
because $\val$ was prepared in a previous view. Then all correct processes
send $\PREPARE(v, k, \hash(\val))$ for a position $k$ by 
$\tl{v} + 2\delta$. Consider now the case when $\leader(v)$ either has
$\val$ in its log at $t$ because it was already proposed in
$v$; or $\leader(v)$ does not have it. In this case, it follows that
$\leader(v)$ sends $\PREPREPARE(v, k, \val)$ to all correct processes
by $t$, which all correct processes
receive by $t+\delta$. When a correct process receives $\PREPREPARE(v,
k, \val)$, it sends $\PREPARE(v, k, \hash(\val))$. Thus, all
correct process send $\PREPARE(v, k, \hash(\val))$ by
$t+\delta$. Since $t\leq\tl{v} +
\max\{\rho, \delta\} + \delta$. Then, in both cases, all correct
processes
send $\PREPARE(v, k, \hash(\val))$ by $\tl{v} +
\max\{\rho, \delta\} + 2\delta$. It then takes the correct
processes at most $2\delta$ to exchange the sequence of $\PREPARE$ and
$\COMMIT$ messages that commit $\val$. Therefore, all correct
processes commit $\val$ by $\tl{v}+\max\{\rho, \delta\} + 4\delta$.
Let $p_i$ be a correct process. Because $\leader(v)$ is
correct, we can show that $\lastdelivered\geq k-1$ at $p_i$ by 
$\tl{v}+\max\{\rho, \delta\} + 4\delta$. If $\lastdelivered=k-1$, then
$p_i$ delivers $\val$ by $\tl{v}+\max\{\rho, \delta\} + 4\delta$. If
$\lastdelivered>k-1$ at $p_i$ by $\tl{v}+\max\{\rho, \delta\} +
4\delta$, then we can show that $p_i$ has already delivered $\val$
before. Since $p_i$ was picked arbitrarily, we can conclude that all
correct processes deliver $\val$ by
$\tl{v}+\max\{\rho, \delta\} + 4\delta$, as required.
\end{proof}

\begin{proof}[Proof of Theorem~\ref{thm:pbft-latency1}]
Let $p_j$ be the correct process that broadcast $\val$, and let $\B$ be defined as in
  Theorem~\ref{thm:smr-sync-correct}. Assume first that $\B>1$. We have $\tam{0} <
\GST$ and $\tm{\B-1}<\GST+\rho$. Then by
Property~\ref{eq:gen-bounded-entry:main},
\begin{equation}
\tl{\B-1} \le \GST+\rho + 2\delta.
\label{eq:lastv}
\end{equation}

If at least one correct process $p_i$ delivers
$\val$ by $\GST+\rho+2\delta+\max\{\rho+\delta,6\Delta\} + 4\Delta$, then by
Lemma~\ref{lem:relbcast}, all correct processes deliver $\val$ by
$\GST+\rho+3\delta+\max\{\rho+\delta,6\Delta\} + 4\Delta$, as required. Assume
now that no correct process $p_i$ delivers $\val$ by
$\GST+\rho+2\delta+\max\{\rho+\delta,6\Delta\} + 4\Delta$. In particular, this
implies that $p_j$ has not delivered $\val$ by $\GST+2\rho + 2\delta$, so that it
retransmits $\val$ between $\GST+ \rho + 2\delta$ and
$\GST+2\rho+2\delta$. Consider a correct process $p_i$ that enters $\B-1$ and 
let $t_i$ be the time when this process receives the
$\BROADCAST(\val)$ retransmission from $p_j$; then
\begin{equation}\label{bound-on-ti}
t_i \le \GST+2\rho+3\delta
\end{equation}
and by~(\ref{eq:lastv}),
\begin{equation}\label{bound-on-ti2}
t_i > \GST+ \rho + 2\delta \ge \te{i}{\B-1}.
\end{equation}
We now obtain:
$$
\begin{array}{r@{}l@{\qquad}l}
&\max\{t_i, \te{i}{\B-1}+\durationrecovery_i(\B-1)\}+{}
\\
& \durationexecution_i(\B-1)\\
\le {} &\max\{t_i, \te{i}{\B-1}+6\Delta\}+4\Delta &
  \mbox{since $\durationrecovery_i(\B-1)\leq 6\Delta$}\\
&& \mbox{and $\durationexecution_i(\B-1)\leq 4\Delta$} \\
\le {} &  \max\{\GST+2\rho+3\delta, \te{i}{\B-1}+6\Delta\} +4\Delta & \mbox{by~(\ref{bound-on-ti})}\\
\le {}&\max\{\GST+2\rho+3\delta, \GST+\rho+2\delta+6\Delta\}+4\Delta & \mbox{by~(\ref{eq:lastv})}\\
= {}&\GST+\rho+2\delta+\max\{\rho+\delta,6\Delta\} + 4\Delta
\end{array}
$$
Then, since we assume that no correct process delivers $\val$ by
$\GST+\rho+2\delta+\max\{\rho+\delta,6\Delta\} + 4\Delta$, 
by~(\ref{bound-on-ti2}) and Lemma~\ref{thm:castro:completeness:latency} we
get $\tae{i}{\B-1}\fdef$. Thus, $p_i$ either
calls $\padvance$ in $\B-1$ or enters a higher view. If at least one correct
process enters a view higher than $\B$, then by
Proposition~\ref{sync:noskip}, $\tm{\B}\fdef$. If all correct processes that
enter $\B-1$ call $\padvance$, then by \progress we get $\tm{\B}\fdef$ as well. By
Lemma~\ref{thm:castro:completeness:latency}, we also have
\begin{multline*}
\tae{i}{\B-1}\leq
\max\{t_i, \te{i}{\B-1}+\durationrecovery_i(\B-1)\} +
\durationexecution_i(\B-1)
\leq{}
\\
\GST+\rho+2\delta+\max\{\rho+\delta,6\Delta\} + 4\Delta
\end{multline*}
for any correct process $p_i$ that enters $\B-1$. Applying the first clause of
Property~\ref{eq:lat-bound1:main}, we get
\begin{equation}\label{e-last-v}
\tl{\B} \le \max\{\taelast{\B-1}, \GST+\rho\}+ \delta
\le \GST+\rho+3\delta+\max\{\rho+\delta,6\Delta\} + 4\Delta.
\end{equation}

Thus, if $\B>1$, then either all correct processes deliver $\val$ by
$\GST+\rho+2\delta+\max\{\rho+\delta,6\Delta\} + 4\Delta$, or $\tm{\B}\fdef$
and~(\ref{e-last-v}) holds. Furthermore, if $\B=1$, then by
Lemma~\ref{lem:pbft-view1-lat}, $\tm{\B}\fdef$ and
$\tl{\B}\leq \GST + \rho + \delta$. We have thus established that either all
correct processes deliver $\val$ by
$\GST+\rho+2\delta+\max\{\rho+\delta,6\Delta\} + 4\Delta$, or $\tm{\B}\fdef$
and~(\ref{e-last-v}) holds. In the latter case, by
Lemma~\ref{lem:all-good:latency}, all correct processes deliver $\val$ by
$\tl{\B}+\max\{\rho, \delta\} + 4\delta$, and by~(\ref{e-last-v}),
$$
\tl{\B}+\max\{\rho, \delta\} + 4\delta \le
\GST+\rho + \max\{\rho+\delta, 6\Delta\} + 4\Delta+\max\{\rho, \delta\} +
7\delta,
$$
as required.
\end{proof}

\begin{lemma}
  Assume that $\B=1$ in Theorem~\ref{thm:bounds}, $\tm{1}\geq\GST$ and $\leader(1)$ is
  correct. If $\durationrecovery_i(1) > 5\delta$ and
  $\durationexecution_i(1) > 4\delta$ at each correct process $p_i$ that enters
  $1$, then no correct process calls $\padvance$ in $v$.
\label{thm:all-good:veq1}
\end{lemma}
We omit the proof of this lemma. It is virtually identical to that of
Lemma~\ref{thm:castro:all-good}, considering the special case of
$v=\B=1$ and the optimization by which in view $1$ the processes do
not exchange $\NEWLEADER$ messages.

\begin{lemma}
  Assume that $\B=1$ in Theorem~\ref{thm:bounds}, $\tm{1}\geq\GST$ and
  $\leader(1)$ is correct. Assume that a correct process $p_j$ broadcast a value
  $\val$ at $t\geq\GST$. If $\durationrecovery_i(1) > 5\delta$ and
  $\durationexecution_i(1) > 4\delta$ at each correct process $p_i$ that enters
  $1$, then all correct processes deliver $\val$ by
  $\max\{t, \tl{1}\} + 4\delta$.
\label{lem:all-good:latency-goodcase}
\end{lemma}
\begin{proof}
By Lemma~\ref{thm:all-good:veq1}, no correct process calls $\padvance$
in $1$. Therefore, by \validitysync, no correct process enters
$2$, and by Proposition~\ref{sync:noskip} the same holds for any
view $>1$. By \entry, all correct processes enter $1$.

When $\leader(1)$ enters view $1$, it sends 
a $\NEWVIEW$ message to all
processes. A process handles $\NEWVIEW$ by $\tl{1}+\delta$ and sets
its $\status$ to $\NORMAL$. The $\leader(1)$ receives $\val$ by
$t+\delta$. The leader will send $\PREPREPARE(1, k, \val)$ when it has
received $\val$ and has $\status=\NORMAL$, i.e., by $\max\{t,
\tl{1}\}+\delta$. All correct processes
receive $\PREPREPARE(1, k, \val)$ by $\max\{t,
\tl{1}\} + 2\delta$. It then takes the correct
processes at most $2\delta$ to exchange the sequence of $\PREPARE$ and
$\COMMIT$ messages that commit $\val$. Therefore, all correct
processes commit $\val$ by $\max\{t,
\tl{1}\} + 4\delta$.
Let $p_i$ be a correct process. Because $\leader(1)$ is
correct, we can show that $\lastdelivered\geq k-1$ at $p_i$ by 
$\max\{t,
\tl{1}\} + 4\delta$. If $\lastdelivered=k-1$, then
$p_i$ delivers $\val$ by $\max\{t,
\tl{1}\} + 4\delta$. In
case $\lastdelivered>k-1$ at $p_i$ by $\max\{t,
\tl{1}\} + 4\delta$, we can show that $p_i$ has already deliver $\val$
before. Since $p_i$ was picked arbitrarily, we can conclude that all
correct processes deliver $\val$ by
$\max\{t,
\tl{1}\} + 4\delta$, as required.
\end{proof}

\begin{proof}[Proof of Theorem~\ref{thm:pbft-latency:good}]
When a correct process starts the protocol, it calls $\padvance$ from view $0$
unless it has already entered a higher view. If all correct processes call
$\padvance$ from view $0$ when they start the protocol, then by \startup we get
$\tm{1}\fdef$. If at least one correct process does not call $\padvance$ from
view $0$ because it has already entered a higher view, then by
Proposition~\ref{sync:noskip} we also get $\tm{1}\fdef$.
Since all correct processes start the protocol after $\GST$, we have
$\tam{0}>\GST$.  Thus, applying the second clause of
Property~\ref{eq:lat-bound1:main}, we get $\tl{1} \le
\taelast{0}+\delta$. Furthermore, by Theorem~\ref{thm:bounds} we get
$\B=1$. Hence, $\tl{\B} \le \taelast{0}+\delta$, as required. By
Lemma~\ref{lem:all-good:latency-goodcase}, all correct processes deliver $\val$
by $\max\{t, \tl{\B}\} + 4\delta \le \max\{t, \taelast{0}+\delta\} + 4\delta$,
as required. 
\end{proof}

\subsection{Space Requirements of \pbft}
\label{sec:pbft-space}

Since the synchronizer is not guaranteed to switch processes between views all
at the same time, a process in a view $v$ may receive a message from a higher
view $v' > v$, which needs to be stored in case the process finally switches to
$v'$. If implemented naively, this would require a process to store unboundedly
many messages. Instead, we allow a process to store, for each message type and
sender, only the message of this type received from this sender that has the
highest view. We can easily adapt the liveness proof of \pbft to take this
into account. In the proof, when we argue that a process handles a
message $m$ in a view $v$ at a given time $t$, we establish before that no
correct process has entered a greater view by $t$. This implies that no
correct process has sent a message for a view $> v$ by $t$. Thus, $m$
will not be discarded at correct processes before $t$, which is enough
to ensure that liveness is not violated. Recall that our synchronizer from
\S\ref{sec:sync} uses only bounded 
space. Thus, \pbft with this synchronizer requires space proportional to the
number of requests submitted by clients. This is appropriate for blockchain
applications, where a process participating in the SMR protocol needs to store
the blockchain and candidate blocks anyway.

\section{PBFT with Leader Rotation Using an SMR Synchronizer}
\label{sec:pbft-rotation}

We now demonstrate how an SMR synchronizer can be used to implement Byzantine
SMR protocols that periodically force a leader
change~\cite{aardvark,spinning,mirbft}. Figures~\ref{fig:pbft-normal}-\ref{fig:pbft-recovery}
give an implementation of such a variant of PBFT, which we call \pbftr. As in
\pbft, processes monitor the behavior of the leader and ask the synchronizer to
advance to another view if they suspect that the leader is faulty. To ensure
that leaders periodically rotate, processes also call $\padvance$ once they
deliver $B$ values proposed by the current leader. Thus, the leader of a view
$v$ is responsible for filling in the slots in the $\cmd$ from $(v-1)\cdot B+1$
to $(v-1)\cdot B + B$.

\begin{figure}[t]
\vspace{3pt}
\begin{tabular}{@{\!}l@{\!\!\!\!\!\!}|@{\ \,}l@{}}
\scalebox{0.96}{%
\begin{minipage}[t]{6.6cm}
\removelatexerror
\vspace*{-11pt}
\begin{algorithm*}[H]
\SetInd{\marginA}{\marginB}
  \setcounter{AlgoLine}{0}

\SetInd{\marginA}{\marginB}
  \SubAlgo{\Fun ${\tt start}()$}{
    \lIf{$\currview = 0$}{\padvance()\label{line:start}}
  }

  \smallskip
 
  \SubAlgo{\textbf{when \timerexecute{} or \timerrecovery{}
      expires}\label{alg:expire-timerexecute}}{ 
    ${\tt stop\_all\_timers}()$\;
    $\padvance()$\; \label{alg:timerexpire-move}
    $\status \leftarrow \BLOCKED$\;
    $\durationexecution \leftarrow \durationexecution + \tau$\; \label{line:rotation:timer-inc1}
    $\durationrecovery \leftarrow \durationrecovery + \tau$\; \label{line:rotation:timer-inc2}
  }

  \smallskip

  \SubAlgo{\Fun ${\tt broadcast}(\val)$\label{alg:broadcast}}{
    \textbf{pre:} $\valid(\val)$\;
    \Send $\langle \BROADCAST(\val) \rangle_i$ \KwTo \all{}
    {\bf periodically until $x$ is \qquad delivered}\label{alg:send-to-all}
  }

  \smallskip

  \SubAlgo{\WhenReceived $\BROADCAST(\val)$\label{alg:receive-broadcast}}{
    \textbf{pre:} $\valid(\val) \wedge \val \not\in \queue \wedge {}$\\
     \nonl $\phantom{\text{{\bf pre:} }} (\forall k.\, k \le \lastdelivered {\implies} {}$\\
     \nonl $\phantom{\text{{\bf pre:} }(} \comcmd[k]\not=\val)$\;
    $\queue \leftarrow \append(\queue, \val)$\label{alg:append-queue}\;
  }

  \smallskip

  \SubAlgo{{\bf when} $p_i = \leader(\currview)  \wedge {}$\\
    \nonl $\phantom{\text{{\bf when} }}\status = \NORMAL \wedge {}$\\
    \nonl $\phantom{\text{{\bf when} }}\nextv \le \currview \cdot \batch
    \wedge {}$\\
    \nonl $\phantom{\text{{\bf when} }}\exists \val \in \queue.\, \forall k.\, \cmd[k]\not=\val$\label{alg:leader-broadcast}}{
    $\val \leftarrow {}$the first value in $\queue$ that \mbox{is not in $\cmd$}\;
    $\stoptimer(\timerbroadcast)$\;
    ${\tt propagate}(\val)$\;
  }

  \smallskip

  \SubAlgo{\textbf{when \timerbroadcast\ expires}}{\label{alg:expire-timerbroadcast}
    ${\tt propagate}(\noop)$\;
  }

    \smallskip

  \SubAlgo{\Fun ${\tt propagate}(\val)$\label{alg:propagate}}{
    \Send $\langle \PREPREPARE(\currview, $ $\nextv, \val) \rangle_i$
    \KwTo \all\;
    $\nextv \leftarrow \nextv + 1$\;\label{alg:increase-next}
    \If{$\nextv \le \currview \cdot \batch$}{%
      $\starttimer(\timerbroadcast, T)$\label{alg:start-timerbroadcast}}
  }

\end{algorithm*}
\vspace*{-6pt}
\end{minipage}}
&
\scalebox{0.96}{%
\begin{minipage}[t]{9cm}
\removelatexerror
\vspace*{-11pt}
\begin{algorithm*}[H]

  \SubAlgo{\WhenReceived $\langle \PREPREPARE(v, k, \val) \rangle_{j}$\label{alg:receive-propose}}{
    \textbf{pre:} $p_j = \leader(v) \wedge \currview = v \wedge {}$\\
    \nonl$\phantom{\text{{\bf pre:} }}\status = \NORMAL \wedge \phase[k] = \START \wedge {}$\\\label{alg:safety-check}
    \nonl$\phantom{\text{{\bf pre:} }}k \le v \cdot \batch \wedge \valid(\val) \wedge (\forall k'.\, \cmd[k']\not=\val)$\;
    $(\cmd, \phase)[k] \leftarrow (\val, \PREPREPARED)$\;
    \Send $\langle \PREPARE(v, k, \hash(\val)) \rangle_i$ \KwTo \all\;
  }

  \smallskip

  \SubAlgo{\WhenReceived $\{\langle \PREPARE(v, k, h) \rangle_j \mid p_j
    \in Q\} = C$ \qquad\qquad {\bf for a quorum $Q$}\label{alg:receive-prepared}}{
    \textbf{pre:} $\currview = v \wedge \phase[k] = \PREPREPARED \wedge {}$\\
    \nonl$\phantom{\text{{\bf pre:} }}{} \status = \NORMAL \wedge \hash(\cmd[k]) = h$\;
    $(\prepcmd, \prepview, \cert, \phase)[k] \leftarrow {}$\\
    \nonl$\quad (\cmd[k], \currview, C, \PREPARED)$\label{alg:assign-cballot}\;
    \Send $\langle \COMMIT(v, k, h) \rangle_i$ \KwTo \all\;
  }

  \smallskip

  \SubAlgo{\WhenReceived $\{\langle \COMMIT(v, k, h) \rangle_j \mid p_j
    \in Q\} = C$ \quad\qquad\qquad {\bf for a quorum $Q$}\label{alg:receive-committed}}{
    \textbf{pre:} $\currview = v \wedge \phase[k] = \PREPARED \wedge {}$\\
    \nonl$\phantom{\text{{\bf pre:} }}\status = \NORMAL \wedge \hash(\prepcmd[k]) = h$\;
    $(\comcmd,\phase)[k] \leftarrow (\cmd[k], \COMMITTED)$\;
    
     \Broadcast $\langle \DECISION(\comcmd[k], k, C)$\; \label{alg:send-decision}
  }

  \smallskip

  \SubAlgo{{\bf when} ${\comcmd[\lastdelivered+1] \not= \bot}$\label{alg:deliver}}{
    $\lastdelivered \leftarrow \lastdelivered+1$\; \label{alg:update-last}
    \If{$\comcmd[\lastdelivered] \not= \noop$}{${\tt deliver}(\comcmd[\lastdelivered])$}\label{alg:deliver-value}
    $\remove(\queue,
    \comcmd[\lastdelivered])$\; \label{alg:remove-queue}
     \If{$\status={\normalfont \NORMAL}$}{
    \uIf{$\lastdelivered = \currview \cdot
      \batch$}{
      ${\tt stop\_all\_timers}()$\; \label{line:stop-deliver}
      $\padvance()$\; \label{line:advance-deliver}
      $\status \leftarrow \BLOCKED$\;}
   \ElseIf{$\lastdelivered \,{>}\, (\currview \,{-}\, 1) \,{\cdot}\, \batch$}{
      $\stoptimer(\timerexecute)$\label{alg:stop-timerexecute2}\;
      $\starttimer(\timerexecute,
      \durationexecution)$\label{alg:start-timerexecute}\;
    }
    }
  }

\smallskip

\SubAlgo{\WhenReceived $\DECISION(\val, k, C)$ \label{alg:receive-decision}}{
  \textbf{pre:} $\comcmd[k] \not= \bot \wedge {}$\\
  \nonl$\phantom{\text{{\bf pre:} }}\exists v.\, \committed(C, v, k,
    \hash(\val))$\;  \label{alg:decision-safetycheck} 
    $\comcmd[k] \leftarrow \val$\;
  }
\end{algorithm*}
\vspace*{-6pt}
\end{minipage}}
\end{tabular}
\vspace*{-2pt}
\caption{Normal protocol operation of \pbftr at a process $p_i$.}
\label{fig:pbft-normal}
\end{figure}

\begin{figure}[t]
\vspace{2pt}
\begin{tabular}{@{}l@{\!\!\!\!\!\!\!}|@{\ }l@{}}
\scalebox{0.96}{%
\begin{minipage}[t]{7.4cm}
\removelatexerror
\vspace*{-12pt}
\begin{algorithm*}[H]

\SetInd{\marginA}{\marginB}

  \SubAlgo{\Upon $\newview(v)$\label{alg:newview}}{
     ${\tt stop\_all\_timers}()$\; \label{line:stoptimers-enterview}
    $\currview \leftarrow v$\;\label{line:set-currentview}
    $\status \leftarrow \RECOVERING$\;
    \Send $\langle \NEWLEADER(\currview, \prepview,$ $\prepcmd, \cert) \rangle_i$
    \KwTo $\leader(\currview)$\; 
    $\starttimer(\timerrecovery,$ $\durationrecovery)$\;\label{line:start-timerrecovery}
  }

\smallskip

\SubAlgo{\WhenReceived $\{\langle \NEWLEADER(v, \vprepview_j, $ $\vprepcmd_j,
    \vcert_j) \rangle_j \mid$ $p_j \in Q\} = M$ \quad\qquad {\bf for a quorum
      $Q$}\label{alg:receive-newleader}}{
    \textbf{pre:} $p_i = \leader(v) \wedge \currview = v \wedge {}$\\
    \nonl $\phantom{\text{{\bf pre:} }} \status = \RECOVERING \wedge{}$\\
    \nonl $\phantom{\text{{\bf pre:} }} \forall m \in M.\, \ValidNewLeader(m))$\; 
    \ForAll{$k$\label{alg:select-proposal}}{
      \If{$\exists p_{j'} \in Q.\, \vprepview_{j'}[k] \not= 0  \wedge {}$
        $(\forall p_{j} \in Q.\, \vprepview_{j}[k] \le \vprepview_{j'}[k])$}{%
        $\phantom{(}\vcmd'[k] \leftarrow \vprepcmd_{j'}[k]$\label{alg:prepared-entries}}
    }
    $\nextv \leftarrow (v-1) \cdot \batch +1$\;\label{alg:set-next}
}
 
\end{algorithm*}
\vspace*{-5pt}
 \begin{tikzpicture}[overlay]
 \node[draw=none, fill=white, thick,minimum width=.5cm,minimum height=.2cm] (b) at (.743,0){};%
 \end{tikzpicture}
\end{minipage}}
&
\scalebox{0.96}{%
\begin{minipage}[t]{8cm}
\removelatexerror
\vspace*{-22pt}
\begin{algorithm*}[H]
\nonl\SubAlgo{}{
  \ForAll{$k = 1..(\nextv-1)$\label{alg:clean-entries}}{
    \If{$\vcmd'[k] = \bot \vee{} $\\
      \nonl \ \,$\exists k'.\, k' \not= k \wedge \vcmd'[k'] = \vcmd'[k] \wedge{}$\\
       \nonl \ \,$\exists p_{j'} \in Q.\, \forall p_{j} \in
       Q.\,$\\
       \nonl \ \,$\vprepview_{j'}[k'] > \vprepview_{j}[k]$}{%
        $\vcmd'[k] \leftarrow \noop$\label{alg:newview-end}%
      }
    }
    \Send $\langle \NEWVIEW(v, \vcmd', M) \rangle_i$ \KwTo \all\; \label{alg:send-newview}
    $\starttimer(\timerbroadcast, T)$\;
  }

\smallskip

  \SubAlgo{\WhenReceived $\langle \NEWVIEW(v, \vcmd', M) \rangle_j=m$\label{alg:receive-newview}}{ 
    \textbf{pre:} $\status = \RECOVERING \wedge {}$\\
    \nonl $\phantom{\text{{\bf pre:} }} \currview = v \wedge\ValidNewState(m)$\; 
    $\stoptimer(\timerrecovery)$\;\label{alg:stop-timerrecovery}
    $\cmd \leftarrow \vcmd'$\; 
    \uIf{$\lastdelivered \geq \currview \cdot \batch$}{
 $\padvance()$\; 
      $\status \leftarrow \BLOCKED$\;
    }\Else{
      \ForAll{$\{k \mid \cmd[k] \not= \bot\}$}{
      $\phase[k] \leftarrow \PREPREPARED$\;
      \Send $\langle \PREPARE(v, k, \hash(\cmd[k])) \rangle_i$ \quad\KwTo
      \all\; 
    }
      $\starttimer(\timerexecute,\durationexecution)$\;\label{line:timer-abovedelivered}
      $\status \leftarrow \NORMAL$\; \label{alg:status-normal}
    }
  }

\end{algorithm*}
\vspace*{-5pt}
\end{minipage}}
\end{tabular}

\caption{View-initialization protocol of \pbftr at a process $p_i$.}
\label{fig:pbft-recovery}
\end{figure}

A process broadcasts a valid value $\val$ using a ${\tt broadcast}$ function
(line~\ref{alg:broadcast}). As in \pbft, this keeps sending the value to all
processes in a $\BROADCAST$ message until the current process delivers the
value. When a process receives a $\BROADCAST$ message with a new value
(line~\ref{alg:receive-broadcast}), it appends the value to a $\queue$ of values
pending to be broadcast. When the leader has new values in its $\queue$ and has
not yet exhausted the range of log slots it is allowed to fill
(line~\ref{alg:leader-broadcast}), it selects the first new value in the queue
and proposes it by calling the ${\tt propagate}$ function
(line~\ref{alg:propagate}). The processes then handle the proposal as in \pbft.
When a process delivers a value, the process removes it from the queue of
pending values (line~\ref{alg:remove-queue}). If this is the value is in the
last slot allocated to the current leader, the process also calls $\padvance$ to
request a leader change (line~\ref{line:advance-deliver}).

A follower monitors the leader's behavior using two timers, $\timerexecute$ and
$\timerrecovery$, which are similar to the corresponding timers in \pbft, but
slightly different. The (single) timer $\timerexecute$ checks that new values
are delivered at regular intervals: a process sets $\timerexecute$ for a
duration determined by $\durationexecution$ when it delivers a value and the
current leader has not yet exhausted the slots allocated to it
(line~\ref{alg:start-timerexecute}); the process stops the timer when it
delivers the next value (line~\ref{line:timer-abovedelivered}). If a correct
leader does not have anything to propose during a fixed time interval $T$, the
leader proposes a $\noop$. The leader ensures this using a timer
$\timerbroadcast$ (lines~\ref{alg:start-timerbroadcast}
and~\ref{alg:expire-timerbroadcast}). The above mechanism protects against a
faulty leader not making {\em any} proposals. However, it allows a faulty leader
to omit {\em some} of the values submitted by clients. The overall protocol
nevertheless protects against censorship because leaders periodically rotate,
and thus the protocol will go through infinitely many views with correct
leaders. Since values to be broadcast are sent to all processes
(line~\ref{alg:send-to-all}) and are handled in the order of arrival
(line~\ref{alg:append-queue}), each value will eventually be proposed by a
correct leader.

Finally, $\timerrecovery$ is used to check that the leader initializes a view
quickly enough. Like in \pbft, a process starts this timer for a duration
determined by $\durationrecovery$ when it enters a new view
(line~\ref{line:start-timerrecovery}). Unlike in \pbft, a process stops the
timer when it receives a $\NEWVIEW$ message with the initial log from the leader
(line~\ref{alg:stop-timerrecovery}). In \pbft the timer is stopped only after
delivering all the values in the initial log, but in \pbftr the deliver of these
values is checked by $\timerexecute$.

The Integrity, External Validity and Ordering properties of \pbftr are proved in
the same way as for \pbft (\S\ref{sec:pbft-safety}).

\subsection{Proof of Liveness for \pbftr}
\label{sec:liveness-rotation}

Assume that \pbftr is used with an SMR synchronizer satisfying the specification
in Figure~\ref{fig:multi-sync-properties}; to simplify the following latency
analysis, we assume $d = 2\delta$, as for the synchronizer in
Figure~\ref{fig:sync}. We now prove that the protocol satisfies the Liveness
property of Byzantine atomic broadcast. First, due to the periodic leader
rotation mechanism in \pbftr, we can prove that it satisfies
Proposition~\ref{lem:live-toy}, stating that processes keep entering views
forever. The proof is similar to the one in \S\ref{sec:sync}, using
the properties of the SMR synchronizer and 
the following lemma, analogous to Lemma~\ref{thm:castro:completeness} in
the proof of \pbft. The lemma shows that a correct process stuck in a view
will eventually call $\padvance$, either because it has delivered a full batch
of values or because one of its timers has expired.
\begin{lemma}
Assume that a correct process $p_i$ enters a view $v$. If $p_i$ never
enters a view higher than $v$, then it eventually calls $\padvance$ in $v$.
\label{proposition:call-advance}
\end{lemma}
\begin{proof}
We prove the proposition by contradiction: assume that $p_i$ does not
call $\padvance$ while in $v$. When a timer expires, a correct process calls
$\padvance$. Thus, no timer expires at $p_i$ while in $v$. This
implies that $p_i$ receives a $\NEWVIEW$ message from $\leader(v)$ and
stops $\timerrecovery$. If $\lastdelivered\geq \currview \cdot \batch$
when $p_i$ handles the $\NEWVIEW$ message, then it calls $\padvance$,
which is impossible. Thus, $p_i$ has $\lastdelivered< \currview \cdot
\batch$ when it handles the $\NEWVIEW$ message. Therefore, it starts
$\timerexecute$. Since no timer expires at $p_i$ 
while in $v$, then $p_i$ must stop $\timerexecute$ before it
expires. The process $p_i$ stops $\timerexecute$ in
lines~\ref{line:stop-deliver},~\ref{alg:stop-timerexecute2}
and~\ref{line:stoptimers-enterview}. We consider each one of these in
turn. If $\timerexecute$ is stopped at
line~\ref{line:stop-deliver}, then $p_i$ calls $\padvance$, which
is impossible. If $\timerexecute$ is stopped at
line~\ref{line:stoptimers-enterview}, then $p_i$ enters a view $>v$,
which is impossible. Therefore, $p_i$ must stop $\timerexecute$ by
executing line~\ref{alg:stop-timerexecute2}.
In this case $p_i$ has delivered a value at
a position $k>(\currview-1) \cdot \batch$. After stopping
$\timerexecute$ at line~\ref{alg:stop-timerexecute2}, $p_i$ starts
$\timerexecute$ again. Every time $p_i$ executes
line~\ref{alg:stop-timerexecute2}, it increases $\lastdelivered$
by one (line~\ref{alg:update-last}). Given that this is the only place
where $\lastdelivered$ is assigned, we can conclude that
$\lastdelivered$ never decreases. Then $p_i$ cannot be restarting
$\timerexecute$ indefinitely. Eventually,
the variable $\lastdelivered$ at $p_i$ will be equal to $\currview
\cdot \batch$, in which case $p_i$ will execute
line~\ref{line:stop-deliver}. But we have established that $p_i$
cannot execute line~\ref{line:stop-deliver}, which reaches a contradiction. Hence, $p_i$
calls $\padvance$ while in $v$, as required.
\end{proof}

\begin{lemma}
  In any execution of \pbftr:
  $\forall v.\, \exists v'.\, v'>v \wedge {\tm{v'}\fdef}$.
\label{proposition:keep-entering}
\end{lemma}
\begin{proof}
Since all correct processes call $\padvance$ at the beginning, by \startup some
correct process eventually enters view $1$. Assume now that the proposition is
false, so that there exists a maximal view $v$ entered by any correct
process. Let $P$ be any set of $f+1$ correct processes and consider an arbitrary
process $p_i \in P$ that enters $v$. Since no correct process enters a view
$v'>v$, by Lemma~\ref{proposition:call-advance}, $p_i$ calls $\padvance$ in
$v$. Since $p_i$ was picked arbitrarily, we have
$\forall p_i \in P.\, {\te{i}{v}\fdef} {\implies} {\ta{i}{v}\fdef}$. Then by
\progress we get $\tm{v+1}\fdef$, which yields a contradiction.
\end{proof}

\begin{lemma}
  In any execution of \pbftr:
  $\forall v.\, v>0 {\implies} {\tm{v}\fdef}$.
\label{proposition:enter-all}
\end{lemma}
\begin{proof}
  Follows from Lemma~\ref{proposition:keep-entering} and Proposition~\ref{sync:noskip}.
\end{proof}

Let $\B'$ be the minimal view such that $\B' \ge \B$ (for the $\B$ from
Figure~\ref{fig:multi-sync-properties}) and $\tm{\B'}\ge \GST$; such a view
exists by Lemma~\ref{proposition:enter-all}. Hence, starting from $\B'$, process
clocks track real time and messages sent by correct processes are delivered
within $\delta$. We denote by $\durationexecution_i(v)$ and
$\durationrecovery_i(v)$ respectively the value of the $\durationexecution$ and
$\durationrecovery$ variable at a correct process $p_i$ while in view $v$. We
now prove a lemma analogous to Lemma~\ref{thm:castro:timers} in the proof of
\pbft. It shows that, in any view $v \ge \B'$ with a correct leader, if the
timeouts at a correct process $p_i$ that enters $v$ are long enough and some
timer expires at $p_i$, then this process cannot be the first to initiate a view
change.
\begin{lemma}
Let $v \geq \B'$ be a view such that $\leader(v)$ is correct, and let
$p_i$ be a correct process that enters $v$. Assume that $\durationexecution_i(v)
> \max\{4\delta, T+3\delta\}$ and $\durationrecovery_i(v) >4\delta$.
If a timer expires at $p_i$ in $v$, then $p_i$ is not the first correct process
to call $\padvance$ in $v$. 
\label{thm:timers}
\end{lemma}
The proof relies on the following technical lemma.

\begin{lemma}
Let $v$ be a view such that $\leader(v)$ is correct and sends
$\NEWVIEW(v, \vcmd',  \_)$. Let $p_i$ be a
correct process that enters $v$. Assume that $p_i$ does not leave $v$ or
call $\padvance$ in $v$ before $t_1$, processes $\NEWVIEW(v, \vcmd',
\_)$ at $t_2$ and receives $m=\PREPREPARE(v,k,\val)$ from $\leader(v)$
at $t_3<t_1$. Then $p_i$ processes $m$ at $t=\max\{t_2, t_3\}$.
\label{lem:accept-proposal}
\end{lemma}
\begin{proof}
The process $p_i$ processes a $\PREPREPARE$ message if the conditions in
line~\ref{alg:safety-check} are satisfied.  Since $\leader(v)$ sends $m$, we
have $p_j=\leader(v)$ and $\valid(\val)$. By line~\ref{alg:set-next} and
line~\ref{alg:leader-broadcast}, $\leader(v)$ only proposes values for positions
between $(v-1)\cdot B +1$ and $v\cdot B$. Then $k\leq v\cdot B$.

By line~\ref{alg:safety-check}, in any view $v'<v$, $p_i$ can only have
accepted proposals for $k'\leq (v-1)\cdot B$. Then, when $p_i$
enters $v$, it has $\phase[k'']=\START$ for all $k''>(v-1)\cdot B$. Furthermore,
by line~\ref{alg:increase-next}, the leader does not 
propose a value for the same position twice. Then it is guaranteed
that $\phase[k]=\START$ from $\te{i}{v}$
until $p_i$ processes $m$ or leaves $v$.
Furthermore, when $p_i$ enters $v$, it sets $\currview =
v$. This holds at least until $p_i$ leaves $v$, which cannot happen
before $t_1$. Thus, we have that $\phase[k]=\START$ and 
$\currview=v$ from $\te{i}{v}$ until $p_i$ processes $m$ or leaves $v$.

When $p_i$
processes $\NEWVIEW(v, \vcmd',
\_)$, it sets $\status=\NORMAL$ and $\cmd=\vcmd'$, which is the initial log of
$\leader(v)$. 
Since, $\leader(v)$ is correct, any value proposed by
$\leader(v)$ is not in its $\cmd$.
Thus, we have that
$\status=\NORMAL$ and $\forall k'.\, \cmd[k']\not=\val$ at $p_i$ from
the moment $p_i$ processes $\NEWVIEW(v, \vcmd',
\_)$, i.e., from $t_2$, until $p_i$ processes $m$ or leaves $v$.

Therefore, after processing  $\NEWVIEW(v, \vcmd',
\_)$ and before processing $m$ or leaving $v$, 
the conditions in line~\ref{alg:safety-check} are satisfied. 
Since $t_3<t_1$, then $p_i$ processes $m$. If $t_3<t_2$, then $p_i$
processes $m$ at $t_2$. Otherwise, it processes it at
$t_3$. Hence, $p_i$ processes $m$ at $t=\max\{t_2, t_3\}$, as required.
\end{proof}

\begin{proof}[Proof of Lemma~\ref{thm:timers}]
Since $v\geq\B'$, we have $\tm{v}\geq\GST$, so that all messages sent by correct
processes after $\tm{v}$ get delivered to all correct processes within $\delta$
and process clocks track real time. We now make a case split on which timer
expires at $p_i$ in $v$. We first consider the case of $\timerrecovery$. By
contradiction, assume that $p_i$ is the first correct process to call
$\padvance$ in $v$.  The process starts its $\timerrecovery$ when it enters
a view (line~\ref{alg:newview}), and hence, at $\tm{v}$ at the earliest. Because
$p_i$ is the first correct process to call $\padvance$ in $v$ and
$\durationrecovery_i(v)>4\delta$ at $p_i$, no correct process calls $\padvance$ in
$v$ until after $\tm{v}+4\delta$. Then by \entry all correct
processes enter $v$ by $\tm{v}+2\delta$.  Furthermore, by
\validitysync no correct process can enter $v+1$ until after
$\tm{v}+4\delta$, and by Proposition~\ref{sync:noskip}, the same holds for any
view $>v$.  Thus, all correct processes stay in $v$ at least until
$\tm{v}+4\delta$.  When a correct process enters a view, it sends a $\NEWLEADER$
message to the view's leader, and when the leader receives a quorum of such
messages, it sends a $\NEWVIEW$ message.  Since $\tl{v}\leq \tm{v}+2\delta$,
$\leader(v)$ is guaranteed to receive $\NEWLEADER$ message from a quorum of
processes and send a $\NEWVIEW$ message to all processes by
$\tm{v}+3\delta$. Thus, all correct processes receive the $\NEWVIEW$ message by
$\tm{v}+4\delta$. In particular, this is the case for $p_i$. Since $p_i$'s
$\timerrecovery$ has not expired by then, the process stops the timer, which
contradicts our assumption.

We now consider the case when $\timerexecute$ expires at $p_i$ in $v$.
We again prove it by contradiction: assume that $p_i$ is the
first correct process to call $\padvance$ in $v$. The process starts its
$\timerexecute$ after handling the leader's $\NEWVIEW(v, \_, \_)$
message if $\lastdelivered< v\cdot \batch$ or
after delivering a value whose position $k$ in $\cmd$ is
such that $(v-1)\cdot \batch < k < v\cdot \batch$. We only consider the
former case; the latter is analogous. Let $t_1$ be the time when $p_i$ starts
the $\timerexecute$ that expires and let $t_2$ be the first time when a correct process
calls $\padvance$ in $v$. We have assumed that $\timerexecute$ expires at $p_i$ in
$v$, $\durationexecution_i(v)>\max\{4\delta, T+3\delta\}$ and $p_i$ is the
first correct process that calls $\padvance$ in $v$. Hence,
$t_2 > t_1+\max\{4\delta, T+3\delta\}$. Then by
\validitysync no correct process can enter $v+1$ by $t_2$,
and by Proposition~\ref{sync:noskip}, the 
same holds for any view $>v$. Furthermore, $t_1\ge \tm{v}$. Then by
\entry all correct processes enter $v$ by $t_1+2\delta$.

All correct processes receive the leader's $\NEWVIEW(v, \_, \_)$ by
$t_1+\delta$. Therefore, they handle it by $t_1+2\delta$: once they
have received it and entered $v$. If a correct process has
$\lastdelivered\geq v\cdot \batch$ by the time it handles the
leader's $\NEWVIEW(v, \_, \_)$, then it calls $\padvance$. We have
established that no correct process calls $\padvance$ until after 
$t_1+\max\{4\delta, T+3\delta\}$. Since all correct processes handle
the leader's $\NEWVIEW(v, \_, \_)$ by
$t_1+2\delta$, then all correct processes have $\lastdelivered < v\cdot \batch$ by the time they handle the
leader's $\NEWVIEW(v, \_, \_)$. Thus, all correct processes send a $\PREPARE$
for all positions $\leq (v-1)\cdot\batch$ by $t_1+2\delta$. It then
takes them at most $2\delta$ to exchange the sequence of $\PREPARE$
and $\COMMIT$ message leading to commit all positions $\leq
(v-1)\cdot\batch$. Since $t_1+4\delta < t_2$, all correct processes 
have $\lastdelivered \ge (v-1)\cdot\batch$ by $t_1+4\delta$.

Assume that $\leader(v)$
sends $\NEWVIEW(v, \_, \_)$ at $t_3$. Since at this time, $\leader(v)$
starts $\timerproposal$, the leader makes a new proposal
no later than $t_3 + T$. Thus, all correct
processes receive the corresponding message $\PREPREPARE(v, k, \_)$ from
$\leader(v)$ by $t_3 + T + \delta < t_1 + T + \delta$.
Since $\leader(v)$ is correct, $k=(v-1)\cdot \batch + 1$ (line~\ref{alg:set-next}).
Let $t_4$ be the time when all correct
processes handle $m$. By Lemma~\ref{lem:accept-proposal}, $t_4\leq
\max\{t_1+2\delta, t_1 + T + \delta\}$. Then it
takes them at most $2\delta$ to exchange the sequence of $\PREPARE$
and $\COMMIT$ message leading to commit. Thus, all correct
processes receive $\COMMIT(v, k, \_)$ by $t_4+2\delta$.
We have earlier established that $t_2 >
t_1+\max\{4\delta, T+3\delta\}$. Then $t_2>\max\{t_4+2\delta, t_1+4\delta\}$. 
Since no correct process leaves $v$ by $t_2$ and by
$t_1+4\delta$ the process $p_i$ has $\lastdelivered\ge
(v-1)\cdot\batch$, then by this time $p_i$ executes the handler in
line~\ref{alg:deliver} for the position $k$.
Since $k=(v-1)\cdot \batch+1$, then $p_i$
executes line~\ref{alg:stop-timerexecute2} and stops $\timerexecute$,
which contradicts our assumption.
\end{proof}

Using Lemma~\ref{thm:timers}, we can establish two key facts necessary to prove
the liveness of PBFT-light, which are stated by the lemma below. Fact 1 is
analogous to Lemma~\ref{thm:castro:all-good} in the proof of \pbft. It
establishes that in any view $v \ge \B'$ where the leader is correct and the
timeouts at all correct processes are high enough, some correct process will
deliver a full batch of values. Fact 2 rules out the scenarios discussed at the
end of \S\ref{sec:pbft}, in which some processes increase their timeouts
sufficiently while others do not. It establishes that in a view $v \ge \B'$ with
a correct leader that does not operate normally (no correct process delivers a
full batch of values), a process $p_i$ with sufficiently high timeouts cannot be
the first one to call $\padvance$. This means some other process with lower
timeouts will have to initiate the view change, and thus increase its timeouts.

\begin{lemma}
  Consider a view $v\geq\B'$ such that $\leader(v)$ is correct.
\begin{enumerate}
\item If $\durationexecution_i(v) > \max\{4\delta, T+3\delta\}$ and
  $\durationrecovery_i(v) > 4\delta$ at each correct processes $p_i$ that enters
  $v$, then some correct process calls $\padvance$ in $v$ due to delivering a
  full batch of values (line~\ref{line:advance-deliver}).
\item Assume that no correct process calls $\padvance$ in $v$ due to delivering
  a full batch of values (line~\ref{line:advance-deliver}). If $p_i$ is a
  correct process that enters $v$ and we have
  $\durationexecution_i(v) > \max\{4\delta, T+3\delta\}$ and
  $\durationrecovery_i(v) > 4\delta$, then $p_i$ cannot be the first correct
  process to call $\padvance$ in $v$.
\end{enumerate}
\label{thm:pbft-facts}
\end{lemma}
\begin{proof}
By Lemma~\ref{proposition:enter-all}, a correct process
eventually enters view $v+1$. Then by \validitysync
there must exist a correct
process that calls $\padvance$ while in $v$.
Let $p_j$ be the first correct process that does so.
This process must call $\padvance$ when: {\em (i)} a timer expires; or
{\em (ii)} it delivers the full batch of values (line~\ref{line:advance-deliver}).

{\em Case 1.} We have $\durationexecution_j(v) > \max\{4\delta, T+3\delta\}$ and
$\durationrecovery_j(v) > 4\delta$. Then by Lemma~\ref{thm:timers}, {\em (i)} is
impossible, so that {\em (ii)} must hold.

{\em Case 2.} By contradiction, assume that $p_i = p_j$.  Since no correct
process calls $\padvance$ in $v$ due to delivering a full batch of values, {\em
  (ii)} is impossible. But since
$\durationexecution_i(v) > \max\{4\delta, T+3\delta\}$ and
$\durationrecovery_i(v) > 4\delta$, by Lemma~\ref{thm:timers}, {\em (i)} is
impossible either.
\end{proof}

\begin{theorem}
  \pbftr satisfies the Liveness property of Byzantine atomic broadcast.\!\!\!\!\!
\label{thm:liveness-pbft-timers}
\end{theorem}
\begin{proof} 
Consider a valid value $x$ broadcast by a correct process.  We first prove that
$x$ is eventually delivered by some correct process.  By contradiction, assume
that $x$ is never delivered by a correct process. A correct process broadcasts a
value until it is delivered (line~\ref{alg:send-to-all}). When a correct process
receives a valid value that is not in its queue, it appends the value
(line~\ref{alg:append-queue}).
Thus, since $x$ is never delivered,
there exists a point in time $t$ starting from which $x$ is always in the queues
of all correct processes. By Lemma~\ref{proposition:enter-all}, every view is
entered by at least one correct process. Let view $v_1$ be the first view such
that $v_1\ge \B'$ and $\tm{v_1}\geq t$.

\setcounter{myclaim}{0}
\begin{myclaim}
In any view $v\geq v_1$ with a correct leader, if a
  correct process calls $\padvance$ due to delivering a full batch of values
  (line~\ref{line:advance-deliver}), then $\leader(v)$ eventually removes $B$
  values preceding $x$ from its queue.
\end{myclaim}
\begin{claimproof}
Consider a view $v\ge v_1$ with a correct leader where a
correct process $p_i$ calls $\padvance$ at line~\ref{line:advance-deliver}. Then
it delivers a full batch of values proposed by the leader of $v$, which are all
distinct from $x$. Since $\leader(v)$ is correct and $x$ is forever in its
queue, we have two options: {\em (i)} all values proposed by the leader in $v$
are distinct from $\noop$ and precede $x$ in the leader's queue; or {\em (ii)}
$\exists k\leq (v-1)\cdot B .\, \cmd[k]=x$ at $\leader(v)$. But the latter is
impossible. Indeed, since $p_i$ delivers a full batch, it delivers all values
whose position in its log is $\leq v\cdot B$. Since the log of the leader and
$p_i$ is the same for those positions when $p_i$ calls $\padvance$, if {\em
  (ii)} holds, then $x$ is delivered. Hence, {\em (i)} must hold.
Since $p_i$ reliably broadcasts committed values (line~\ref{alg:send-decision}),
the protocol guarantees that if a correct process delivers a value, all correct
processes eventually do. Thus, $\leader(v)$ will eventually deliver these values
and remove them from its queue.
\end{claimproof}

\begin{myclaim}
  In any view $\ge v_1$ with a correct leader, at least one correct process
  calls $\padvance$ because one of its timers expires
  (line~\ref{alg:timerexpire-move}), and no correct process calls $\padvance$
  because it delivers a full batch of values (line~\ref{line:advance-deliver}).
\end{myclaim}
\begin{claimproof}
Assume that the claim does not hold.  By
Lemma~\ref{proposition:enter-all} and since leaders rotate round-robin across
views, there are infinitely many views $\geq v_1$ with a correct leader that are
entered by a correct process. Furthermore, by \validitysync,
at least one correct process calls $\padvance$ in each of these. Since we assume
that the claim is false, there is a correct process $p_i$ that leads an infinite
number of views $\ge v_1$ in which some correct process calls $\padvance$ at
line~\ref{line:advance-deliver}. By Claim 1, for each such view $p_i$ eventually
removes $B$ values preceding $x$ from its queue. In views $\ge v_1$ no new
values are added to $p_i$'s queue before $x$. Hence, there exists a view
$\ge v_1$ in which $x$ is proposed and a correct process calls $\padvance$ at
line~\ref{line:advance-deliver}, thus delivering $x$. This reaches a
contradiction.
\end{claimproof}

\begin{myclaim}
  Every correct process calls the timer expiration handler
  (line~\ref{alg:expire-timerexecute}) infinitely often.
\end{myclaim}
\begin{claimproof}
Assume this is not the case and let $C_{\rm fin}$ and
$C_{\rm inf}$ be the sets of correct processes that call the timer expiration
handler finitely and infinitely often, respectively. Then
$C_{\rm fin} \not= \emptyset$, and by Claim~2, $C_{\rm inf} \not= \emptyset$.
The values of $\durationexecution$ and $\durationrecovery$ 
increase unboundedly at processes from $C_{\rm inf}$, and do not change after
some view $v_2$ at processes from $C_{\rm fin}$.
By Lemma~\ref{proposition:enter-all} and since leaders rotate round-robin, there
exists a view $v_3\geq \max\{v_2, v_1\}$ with a correct leader such that for any
process $p_i \in C_{\rm inf}$ that enters $v_3$ we have
$\durationexecution_i(v_3)> \max\{4\delta, T+3\delta\}$ and
$\durationrecovery_i(v_3)> 4\delta$. By Claim~2, at least one correct process
calls $\padvance$ because one of its timers expires in $v_3$; let $p_l$ be the
first process to do so.  Since $v_3 \ge v_2$, this cannot be a process from
$C_{\rm fin}$, since none of these processes can increase their timers in
$v_3$. Hence, $p_l\in C_{\rm inf}$, which contradicts
Lemma~\ref{thm:pbft-facts}(2).
\end{claimproof}

We now prove that $x$ is delivered by a correct process. By Claim~3 and
Lemma~\ref{proposition:enter-all}, there exists a view $v_4\geq v_1$ with a
correct leader such that for any correct process $p_i$ that enters $v_4$ we have
$\durationexecution_i(v_4) > \max\{4\delta, T+3\delta\}$ and
$\durationrecovery_i(v_4) > 4\delta$. By Lemma~\ref{thm:pbft-facts}(1), some
correct process calls $\padvance$ in $v_4$ due to delivering a full batch of
values (line~\ref{line:advance-deliver}). This contradicts Claim~2 and thus
proves that $x$ is delivered by a correct process. By reliably broadcasting
committed values (line~\ref{alg:send-decision}), the protocol guarantees that if
a correct process delivers a value, then all correct eventually do. From here
the Liveness property follows.
\end{proof}

\subsection{Latency Bounds for \pbftr}
\label{app:latency}

Assume that \pbftr is used with our SMR synchronizer in
Figure~\ref{fig:sync}. We now quantify its latency using the bounds for the
synchronizer in Theorem~\ref{thm:smr-sync-correct}. We again assume the
existence of a known upper bound $\Delta$ on the maximum value of the
post-$\GST$ message delay. We also modify the protocol in
Figure~\ref{fig:pbft-castro-normal} so that in lines
\ref{line:rotation:timer-inc1}-\ref{line:rotation:timer-inc2} it does not
increase $\durationrecovery$ and $\durationexecution$ above $4\Delta$ and
$\max\{4\Delta, T+3\Delta\}$, respectively. This corresponds to the bounds in
Lemma~\ref{thm:pbft-facts}(1) and preserves the protocol liveness.

We establish latency bounds for two scenarios key to the protocol's
performance. Our first bound considers the case when the protocol starts during
the asynchronous period, before $\GST$. We quantify how quickly after $\GST$ the
protocol enters the first functional view $\B$ in which a correct leader can
propose a full batch of $B$ values that will be delivered by all correct
processes. This view $\B$ is the same as the one in \entry for our synchronizer,
defined by Theorem~\ref{thm:smr-sync-correct}. For simplicity, we assume that
timeouts are high enough at $\GST$.
\begin{theorem}
  Assume that all correct processes start executing \pbftr before $\GST$ and
  that at $\GST$ each of them has $\durationrecovery > 4\delta$ and
  $\durationexecution > \{4\delta, T+3\delta\}$. Let $\B$ be defined as in
  Theorem~\ref{thm:smr-sync-correct}. Then
  $\tl{\B} \le \GST + \rho + 4\Delta+ B \cdot \max\{4\Delta, T+3\Delta\} +
  3\delta$, and if $\leader(\B)$ is correct, it proposes $B$ values in $\B$ that
  are delivered by all correct processes.
\label{thm:pbftr-latency1}
\end{theorem}
Intuitively, the bound in the theorem captures worst-case scenarios in which
some correct processes may need to spend up to time
$4\Delta+ B \cdot \max\{4\Delta, T+3\Delta\}$ in a non-functional view $\B-1$,
e.g., to commit $B$ $\noop$ values generated by a Byzantine leader. The theorem
shows that \pbftr, like \pbft, recovers after a period of asynchrony in bounded
time.
We now prove Theorem~\ref{thm:pbftr-latency1}. 
The following lemma bounds the latency of entering $\B=1$.
\begin{lemma}
Assume that all correct processes
starts executing \pbftr before $\GST$.
If $\B = 1$, and some correct process
enters $\B$, then any correct process that enters $\B$ will do that 
no later than at $\GST + \rho + \delta$.
\label{lem:pbftr-view1-lat}
\end{lemma}
\begin{proof}
When a correct process starts the protocol, it calls $\padvance$ from view $0$
unless it has already entered a higher view. Therefore,
$\taelast{0}<\GST$.
Applying the first clause of
Property~\ref{eq:lat-bound1:main}, we get $\tl{1} \le
\max(\taelast{0}, \GST+\rho) + \delta = \GST + \rho + \delta$,
as required.
\end{proof}

We next consider the case of $\B > 1$.
\begin{lemma}
  Assume that all correct process starts executing \pbftr before
  $\GST$ and at $\GST$
  each correct process has $\durationrecovery > 4\delta$ and
  $\durationexecution > \max\{4\delta, T+3\delta\}$.
  Then, if $\B > 1$, and some correct process enters $\B$, then 
$\tl{\B} \le \GST + \rho + 4\Delta+ B \cdot \max\{4\Delta, T+3\Delta\}
+ 3\delta$.
\label{lem:pbftr-viewV-lat}
\end{lemma}
\begin{proof}
Since some correct process enters $\B > 1$, by Proposition~\ref{sync:noskip},
some correct process enters the view $\B-1$ as well. Let $p_i$ be a correct
process that enters $\B-1$. We show that
\begin{equation}
\label{eq:pbftr-tai}
\tae{i}{\B-1} \le \te{i}{\B-1} + 4\Delta+ B \cdot \max\{4\Delta, T+3\Delta\}.
\end{equation}
If $p_i$ enters a view  $>\B-1$ before $\te{i}{\B-1} +  4\Delta+ B \cdot \max\{4\Delta, T+3\Delta\}$, then~(\ref{eq:pbftr-tai}) holds. Suppose now that $p_i$ does not
enter a view $>\B-1$ before $\te{i}{\B-1} +  4\Delta+ B \cdot \max\{4\Delta, T+3\Delta\}$. 

By the structure of the code, at $\te{i}{\B-1}$,
$p_i$ starts $\timerrecovery$
for the duration $\durationrecovery_i(\B-1)$. If $\timerrecovery$ expires
before $p_i$ receives $\NEWVIEW(\B-1,\_,\_)$ from $\leader(\B-1)$, $p_i$
attempts to advance from $\B-1$. Since $\durationrecovery_i(\B-1)\leq
4\Delta$, (\ref{eq:pbftr-tai}) holds. Otherwise, $p_i$ stops
$\timerrecovery$. Assume that $p_i$ has $\lastdelivered\geq
\currview\cdot B$ when it stops $\timerrecovery$. Then, $p_i$ calls $\padvance$ by
$\te{i}{\B-1}+\durationrecovery_i(\B-1)$. Since $\durationrecovery_i(\B-1)\leq
4\Delta$, (\ref{eq:pbftr-tai}) holds. Assume now that $p_i$ has $\lastdelivered<
\currview\cdot B$ when it stops $\timerrecovery$. Therefore, it starts
$\timerexecute$ for the duration $\durationexecution_i(\B-1)$. If
$\timerexecute$ expires, then $p_i$ calls $\padvance$ by 
$$
\te{i}{\B-1}+\durationrecovery_i(\B-1)+\durationexecution_i(\B-1).
$$
Since $\durationrecovery_i(\B-1)\leq
4\Delta$ and $\durationexecution_i(\B-1)\leq \max\{4\Delta, T+3\Delta\}$,
(\ref{eq:pbftr-tai}) holds. If $p_i$ stops $\timerexecute$, then 
it delivers a value at a position $>(\currview-1)\cdot B$. Then $p_i$
restarts and stops $\timerexecute$ every time a new value
is delivered, until it delivers the value at position $\currview\cdot
B$ or $\timerexecute$ expires. In both cases, $p_i$
attempts to advance from $\B-1$ by
$$
\te{i}{\B-1}+\durationrecovery_i(\B-1)+B\cdot\durationexecution_i(\B-1).
$$
Since $\durationrecovery_i(\B-1)\leq
4\Delta$ and $\durationexecution_i(\B-1)\leq \max\{4\Delta, T+3\Delta\}$,
(\ref{eq:pbftr-tai}) holds. 

Since $p_i$ was picked arbitrarily, we can conclude that every correct process
that enters $\B-1$ either attempts to advance from it or enters a higher view no
later than at $\tl{\B-1} + 4\Delta+ B \cdot \max\{4\Delta, T+3\Delta\}$, and
therefore,~(\ref{eq:pbftr-tai}) holds. By
Property~\ref{eq:gen-bounded-entry:main},
$$
\tl{\B-1} \le \max(\tm{\B-1}, \GST+\rho) + 2\delta.
$$
Since by the definition of $\B$, $\tm{\B-1} \le \GST+\rho$, we have
$$
\tl{\B-1} \le \GST + \rho + 2\delta.
$$
Thus,
\begin{multline*}
\taelast{\B-1} \le \tl{\B-1} + 4\Delta+ B \cdot \max\{4\Delta,
T+3\Delta\} \le{}
\\
\GST + \rho + 2\delta + 4\Delta+ B \cdot \max\{4\Delta, T+3\Delta\}.
\end{multline*}
We can now apply the first clause of 
Property~\ref{eq:lat-bound1:main} to obtain
\begin{align*}
  \tl{\B} & \le \max(\taelast{\B-1}, \GST + \rho) + \delta
                \\
                & \le \max(\GST + \rho + 2\delta + 4\Delta+ B \cdot
                  \max\{4\Delta, T+3\Delta\}, \GST + \rho) + \delta 
\\
& \le \GST + \rho + 4\Delta+ B \cdot \max\{4\Delta, T+3\Delta\} + 3\delta,
\end{align*}
as required.
\end{proof}

\begin{proof}[Proof of Theorem~\ref{thm:pbftr-latency1}]
By Lemma~\ref{proposition:enter-all}, some correct process eventually
enters $\B$.
By the theorem's premise, all correct processes start the protocol
before $\GST$.
Thus, by Lemma~\ref{lem:pbftr-view1-lat},
$$
\tl{1} \le \GST + \rho + \delta
< \GST + \rho + 4\Delta+ B \cdot \max\{4\Delta, T+3\Delta\} + 3\delta.
$$
And, if $\B > 1$, then by Lemma~\ref{lem:pbftr-viewV-lat},
$$
\tl{\B} \le \GST + \rho + 4\Delta+ B \cdot \max\{4\Delta, T+3\Delta\} + 3\delta.
$$
Since the timeout durations are monotone, the theorem's assumption about timeout durations
implies that for any process $p_i$ that enters $\B$
$$
  \durationrecovery_i(\B) > 4\delta\, \wedge\,
  \durationexecution_i(\B) > \max\{4\delta, T+3\delta\}.
$$
By Lemma~\ref{thm:pbft-facts}(1), some correct process calls $\padvance$
in $\B$ due to delivering a full batch of $B$ values proposed
by $\leader(\B)$. By reliably broadcasting committed values,
the protocol guarantees that all these values will be eventually delivered
by all correct processes, as required.
\end{proof}

Our next bound assumes that the protocol executes during a synchronous period
and quantifies how quickly it recovers after encountering a view $v$ with a
faulty leader. For simplicity, we assume that this leader is initially crashed
and the views $<v$ operated normally.
\begin{theorem}
  Assume that all correct processes start executing \pbftr after $\GST$, and
  consider a view $v$ such that $\leader(v)$ is initially crashed. Suppose that
  initially $\durationrecovery > 4\delta$ and
  $\durationexecution > \max\{4\delta, T+3\delta\}$ and, in each view $v' < v$,
  each correct process calls $\padvance$ due to delivering a full batch of
  values proposed in $v'$. Then $\tl{v+1} \le \tl{v} + R + \delta$, where $R$ is
  the initial value of $\durationrecovery$. Furthermore, if $\leader(v+1)$ is
  correct, then it proposes $B$ values in $v+1$ that are delivered by all
  correct processes.
\label{thm:pbftr-latency2}
\end{theorem}
The bound established by the theorem illustrates the benefits of how \pbftr (as
well as PBFT) manages timeouts. Since processes do not increase timeouts in good
views with correct leaders, they pay a minimal latency penalty once they
encounter a bad leader.

We now prove Theorem~\ref{thm:pbftr-latency2}. The following lemma bounds
the latest time by which a correct process process can enter view $v+1$
assuming the leader of view $v>0$ is initially crashed.
\begin{lemma}
  Assume that all correct processes start executing \pbftr after $\GST$,
  and consider a view $v > 0$ such that $\leader(v)$ is initially
  crashed. Suppose that any correct processes $p_i$ that enters
  $v$ has $\durationrecovery_i(v) = R$. If a correct process enters $v+1$, then 
  $\tl{v+1}\le\tl{v} + R + \delta$.
\label{lem:pbft-lat2}
\end{lemma}
\begin{proof}
Since some correct process enters $v+1$, by Proposition~\ref{sync:noskip},
$\tm{v}\fdef$. Consider a correct process $p_i$
that enters $v$, and assume that $p_i$ does not enter any views $>v$
before $\te{i}{v} + R$. By the protocol, $p_i$ starts $\timerrecovery$ at
$\te{i}{v}$ to await $\NEWVIEW(v,\_,\_)$ from $\leader(v)$. 
Since $p_i$ starts executing after $\GST$, its
local clock advances at the same rate as real time. Thus, given that
$\leader(v)$ is initially crashed, the $p_i$'s  $\timerrecovery$ 
will expire at $\te{i}{v} + \durationrecovery_i(v) = \te{i}{v} + R$.
Hence, $p_i$ calls $\padvance$ at $\te{i}{v} + R$. Therefore, 
all correct processes that enter $v$ either attempt to advance
from $v$, or enter a view $>v$ no later than at $\tl{v} + R$. Hence,
$$
\taelast{v} \le \tl{v} + R.
$$
Since all correct processes start executing the protocol after $\GST$,
$\tam{0} \ge \GST$, and therefore, by Property~\ref{eq:lat-bound1:main},
$$
\tl{v+1} \le \taelast{v} + \delta \le \tl{v} + R + \delta,
$$
as required.
\end{proof}

\begin{proof}[Proof of Theorem~\ref{thm:pbftr-latency2}]
Since in each view $v' < v$ all correct processes deliver $B$ values proposed in $v'$,
by the structure of the code, no correct process increases the durations of any 
of its timers. Thus, $\durationrecovery_i(v) = R$ for any correct
processes $p_i$ that enters $v$,
where $R$ is the initial value of $\durationrecovery$. Therefore, 
by Lemma~\ref{lem:pbft-lat2}, $\tl{v+1}\le\tl{v} + R + \delta$. Suppose that
$\leader(v+1)$ is correct. Then by the theorem's assumption about the
initial timeout durations, and since all timeout durations are
monotone, we have that for any correct process $p_i$ that enters $v$
$$
  \durationrecovery_i(v+1) > 4\delta\, \wedge\,
  \durationexecution_i(v+1) > \max\{4\delta, T+3\delta\}.
$$
Hence, by Lemma~\ref{thm:pbft-facts}(1), some correct process calls $\padvance$
in $v+1$ due to delivering a full batch of $B$ values proposed
by $\leader(v+1)$. By reliably broadcasting committed values,
the protocol guarantees that all these values will be eventually delivered
by all correct processes, as required.
\end{proof}

\section{A HotStuff-like Protocol Using an SMR Synchronizer}
\label{sec:hotstuff}

\begin{figure}[!t]
\vspace{3pt}
\begin{tabular}{@{\!}l@{\!\!\!\!\!\!\!}|@{\ }l@{}}
\scalebox{0.96}{%
\begin{minipage}[t]{6.8cm}
\removelatexerror
\vspace*{-11pt}
\begin{algorithm*}[H]
\SetInd{\marginA}{\marginB}
  \setcounter{AlgoLine}{0}

\SetInd{\marginA}{\marginB}
  \SubAlgo{{\bf when the process starts}}{
    \padvance();
  }

  \smallskip

  \SubAlgo{\textbf{when \timerexecute{} or  \timerrecovery{} expires}}{ 
    ${\tt stop\_all\_timers}()$\;
    $\padvance()$\;
    $\status \leftarrow \BLOCKED$\;
    $\durationexecution \leftarrow \durationexecution + \tau$\;
    $\durationrecovery \leftarrow \durationrecovery + \tau$\;
  }

  \smallskip

  \SubAlgo{\Fun ${\tt broadcast}(\val)$}{
    \textbf{pre:} $\valid(\val)$\;
    \Send $\langle \BROADCAST(\val) \rangle_i$ \KwTo \all{}
    {\bf periodically until $x$ is delivered}
  }

  \smallskip

  \SubAlgo{\WhenReceived $\BROADCAST(\val)$}{
    \textbf{pre:} $\valid(\val) \wedge \val \not\in \queue \wedge {}$\\
     \nonl $\phantom{\text{{\bf pre:} }} (\forall k.\, k \le \lastdelivered {\implies} {}$\\
     \nonl $\phantom{\text{{\bf pre:} }(} \comcmd[k]\not=\val)$\;
    $\queue \leftarrow \append(\queue, \val)$\;
  }

  \smallskip

  \SubAlgo{{\bf when} $\status = \NORMAL \wedge {}$\\
    \nonl $\phantom{\text{\bf when }} p_i = \leader(\currview) \wedge {}$\\
    \nonl $\phantom{\text{\bf when }} \nextv \le \currview \cdot \batch \wedge {}$\\
    \nonl $\phantom{\text{\bf when }} \exists \val \in \queue.\, \forall k.\, \cmd[k]\not=\val$}{
    $\val \leftarrow {}$the first value in $\queue$ that \mbox{is not in $\cmd$}\;
    $\stoptimer(\timerbroadcast)$\;
    ${\tt propagate}(\val)$\;
  }

  \smallskip

  \SubAlgo{\textbf{when \timerbroadcast\ expires}}{
    ${\tt propagate}(\noop)$\;
  }

    \smallskip

  \SubAlgo{\Fun ${\tt propagate}(\val)$}{
    \Send $\langle \PREPREPARE(\currview, $ $\nextv, \val) \rangle_i$
    \KwTo \all\;
    $\nextv \leftarrow \nextv + 1$\;
    \If{$\nextv \le \currview \cdot \batch$}{%
      $\starttimer(\timerbroadcast, T)$}
  }
\smallskip

  \SubAlgo{\WhenReceived $\langle \PREPREPARE(v, k, \val) \rangle_{j}$}{
    \textbf{pre:} $p_j = \leader(v) \wedge \currview = v \wedge {}$\\
    \nonl$\phantom{\text{{\bf pre:} }}\status = \NORMAL \wedge {}$\\
    \nonl$\phantom{\text{{\bf pre:} }}\phase[k] = \START \wedge {}$\\
    \nonl$\phantom{\text{{\bf pre:} }}k \le v \cdot \batch \wedge
    \valid(\val) \wedge {}$\\
    \nonl$\phantom{\text{{\bf pre:} }}(\forall k'.\, \cmd[k']\not=\val)$\;
    $(\cmd, \phase)[k] \leftarrow (\val, \PREPREPARED)$\;
    \Send $\langle \PREPARE(v, k, \hash(\val)) \rangle_i$ \KwTo \all\;
  }

\end{algorithm*}
\vspace*{-6pt}
\end{minipage}}
&
\scalebox{0.96}{%
\begin{minipage}[t]{8.8cm}
\removelatexerror
\vspace*{-11pt}
\begin{algorithm*}[H]
  
  \SubAlgo{\WhenReceived $\{\langle \PREPARE(v, k, h) \rangle_j \mid p_j
    \in Q\} = C$ \quad\qquad\qquad {\bf for a quorum $Q$}}{\label{alg:hotstuff:prepare}
    \textbf{pre:} $\currview = v \wedge \phase[k] = \PREPREPARED \wedge{}$\\
    \nonl$\phantom{\text{{\bf pre:} }}{} \status = \NORMAL \wedge \hash(\cmd[k]) = h$\;
    $(\prepcmd, \prepview, \cert, \phase)[k] \leftarrow {}$\\
    \nonl$\quad (\cmd[k], \currview, C, \PREPARED)$\;
    \Send $\langle \PRECOMMIT(v, k, h) \rangle_i$ \KwTo \all\; \label{alg:hotstuff:sendprecommit}
  }

  \smallskip

  \SubAlgo{\WhenReceived $\{\langle \PRECOMMIT(v, k, h) \rangle_j \mid p_j
    \in Q\} = C$ \qquad\qquad {\bf for a quorum $Q$}}{\label{alg:hotstuff:precommit}
    \textbf{pre:} $\currview = v \wedge \phase[k] = \PREPARED \wedge{} $\\
    \nonl$\phantom{\text{{\bf pre:} }}{} \status = \NORMAL \wedge \hash(\prepcmd[k]) = h$\;
    $(\lockview, \phase)[k] \leftarrow (\currview, \PRECOMMITTED)$\;\label{hotstuff:lock}
    \Send $\langle \COMMIT(v, k, h) \rangle_i$ \KwTo \all\;
  }

  \smallskip

  \SubAlgo{\WhenReceived $\{\langle \COMMIT(v, k, h) \rangle_j \mid p_j
    \in Q\} = C$ \quad\qquad\qquad {\bf for a quorum $Q$}}{\label{alg:hotstuff:commit}
    \textbf{pre:} $\currview = v \wedge{}$\\
    \nonl$\phantom{\text{{\bf pre:} }}\phase[k] = \PRECOMMITTED \wedge {}$\\
    \nonl$\phantom{\text{{\bf pre:} }}\status = \NORMAL \wedge \hash(\prepcmd[k]) = h$\;
    $(\comcmd,\phase)[k] \leftarrow (\cmd[k], \COMMITTED)$\;
    
     \Broadcast $\langle \DECISION(\comcmd[k], k, C)$\;
  }

  \smallskip

  \SubAlgo{{\bf when} $\comcmd[\lastdelivered+1] \not= \bot
    \wedge{}$\\
    \nonl$\phantom{\text{{\bf when} }}\status = \NORMAL$}{
    $\lastdelivered \leftarrow \lastdelivered+1$\;
    \If{$\comcmd[\lastdelivered] \not= \noop$}{${\tt deliver}(\comcmd[\lastdelivered])$}
   $\remove(\queue, \comcmd[\lastdelivered])$\;
    \If{$\status={\normalfont \NORMAL}$}{
    \uIf{$\lastdelivered = \currview \cdot
      \batch$}{
      ${\tt stop\_all\_timers}()$\;
      $\padvance()$\;
      $\status \leftarrow \BLOCKED$\;
    }
   \ElseIf{$\lastdelivered \hspace{1pt}{>}\hspace{1pt} (\currview
     \hspace{1pt}{-}\hspace{1pt} 1) \hspace{1pt}{\cdot}\hspace{1pt} \batch$}{
      $\stoptimer(\timerexecute)$\;
      $\starttimer(\timerexecute,
      \durationexecution)$\;
      }
}
  }

\smallskip

\SubAlgo{\WhenReceived $\DECISION(\val, k, C)$}{
  \textbf{pre:} $\comcmd[k] \not= \bot \wedge{}$\\
  \nonl $\phantom{\text{{\bf pre:} }} \exists v.\, \committed(C, v, k, \hash(\val))$\;
    $\comcmd[k] \leftarrow \val$\;
  }
\end{algorithm*}
\vspace*{-6pt}
\end{minipage}}
\end{tabular}
\vspace*{-2pt}
\caption{Normal protocol operation of HotStuff at a process $p_i$.}
\label{fig:hotstuff-normal}
\end{figure}

\begin{figure}[t]
\vspace{2pt}
\begin{tabular}{@{\!}l@{\!\!\!\!\!\!}|@{\ \ }l@{}}
\scalebox{0.96}{%
\begin{minipage}[t]{7.4cm}
\removelatexerror
\vspace*{-12pt}
\begin{algorithm*}[H]

\SetInd{\marginA}{\marginB}

  \SubAlgo{\Upon $\newview(v)$}{
     ${\tt stop\_all\_timers}()$\;
    $\currview \leftarrow v$\;
    $\status \leftarrow \RECOVERING$\;
    \Send $\langle \NEWLEADER(\currview,\prepview, $ $\prepcmd, \cert) \rangle_i$
    \KwTo $\leader(\currview)$\; 
    $\starttimer(\timerrecovery,$ $\durationrecovery)$\;
  }

\smallskip

\SubAlgo{\WhenReceived $\{\langle \NEWLEADER(v,\vprepview_j, $ $\vprepcmd_j,
    \vcert_j) \rangle_j \mid$ $p_j \in Q\} = M$ \quad\qquad {\bf for a quorum
      $Q$}}{
    \textbf{pre:} $p_i = \leader(v) \wedge \currview = v \wedge {}$\\
    \nonl $\phantom{\text{{\bf pre:} }} \status = \RECOVERING \wedge{}$\\
    \nonl $\phantom{\text{{\bf pre:} }} \forall m \in M.\, \ValidNewLeader(m)$\; 
    \ForAll{$k$}{
      \If{$\exists p_{j'} \in Q.\, \vprepview_{j'}[k] \not= 0 \wedge{}$
        $\forall p_{j} \in Q.\,\vprepview_{j}[k] \le \vprepview_{j'}[k]$}{%
        $\vcmd'[k] \leftarrow \vprepcmd_{j'}[k]$\;
      $\vprepview'[k] \leftarrow \vprepview_{j'}[k]$\;
    $\vcert'[k] \leftarrow \vcert_{j'}[k]$}
  $\nextv \leftarrow (v-1) \cdot \batch +1$\;
        \Send $\langle \NEWVIEW(v, \vcmd', \vprepview', $ \quad$\vcert') \rangle_i$ \KwTo \all\;
    $\starttimer(\timerbroadcast, T)$\;
    }
}
 
\end{algorithm*}
\vspace*{-5pt}
\end{minipage}}
&
\scalebox{0.96}{%
\begin{minipage}[t]{8cm}
\removelatexerror
\vspace*{-12pt}
\begin{algorithm*}[H]

  \SubAlgo{\WhenReceived $\langle \NEWVIEW(v, \vcmd', \vprepview', \vcert') \rangle_j = m$}{ 
    \textbf{pre:} $\status = \RECOVERING \wedge {}$\\
    \nonl $\phantom{\text{{\bf pre:} }} \currview = v \wedge \ValidNewState(m)$\label{hotstuff:safety-check}\; 
    $\stoptimer(\timerrecovery)$\;
    $\cmd \leftarrow \vcmd'$\label{hotstuff:assign-log}\;
    \ForAll{$k = 1..(v-1) \cdot \batch$}{
      \If{$\cmd'[k] = \bot \vee {}$\\
        \nonl\ \,$\exists k'.\, k' \not= k \wedge 
        \cmd'[k'] = \cmd'[k] \wedge{}$ $\vprepview'[k'] > \vprepview[k]$}{%
        $\cmd'[k] \leftarrow \noop$\label{hotstuff:assign-nop}%
        }
      }
        \uIf{$\lastdelivered \geq \currview \cdot \batch$}{
           $\padvance()$\; 
      $\status \leftarrow \BLOCKED$\;
      }\Else{
        \ForAll{$k = 1..(v-1) \cdot \batch$}{
       $\phase[k] \leftarrow \PREPREPARED$\;
      \Send $\langle \PREPARE(v, k, \hash(\cmd'[k])) \rangle_i$ \KwTo
      \all\; 
    }
    $\starttimer(\timerexecute,$\quad$\durationexecution)$\;
    $\status \leftarrow \NORMAL$\; 
        }
   
  }  
\end{algorithm*}
\vspace*{-5pt}
\end{minipage}}
\end{tabular}

\caption{View-initialization protocol of HotStuff at a process $p_i$.}
\label{fig:hotstuff-recovery}
\end{figure}

\begin{figure}[t]
\small
\begin{minipage}{15cm}
\begin{gather*}
\accepted(C, v, k, h)
\iff
\exists Q.\, 
\quorum(Q) \wedge C = \{\langle \PREPARE(v, k, h) \rangle_j \mid p_j \in Q\}
\\
\committed(C, v, k, h)
\iff
\exists Q.\, 
\quorum(Q) \wedge C = \{\langle \COMMIT(v, k, h) \rangle_j \mid p_j \in Q\}
\\
\begin{array}{@{}l@{}}
\ValidNewLeader(\langle \NEWLEADER(v, \vprepview, \vprepcmd, \vcert) \rangle_{\_}) \iff {}
\ms
\quad \forall k.\, (\vprepview[k] > 0 {\implies} 
\vprepview[k] < v \wedge 
\accepted(\vcert[k], \vprepview[k], k, \vprepcmd[k]))
\end{array}
\\
\begin{array}{@{}l@{}}
\ValidNewState(\langle \NEWVIEW(v, \vcmd, \vprepview, \vcert) \rangle_i) \iff
\ms
\quad  p_i = \leader(v) \wedge (\forall k.\, \lockview[k] > 0 {\implies} \vcmd[k] \not=  \bot) \wedge{}
\ms
\quad  (\forall k.\, \vcmd[k]\not=\bot {\implies} 
  v>\vprepview[k]>\lockview[k] \wedge {}
  \ms
  \quad \accepted(\vcert[k], \vprepview[k], \hash(\vcmd[k])))
\end{array}
\end{gather*}
\end{minipage}%
\caption{Auxiliary predicates for HotStuff.}
\label{fig:hotstuffpreds}
\end{figure}

In this section we demonstrate how an SMR synchronizer can be used to implement
Byzantine SMR protocols following the approach of HotStuff~\cite{hotstuff},
which reduces the communication complexity of leader
change. Figures~\ref{fig:hotstuff-normal}-\ref{fig:hotstuffpreds} present a
corresponding modification of \pbftr, which we call \hotstuff. For brevity, we
eschew the use of threshold signatures, which can reduce the communication
complexity even further. \hotstuff also excludes optimizations from HotStuff
related to maintaining a hash-chain, but these can be added easily.

\hotstuff adds an extra message exchange to the normal path of \pbftr, in
between the ones for $\PREPARE$ and $\COMMIT$ messages. When a process gathers a
set of $\PREPARE(v, k, \hash(\val))$ messages for a value $\val$ from a quorum
(line~\ref{alg:hotstuff:prepare}), it disseminates a
$\PRECOMMIT(v, k, \hash(\val))$ message
(line~\ref{alg:hotstuff:sendprecommit}). The process then waits until it gathers
a quorum of matching $\PRECOMMIT$ messages for the value $\val$
(line~\ref{alg:hotstuff:precommit}) and disseminates the corresponding $\COMMIT$
message. At this point the process also becomes {\em locked} on $\val$ at
position $k$ in view $v$, which is recorded by setting the position $k$ of an
array $\lockview$ to $\currview$ (line~\ref{hotstuff:lock}).
From this point on, the process will not accept a proposal of a different value
at position $k$ from a leader of a future view, unless the leader can convince
the process that no decision was reached in $\currview$ at position $k$. To this
end, we also modify the view-change protocol of \pbftr. In \hotstuff the leader
of a view $v$ does not forward the set of $\NEWLEADER$ messages used to compute
the view's initial state in its $\NEWVIEW$ message; this reduces the
communication complexity. Instead, a follower checks that the leader's proposal
is safe using a modified $\ValidNewState$ predicate. This checks that, if a
process has previously locked on a value at a position $k$, then either the
leader proposes the same value for that position, or its proposal is justified
by a prepared certificate from a higher view than the lock. In the latter case
the process can be sure that no decision was reached at position $k$ in the view
it is locked on. Finally, \hotstuff delegates the task of filtering out
duplicates to the followers: a follower cannot check that the leader
filters out duplicates correctly without receiving the set of
$\NEWLEADER$ messages used by the leader to compute
the view's initial state.

\subsection{Proof of Safety for \hotstuff}
\label{sec:hotstuff-safety}

Propositions~\ref{lemma:pbft:committed-prepared},~\ref{lemma:pbft:validityliveness}
and~\ref{lemma:pbft:singlecmd} established for \pbft still hold for
\hotstuff. External Validity can be proved similarly. We next prove Ordering and
Integrity.

\begin{proposition}
\label{lemma:hotstuff:view-increase}
The variables $\currview$, $\prepview[k]$ and $\lockview[k]$ (for any $k$) at a
correct process never decrease and we always have
$\lockview[k] \le \prepview[k] \le \currview$.
\end{proposition}

\begin{lemma}
\label{lemma:hotstuff:nodupl}
At a correct process we always have
$$
\forall k, k'.\, \cmd'[k] = \cmd'[k'] \not\in \{\bot, \noop\} {\implies} k=k'.
$$
\end{lemma}
\begin{proof} 
  Analogous to that of Lemma~\ref{lemma:pbft:nodupl}.
\end{proof}

\begin{corollary}
\label{cor:hotstuff:nodupl}
\begin{align*}
\forall \val, v, k, k', C, C'.\, &
\accepted(C, v, k, \hash(\val)) \wedge
\accepted(C', v, k', \hash(\val)) \wedge {}
\\
  &
  \wf(C) \wedge \wf(C') \wedge x \not= \noop
{\implies} k = k'.
\end{align*}
\end{corollary}
\begin{proof} 
  Analogous to that of Corollary~\ref{cor:pbft:nodupl}.
\end{proof}

\begin{lemma}
\label{lemma:hotstuff:main}
Fix $k$, $v$, $v'$, $C$ and $\val$, and assume
$$
\committed(C, v, k, \hash(\val)) \wedge \wf(C) \wedge v' > v.
$$
Then
\begin{itemize}
\item
$\forall C', \val'.\, 
\accepted(C', v', k, \hash(\val')) \wedge
\wf(C') {\implies} \val = \val'$.
\item
$\forall C', k'.\, \val \not= \noop \wedge
\accepted(C', v', k', \hash(\val)) \wedge \wf(C') {\implies} k = k'$.
\end{itemize}
\end{lemma}
\begin{proof}
We prove the statement of the lemma by induction on $v'$. Assume this holds for
all $v' < v^*$; we now prove it for $v' = v^*$. Thus, we have 
\begin{gather}
  \label{hotstuff:hyp2}
\forall C'', \val'', v''.\, v < v'' < v' \wedge \accepted(C'', v'',
k, \hash(\val'')) \wedge \wf(C'') {\implies} \val = \val'';
\\
\label{hotstuff:hyp4}
\forall C'', k'', v''.\, v < v'' < v' \wedge \val \not= \noop \wedge \accepted(C'', v'',
k'', \hash(\val)) \wedge \wf(C'') {\implies} k = k''.
\end{gather}

Assume that $\accepted(C', v', k, \hash(\val'))$ and $\wf(C')$. Since
$\committed(C, v, k, \hash(x))$, a quorum $Q$ of processes sent
$\COMMIT(v, k, \hash(x))$. Since $\accepted(C', v', k, \hash(\val'))$, a quorum
$Q'$ of processes sent $\PREPARE(v', k, \hash(\val'))$. The quorums $Q$ and $Q'$
have to intersect in some correct process $p_{i}$, which has thus sent both
$\COMMIT(v, k, \hash(\val))$ and $\PREPARE(v', k, \hash(\val'))$. Since $v< v'$,
this process $p_{i}$ must have sent the $\COMMIT$ message before the $\PREPARE$
message. Before sending the former, the process set $\lockview[k]$ to $v$
(line~\ref{hotstuff:lock}) and had $\prepcmd[k] = \cmd[k] = \val$. Assume
towards a contradiction that $\val \not= \val'$. Let $v''$ be the first view
after $v$ when $p_i$ assigned $\cmd[k]$ to some $\val'' \not= \val$, so that
$v < v'' \le v'$. Then $p_i$ must have assigned $\cmd[k]$ to $\val''$ at either
line~\ref{hotstuff:assign-log} or line~\ref{hotstuff:assign-nop}. When this
happened, $p_i$ had $\cmd[k] = \prepcmd[k] = \val$ and, by
Proposition~\ref{lemma:hotstuff:view-increase}, $\lockview[k] \ge v > 0$.

By the $\ValidNewState$ check (line~\ref{hotstuff:safety-check}), the leader of
$v''$ must have provided a well-formed prepared certificate $C''$ such that
$\accepted(C'', v''', k, \hash(\val''))$ for $v'''$ such that
$$
v \le \lockview[k] < v''' \le v'' \le v'.
$$
If $p_i$ assigned $\cmd[k]$ to $\val''$ at line~\ref{hotstuff:assign-log}, then
by~(\ref{hotstuff:hyp2}) we get $\val'' = \val$, and above we assumed
$\val'' \not= \val$: a contradiction.  If $p_i$ assigned $\cmd[k]$ to $\val''$
at line~\ref{hotstuff:assign-nop} due to a duplicate value at a position
$k'\not=k$, then $\val'' = \noop$ and $\vcmd[k'] = \val \not= \bot$.
Hence, the leader of $v''$ had to also provide a well-formed prepared
certificate $C''$ such that $\accepted(C'', v'''_0, k', \hash(\val))$ for $v'''_0$
such that
$$
v \le \lockview[k] \le v''' < v'''_0 < v'' \le v'.
$$
Since $\val \not= \val'' = \noop$, by~(\ref{hotstuff:hyp4}) we get $k = k'$: a
contradiction.
Since we reach a contradiction in both cases, we must have $\val = \val'$, as
required.

Assume now that $\val \not= \noop$, $\accepted(C', v', k', \hash(\val))$ and
$\wf(C')$. Since $\committed(C, v, k, \hash(\val))$, a quorum $Q$ of processes
sent $\COMMIT(v, k, \hash(x))$. Since $\accepted(C', v', k', \hash(\val))$, a
quorum $Q'$ of processes sent $\PREPARE(v', k', \hash(\val))$. The quorums $Q$
and $Q'$ have to intersect in some correct process $p_{i}$, which has thus sent
both $\COMMIT(v, k, \hash(\val))$ and $\PREPARE(v', k', \hash(\val))$. When
$p_i$ sent the latter it must have had $\cmd[k'] = \val$. As before, we can also
show that at this moment $p_i$ had $\cmd[k] = \val$. By
Lemma~\ref{lemma:hotstuff:nodupl}, the process starts the view $v'$ with a log
without duplications (except $\noop$s), and does not add duplicate entries due
to the check at line~\ref{alg:castro:safety-check}. Hence, we must have
$k' = k$, as required.
\end{proof}

\begin{corollary}
\hotstuff satisfies Ordering and Integrity.
\end{corollary}
\begin{proof}
The same as the proofs of Corollaries~\ref{thm:pbft:agreement}
and~\ref{thm:pbft-integrity}, but using the corresponding lemmas for \hotstuff
instead of \pbft.
\end{proof}

\subsection{Proof of Liveness for \hotstuff}
\label{sec:liveness-hotstuff}

The proof of liveness is virtually identical to the one for \pbftr with two
exceptions. First, since \hotstuff has an extra phase in its normal path, the
duration of $\timerexecute$ in Lemmas~\ref{thm:timers} and~\ref{thm:pbft-facts}
has to be $>\max\{5\delta, T+4\delta\}$. Second, we need to show that in a view
$v$ with a correct leader, if a correct process $p_i$ receives a $\NEWVIEW$
message $m$ from the leader of $v$, then $\ValidNewState(m)$ holds at
$p_i$. This fact is then used in the proof of Lemma~\ref{thm:timers} to show
that a process accepts the leader's $\NEWVIEW$ message for the view $v$ once it
receives the message and enters $v$. The following lemma states this fact.

\begin{lemma}
\label{lem:hotstuffproposal}
Let $v \geq \B'$ be a view such that $\leader(v)$ is correct, and let $p_i$ be a
correct process that enters $v$. If $p_i$ receives the leader's $\NEWVIEW$
message while in $v$, then $\ValidNewState(m)$ holds at $p_i$.
\end{lemma}
\begin{proof}
The lemma trivially holds if $p_i$ is not locked on a value at any position of
its $\prepcmd$ array when receiving the $\NEWVIEW$ message from the leader of
$v$. We now consider the case when $p_i$ is locked on a value in at least one
position when receiving the $\NEWVIEW$ message. Let $k$ be one of the locked
positions and let $\val = p_i.\prepcmd[k]$ be the value locked and
$v_0 = p_i.\lockview[k] < v$ be the corresponding view. Since $p_i$ locked
$\val$ in $v_0$, it must have previously received messages
$\PRECOMMIT(v_0, k, \hash(\val))$ from a quorum of processes, at least $f+1$ of
which have to be correct. The latter processes must have prepared the value
$\val$ in view $v_0$ at position $k$. When each of these $f+1$ correct processes
enters view $v$, it has $\currview \ge v_0$ and thus sends the corresponding
value and its prepared certificate for the position $k$ in the
$\NEWLEADER(v, \ldots)$ message to $\leader(v)$. The leader is guaranteed to
receive at least one of these messages before making a proposal, since it only
does this after receiving at least $2f+1$ $\NEWLEADER$ messages. Hence, the
leader proposes a value $\val'$ for the position $k$ with a prepared certificate
formed at some view $v' \ge v_0$. Furthermore, if $v' = v_0$, then by
Proposition~\ref{lemma:pbft:singlecmd} we have that $\val' = \val$ and $\val$ is
the only value that can be locked at $k$ and $v_0$ by $p_i$. Thus, $p_i$ will
accept the leader's proposal for the position $k$. Since $k$ was picked
arbitrarily, then $p_i$ will accept the leader's proposal for any locked
position. Hence, the leader's $\NEWVIEW$ message will satisfy $\ValidNewState$
at $p_i$, as required.
\end{proof}

\section{A Liveness Bug in the Byzantine Consensus of Cachin et
  al.~\cite{cachin-book}}
\label{sec:bug}

Cachin et al.'s book~\cite[\S{}5.6.4]{cachin-book} includes an implementation of
Byzantine consensus using an abstraction that, similarly to our SMR
synchronizers, allows clients to flag potential protocol-specific failures. We
now show that their implementation violates the Termination property of
Byzantine consensus, requiring that every correct process eventually decides
some value. Cachin et al.'s Byzantine Consensus algorithm (Alg. 5.19) relies on
a Byzantine Epoch-Change abstraction (Alg. 5.15) which guarantees that
eventually all correct processes enter the same epoch with a correct leader. The
Epoch-Change itself is implemented using a Byzantine Eventual Leader Detector
(Alg. 2.10), which outputs the leader for correct processes to follow. The
Leader Detector considers the current leader faulty if more than $f$ correct
processes have ``complained'' about its behavior via a special call. In
Byzantine Consensus a process complains about the current leader if it fails to
observe a decision within a given time duration. This, however, results in a
problem if some correct processes stop complaining while others are still
unhappy.

To see this, consider an execution of Byzantine Consensus (Alg. 5.19) where all
correct processes enter an epoch $e$ with a faulty leader. The leader may make a
valid proposal to $f+1$ correct processes and withhold it from the remaining $f$
correct processes. Since the consensus algorithm operates based on quorums of
$2f+1$ processes, the faulty processes may then execute the algorithm so that
only the $f+1$ correct processes that received the leader's proposal decide in
epoch $e$. These $f+1$ correct processes will then stop complaining. For the
remaining $f$ correct processes to decide, they need to switch to a new epoch
with a correct leader. But they will not be able to achieve this unless faulty
processes cooperate, because to nominate a different leader, the Leader Detector
requires more than $f$ processes to complain (Alg. 2.10). This breaks the
Eventual Leadership property of Epoch-Change (Module 5.12) and prevents the
remaining $f$ correct processes from ever reaching a decision, violating the
Termination property of Byzantine Consensus. This bug has been confirmed by one
of the book's authors~\cite{cristian-personal}. It is similar to a bug recently
discovered in PBFT's read-only request optimization~\cite{alysson-reads}.

The bug can be easily fixed by reliably broadcasting decisions, as we do in
\pbft (\S\ref{sec:pbft}). But even with this fix, the Eventual Leadership
property of Epoch-Change will remain broken, since correct processes will be
allowed to remain forever in a view with a faulty leader. This shows that 
the specifications of abstractions proposed in~\cite{cachin-book} are not well-suited for
implementing a live Byzantine Consensus. These abstractions are also
underspecified: the implementation of Byzantine Eventual Leader Detector
(Alg. 2.10) will only satisfy its specification (Modules 2.10) if clients use it
in a particular way that has not been formalized. Informally, the authors
require that correct processes eventually cease to complain against a correct
leader and in their proof sketches justify that this will happen because the
processes ``wait long enough for the leader to achieve its goal''. But this
argument is circular, since for processes to stop complaining they need to get
consensus decisions, and for this the Leader Detector has to nominate a correct
leader. Such circular reasoning is known to be unsound for liveness
properties\footnote{M. Abadi and L. Lamport. Conjoining specifications. ACM
  Trans. Program. Lang. Syst., 17(3):507-534, 1995.}.

A part of Cachin et al.'s consensus protocol was also used in the BFT-SMaRt
protocol of Bessani et al.~\cite{bftsmart,bftsmart-thesis}. This was based on an
abstraction of {\em validated and provable (VP) consensus}, which allows its
clients to control when to change the leader via a special {\em VP-Timeout}
call. The overall BFT-Smart protocol appears to be correct, but its liveness
proof sketch suffers from issues with rigor similar to those of Cachin et
al.'s. In particular, the VP-Consensus abstraction is underspecified: the
authors require VP-Consensus to satisfy the Termination property, but to ensure
this, its clients have to use {\em VP-Timeout} in particular ways that have not
been formalized. In particular, clients have to leave enough time in between
{\em VP-Timeout} calls for consensus to decide if the leader is correct; this in
its turn requires knowledge about the time necessary for such a decision. Thus,
Lemma A2 in~\cite[\S{}A]{bftsmart-thesis} cannot hold for an arbitrary
VP-Consensus implementation.

\fi

\end{document}